\documentclass[12pt]{article}
\usepackage{a4wide}
\usepackage{amssymb,amsmath}
\usepackage{graphicx}

\begin{document}
{\renewcommand{\thefootnote}{\fnsymbol{footnote}}
\begin{center}
{\LARGE  Emergent modified gravity:\\[2mm] Polarized Gowdy model on a torus}\\
\vspace{1.5em}
Martin Bojowald\footnote{e-mail address: {\tt bojowald@psu.edu}}
and Erick I.\ Duque   \footnote{e-mail address: {\tt eqd5272@psu.edu}} 
\\
\vspace{0.5em}
Institute for Gravitation and the Cosmos,\\
The Pennsylvania State
University,\\
104 Davey Lab, University Park, PA 16802, USA\\
\vspace{1.5em}
\end{center}
}

\setcounter{footnote}{0}

\begin{abstract}
  New covariant theories of emergent modified gravity exist not only in
  spherically symmetric models, as previously found, but also in polarized
  Gowdy systems that have a local propagating degree of freedom.  Several
  explicit versions are derived here, depending on various modification
  functions. These models do not have instabilities from higher time
  derivatives, and a large subset is compatible with gravitational waves and
  minimally coupled massless matter fields travelling at the same
  speed. Interpreted as models of loop quantum gravity, covariant Hamiltonian
  constraints derived from the covariance conditions found in polarized Gowdy
  systems are more restricted than those in spherical symmetry, requiring new
  forms of holonomy modifications with an anisotropy dependence that has not
  been considered before.  Assuming homogeneous space, the models provide
  access to the full anisotropy parameters of modified Bianchi I dynamics, in
  which case different fates of the classical singularity are realized
  depending on the specific class of modifications.
\end{abstract}

%\tableofcontents

\section{Introduction}

The canonical formulation of spherically symmetric general relativity has
recently been shown \cite{Higher,HigherCov} to allow a larger class of
modifications than is suggested by the more common setting of covariant action
principles. In this framework of emergent modified gravity, it is possible to
couple perfect fluids \cite{EmergentFluid} as well as scalar matter
\cite{EmergentScalar,SphSymmMinCoup} to the new space-time geometries,
including local degrees of freedom in the latter case. Here, we show that it
is also possible to extend spherical symmetry to a polarized Gowdy symmetry
that includes local gravitational degrees of freedom. This extension makes it
possible to study properties of gravitational waves in this new set of
covariant space-time theories.

Building on previous canonical developments, starting with the classic
\cite{Regained} and using more recent contributions \cite{EffLine}, emergent
modified gravity constructs consistent gravitational dynamics and
corresponding space-time geometries by modifying the Hamiltonian constraint of
general relativity and implementing all covariance conditions. A candidate for
the spatial metric of a space-time geometry is provided by the structure
function in the Poisson bracket of two Hamiltonian constraints, which is
required to be proportional to the diffeomorphism constraint as one of the
consistency conditions. Canonical gauge transformations of the candidate
spatial metric must then agree with coordinate transformations in a compatible
space-time geometry, forming the second consistency condition that had been
formulated for the general case and analyzed for the first time in
\cite{HigherCov}. These constructions allow for the possibility that the
spatial metric (or a triad) is not one of the fundamental fields of a
phase-space formulation. It is derived from Hamilton's equations generated by
the constraints and not presupposed, giving it the status of an emergent
geometrical object. This feature is the main difference with standard action
principles in metric or other formulations and makes this approach to modified
gravity more general than previous constructions. Examples of new physical
implications include the possibility of non-singular black-hole solutions
\cite{SphSymmEff,SphSymmEff2}, covariant MOND-like effects \cite{HigherMOND},
and new types of signature change \cite{EmergentSig}.

The constructions of the present paper lead to the first model of emergent
modified gravity that does not obey spherical symmetry. Nevertheless,
spherical symmetry can be realized as a special case, allowing us to draw
conclusions about how generic specific features seen in the more symmetric
context are within a broader setting. An example of interest is the form of
holonomy-type modifications that are often of used in order to model potential
effects from loop quantum gravity. The specific form of these modifications
within a consistent and covariant set of equations is restricted compared with
what had been assumed previously in loop constructions. The extension to
polarized Gowdy models performed here shows that compatible modifications
require significant deviations from what might be suggested by loop quantum
gravity. In particular, the holonomy length for strictly periodic
modifications of the extrinsic-curvature dependence does not directly depend
on the volume or area of a symmetry orbit (all of space in a homogeneous
cosmological model or a sphere at constant radius in black-hole models), but
rather on its anisotropy paremeters. General covariance therefore rules out
the possibility that the holonomy length decreases as space or a spherical
orbit expands, which would be a prerequisite to a nearly constant discreteness
scale that does not increase to macroscopic sizes as the universe
expands. Nevertheless, additional modification functions can be used in order
to implement a dynamical suppression of holonomy modifications on classical
scales, as discussed in detail in \cite{EmergentMubar} for spherically
symmetric models. The traditional picture of models of loop quantum gravity
therefore has to be corrected in order to be compatible with a consistent
space-time geometry. Emergent modified gravity guides the way to a new
understanding by a systematic classification of possible space-time
modifications in canonical form.

In addition, the new gravitational models found here are important in their
own right because they have covariant equations with modifications that do not
require higher-derivative terms and corresponding instabilities
\cite{OstrogradskiProblem}. They are therefore potential alternatives to
general relativity that could be used in comparisons with observations,
provided the symmetry assumptions can be relaxed further. Polarized Gowdy
symmetries constitute a first step in this direction, giving access to some
properties of gravitational waves. In particular, we show that there is a
class of modifications that implies the same propagation speed for
gravitational waves and massless scalar matter travelling on the same
background.

Unlike spherically symmetric models, which have a spatially homogeneous subset
of Kantowski-Sachs models with a single anisotropy parameter, polarized Gowdy
models give full access to the Bianchi I model with two anisotropy
parameters. It is therefore possible to perform a more complete analysis of
the big-bang singularity, which may be avoided depending on the type of
modifications used. As a characteristic property, the classical Kasner
exponents are preserved at large volume, and a non-singular transition from
collapse to expansion happens at the same time for all three spatial
directions. We will present a detailed analysis of these questions in
Section~\ref{sec:Homogeneous spacetime}, after a brief review of canonical and
emergeny modified gravity in Section~\ref{s:Classical} and their application to
polarized Gowdy models in Sections~\ref{sec:Linear combination} and
\ref{s:General} with a summary of different classes of modifications in
Section~\ref{s:Classes}. Implications for covariant holonomy modifications in
models of loop quantum gravity can be found in Section~\ref{s:Hol}.

\section{Classical theory}
\label{s:Classical}

The classical polarized Gowdy system \cite{Gowdy} is defined by space-time
line elements of the form
\begin{equation}
    {\rm d} s^2 = - N^2 {\rm d} t^2 + q_{\theta \theta} ( {\rm d} \theta + N^\theta {\rm d} t )^2
    + q_{x x} {\rm d} x^2
    + q_{y y} {\rm d} y^2
  \label{eq:ADM line element - Gowdy - classical}
\end{equation}
with functions $N$, $N^{\theta}$ and $q_{ab}$ depending only on $t$ and
$\theta$. All three spatial coordinates $x$, $y$ and $\theta$ take values in
the range $[ 0 , 2 \pi)$ for the torus model with spatial slices
$\Sigma\cong T^3 = S^1 \times S^1 \times S^1$.

Equivalently, the spatial
metric components $q_{ab}$ can be parameterized by
\begin{equation} \label{qE}
    q_{\theta \theta} = \frac{E^x E^y}{\varepsilon}
    \ , \hspace{1cm}
    q_{x x} = \frac{E^y}{E^x} \varepsilon
    \ , \hspace{1cm}
    q_{y y} = \frac{E^x}{E^y} \varepsilon
\end{equation}
using the components $E^x$, $E^y$ and $\varepsilon$ of a densitized triad
\begin{equation}
  E_i^a \sigma_i\frac{\partial}{\partial x^a}= \varepsilon
  \sigma_3\frac{\partial}{\partial \theta}+
  E^x\sigma_1\frac{\partial}{\partial x}+ E^y\sigma_2 \frac{\partial}{\partial
    y}
\end{equation}
with Pauli matrices $\sigma_i$.  For some purposes, it is conventional to
write the metric in the diagonal case ($N^{\theta}=0$) in the form
\begin{eqnarray}
    {\rm d} s^2 = e^{2 a} ( - {\rm d} T^2 +  {\rm d} \theta^2 )
    + T \left( e^{2 W} {\rm d} x^2
    + e^{-2 W} {\rm d} y^2 \right)
    \label{eq:ADM line element - Gowdy - classical - conventional variables}
\end{eqnarray}
with a new time coordinate $T$.  This conventional metric is associated to the
canonical metric in a gauge defined by $\varepsilon = T$ and $N=\sqrt{q_{\theta\theta}}$, 
identifying $N^2=q_{\theta \theta} =: e^{2 a}$ and
$W = \ln \sqrt{E^y / E^x}$. If $E^x=E^y$ or $W=0$, the geometry has an
additional rotational symmetry.

\subsection{Canonical formulation}

The densitized-triad components are canonically conjugate to components of
extrinsic curvature, implying canonical pairs  $(K_x , E^x)$, $(K_y , E^y)$,
and $(\mathcal{A} , \varepsilon)$ and
the symplectic structure
\begin{eqnarray}
    \Omega = \frac{1}{\tilde{\kappa}} \int {\rm d} \theta \left( {\rm d} K_x
  \wedge {\rm d} E^x + {\rm d} K_y \wedge {\rm d} E^y + {\rm d} \varepsilon
  \wedge {\rm d} \mathcal{A} \right) 
\end{eqnarray}
with $\tilde{\kappa} = \kappa / (4 \pi^2) = 2 G / \pi$ in terms of Newton's
constant $G$. We will work in units such that $\tilde{\kappa}=1$.

The
classical Hamiltonian and diffeomorphism constraints with a cosmological
constant are \cite{EinsteinRosenAsh}
\begin{eqnarray}\label{eq:Constraints - Gowdy - Classical}
    H &=& -  \frac{1}{\sqrt{E^x E^y \varepsilon}} \bigg[ 
    - \varepsilon E^x E^y \Lambda
    + K_x E^x K_y E^y
    + \left( K_x E^x + K_y E^y \right) \varepsilon \mathcal{A}
    \notag\\
    &&\hspace{1cm}
    + \frac{1}{2} \frac{\varepsilon^2}{E^x E^y} (E^x)' (E^y)'
    + \frac{1}{2} \frac{\varepsilon}{E^x} (E^x)' \varepsilon'
    + \frac{1}{2} \frac{\varepsilon}{E^y} (E^y)' \varepsilon'
    \notag\\
    &&\hspace{1cm}
    - \frac{1}{4} \frac{\varepsilon^2}{(E^y)^2} ((E^y)')^2
    - \frac{1}{4} \frac{\varepsilon^2}{(E^x)^2}((E^x)')^2
    - \frac{1}{4} (\varepsilon')^2
    - \varepsilon \varepsilon''
    \bigg] 
    \nonumber\\
    &=&- \frac{1}{\sqrt{E^x E^y \varepsilon}} \left( - \varepsilon E^x E^y \Lambda
    + E^x K_x E^y K_y
    + \left(E^x K_x + E^y K_y\right) \varepsilon \mathcal{A} \right)
    \nonumber\\
    &&
    - \frac{1}{4} \frac{1}{\sqrt{E^x E^y \varepsilon}} \left( (\varepsilon')^2 - 4 \left(\varepsilon (\ln \sqrt{E^y/E^x} )'\right)^2 \right)
    + \left(\frac{\sqrt{\varepsilon} \varepsilon'}{\sqrt{E^x E^y}}\right)'
    \ ,
    \label{eq:Hamiltonian constraint - Gowdy - Classical}
\end{eqnarray}
and
\begin{eqnarray}
    H_\theta &=& E^x K_x'
    + E^y K_y'
    - \mathcal{A} \varepsilon'
    \label{eq:Diffeomorphism constraint - Gowdy - Classical}
\end{eqnarray}
where the primes are $\theta$ derivatives. 
The smeared constraints have Poisson brackets
\begin{eqnarray}
    \{ H_\theta [N^\theta] , H_\theta [M^\theta] \}&=& - H_\theta [ M^\theta (N^\theta)' - N^\theta (M^\theta)']
    \ , 
    \label{eq:H_t,H_t bracket} \\
    \{ H [N] , H_r [M^\theta] \}&=& - H[M^\theta N'] 
    \ ,
    \label{eq:H,H_theta bracket}
    \\
    \{ H [N] , H[M] \}&=& - H_\theta \left[ q^{\theta \theta} \left( M N' - N M' \right)\right]
    \label{eq:H,H bracket}
\end{eqnarray}
of hypersurface-deformation form, with structure function
$q^{\theta \theta} = \varepsilon/(E^x E^y)$ directly given by the inverse
metric component in the inhomogeneous direction.
From general properties of
canonical gauge systems \cite{LapseGauge,CUP} it then follows that the gauge
transformations for the lapse function $N$ and shift vector $N^{\vartheta}$ are given by
\begin{equation}
    \delta_\epsilon N = \dot{\epsilon}^0 + \epsilon^\theta N' - N^\theta (\epsilon^0)'
    \label{eq:Gauge transformation for lapse - Gowdy}
  \end{equation}
  and
  \begin{equation}
    \delta_\epsilon N^\theta = \dot{\epsilon}^\theta + \epsilon^\theta (N^\theta)' - N^\theta (\epsilon^\theta)' + q^{\theta \theta} \left(\epsilon^0 N' - N (\epsilon^0)' \right)\,.
    \label{eq:Gauge transformation for shift - Gowdy}
\end{equation}

In the classical theory it is clear that the inverse of the structure
function, $q_{\theta\theta}=1/q^{\theta\theta}$, obeys a covariance condition
as a component of the space-time metric. More generally \cite{HigherCov},
covariance conditions can be directly formulated for phase-space functions
such as a structure function in a modified theory. They implement the general
condition that gauge transformations of any candidate space-time metric
component, generated by the canonical constraints, must be of the form of a
Lie derivative by a space-time vector field. Using explicit expressions for
gauge transformations generated by the constraints on a given phase space,
this general condition can be written as a set of partial differential
equations that the constraints have to obey.

These covariance conditions, derived in \cite{HigherCov} for spherical
symmetry, are more complicated for polarized Gowdy models because the
phase space is larger and the line element is a more complicated function of
the phase-space variables.  For the homogeneous component $q_{x x}$ we obtain
the conditions
\begin{equation}
    \frac{1}{E^y} \left( \frac{\partial H}{\partial K_y'} - 2
      \left(\frac{\partial H}{\partial K_y''}\right)' \right) 
    - \frac{1}{E^x} \left( \frac{\partial H}{\partial K_x'} - 2
      \left(\frac{\partial H}{\partial K_x''}\right)' \right) 
    + \frac{1}{\varepsilon} \left( \frac{\partial H}{\partial \mathcal{A}'} -
      2 \left(\frac{\partial H}{\partial \mathcal{A}''}\right)' \right)
    \bigg|_{\rm O.S.} 
    = 0
\end{equation}
and
\begin{equation}
    \frac{1}{E^y} \frac{\partial H}{\partial K_y''}
    - \frac{1}{E^x} \frac{\partial H}{\partial K_x''}
    + \frac{1}{\varepsilon} \frac{\partial H}{\partial \mathcal{A}''}
    \bigg|_{\rm O.S.} = 0
    \ ,
\end{equation}
where ``O.S.'' indicates that the equations are required to hold on-shell,
when constraints and equations of motion are satisfied. For our modified
constraints,  we assume
spatial derivatives up to second order; otherwise there would be additional
terms in these equations. The $x \leftrightarrow y$ exchange symmetry of
the constraint allows us to simplify these on-shell conditions to
\begin{equation}
    \frac{\partial H}{\partial \mathcal{A}'}
    = \frac{\partial H}{\partial \mathcal{A}''}
    = \frac{1}{E^y} \frac{\partial H}{\partial K_y'} - \frac{1}{E^x} \frac{\partial H}{\partial K_x'}
    = \frac{1}{E^y} \frac{\partial H}{\partial K_y''} - \frac{1}{E^x} \frac{\partial H}{\partial K_x''}
    = 0
    \ ,
    \label{eq:Covariance condition on homogeneous components - Gowdy}
\end{equation}
which is clearly satisfied by the classical constraint even off-shell. The
same condition is obtained from the other homogeneous component, $q_{y y}$.

For the inhomogeneous component, the covariance condition reads
\begin{equation}
    \frac{\partial \left(\{ q^{\theta \theta} , H[\epsilon^0]
        \}\right)}{\partial (\epsilon^0)'} \bigg|_{\text{O.S.}} 
    = \frac{\partial \left(\{ q^{\theta \theta} , H[\epsilon^0]
        \}\right)}{\partial (\epsilon^0)''} \bigg|_{\text{O.S.}} 
    = \dotsi
    = 0
    \ ,
\label{eq:Covariance condition on inhomogeneous component - Gowdy}
\end{equation}
which is also satisfied by the classical constraint because it does not
contain any derivatives of $K_x$, $K_y$, or $\mathcal{A}$ that would introduce
a dependence of the Poisson brackets on spatial derivatives of $\epsilon^0$
upon integrating by parts. For this result, it is important to use the
classical property that the structure function $q^{\theta\theta}$ is
independent of the canonical variables conjugate to the triad components. This
property is no longer required in emergent modified gravity.

The gauge transformations of lapse and shift, (\ref{eq:Gauge transformation
  for lapse - Gowdy}) and (\ref{eq:Gauge transformation for shift - Gowdy}),
and the realization of the covariance conditions, \eqref{eq:Covariance
  condition on homogeneous components - Gowdy} and \eqref{eq:Covariance
  condition on inhomogeneous component - Gowdy}, ensure that the space-time
line element \eqref{eq:ADM line element - Gowdy - classical} is invariant, or
the space-time metric $g_{\mu\nu}$ is covariant in the sense that the
canonical gauge transformations of the metric reproduce space-time diffeomorphisms
on-shell: We have
\begin{equation}
    \delta_\epsilon g_{\mu \nu} \big|_{\text{O.S.}} =
    \mathcal{L}_\xi g_{\mu \nu} \big|_{\text{O.S.}}
    \ ,
\end{equation}
where the gauge functions, $(\epsilon^0,\epsilon^\theta)$, on the
left-hand side are related to the 2-component vector generator,
$\xi^\mu = (\xi^t,\xi^\theta)$, of the diffeomorphism on the right-hand side
by
\begin{equation}
    \xi^\mu = \epsilon^0 n^\mu + \epsilon^\theta s^\mu = \xi^t t^\mu + \xi^\theta s^\mu
  \end{equation}
  with components
  \begin{equation}
    \xi^t = \frac{\epsilon^0}{N}
    \quad , \quad
    \xi^\theta = \epsilon^\theta - \frac{\epsilon^0}{N} N^\theta
    \ .
    \label{eq:Diffeomorphism generator projection - Gowdy}
\end{equation}

\subsection{New variables}

It is convenient to perform the canonical transformation
\begin{eqnarray}
    P_{\Bar{W}} = K_x E^x - K_y E^y
    \quad &,&\quad
    \Bar{W} = \ln \sqrt{\frac{E^y}{E^x}}
    \nonumber\\
    \Bar{a} = \sqrt{E^x E^y}
    \quad &,&\quad
    K = \frac{K_x E^x + K_y E^y}{\sqrt{E^x E^y}}
    \label{eq:Gowdy new variables}
\end{eqnarray}
with $\Bar{W}$ and $K$ as the configuration variables, and $P_{\bar{W}}$ and
$\bar{a}$ their respective conjugate momenta. The canonical pair
$(\mathcal{A},\varepsilon)$ is left unchanged by this transformation.

The diffeomorphism constraint in these variables is form-invariant,
\begin{eqnarray}
    H_\theta
    &=&
    E^x K_x'
    + E^y K_y'
    - \mathcal{A} \varepsilon'
    \nonumber \\
    &=&
    \frac{1}{2} E^x \left(\frac{P_{\Bar{W}} + K \sqrt{E^x E^y}}{E^x}\right)'
    + \frac{1}{2} E^y \left(\frac{- P_{\Bar{W}} + K \sqrt{E^x E^y}}{E^y}\right)'
    - \mathcal{A} \varepsilon'
    \nonumber \\
    &=&
    \Bar{a} K'
    + P_{\Bar{W}} \Bar{W}'
    - \mathcal{A} \varepsilon'
    \,,
    \label{eq:Diffeomorphism constraint - Gowdy - Classical - new variables}
\end{eqnarray}
while the Hamiltonian constraint (\ref{eq:Hamiltonian constraint - Gowdy -
  Classical}) reads
\begin{eqnarray}
    H
    &=&
    - \frac{1}{\sqrt{E^x E^y \varepsilon}} \left( - \varepsilon E^x E^y \Lambda
    + E^x K_x E^y K_y
    + \left(E^x K_x + E^y K_y\right) \varepsilon \mathcal{A} \right)
       \label{eq:Hamiltonian constraint - Gowdy - Classical - new variables}
\\
    &&
    - \frac{1}{4} \frac{1}{\sqrt{E^x E^y \varepsilon}} \left( (\varepsilon')^2 - 4 \left(\varepsilon (\ln \sqrt{E^y/E^x} )'\right)^2 \right)
    + \left(\frac{\sqrt{\varepsilon} \varepsilon'}{\sqrt{E^x E^y}}\right)'
    \nonumber\\
    &=& - \sqrt{\varepsilon} \left[
    \Bar{a} \left( - \Lambda
    + \frac{K^2}{4 \varepsilon}
    - \frac{1}{4 \varepsilon} \frac{P_{\Bar{W}}^2}{\Bar{a}^2}
    + K \frac{\mathcal{A}}{\Bar{a}} \right)
    - \varepsilon \frac{(\Bar{W}')^2}{\Bar{a}}
    - \frac{1}{4 \varepsilon} \frac{(\varepsilon')^2}{\Bar{a}}
    + \frac{\Bar{a}' \varepsilon'}{\Bar{a}^2}
    - \frac{\varepsilon''}{\Bar{a}}
    \right]
    \nonumber\\
    &=& - \sqrt{\varepsilon} \left[ 
    \Bar{a} \left( - \Lambda
    + \frac{K^2}{4 \varepsilon}
    + K \frac{\mathcal{A}}{\Bar{a}} \right)
    - \frac{1}{4 \varepsilon} \frac{(\varepsilon')^2}{\Bar{a}}
    + \frac{\Bar{a}' \varepsilon'}{\Bar{a}^2}
    - \frac{\varepsilon''}{\Bar{a}}
    \right]
    + \frac{\sqrt{q^{\theta\theta}}}{2} \left(
    \frac{P_{\Bar{W}}^2}{2 \varepsilon} 
    + 2 \varepsilon (\Bar{W}')^2 \right)\nonumber
 \end{eqnarray}
in these new variables. The first parenthesis resembles the Hamiltonian
constraint of a spherically symmetric model, while the last term expresses
$\Bar{W}$ in the form of a scalar field. This relationship will be discussed
in more detail in Section~\ref{s:SphSymm}.

The space-time metric
\begin{equation}
    {\rm d} s^2 = - N^2 {\rm d} t^2 + q_{\theta \theta} ( {\rm d} \theta + N^\theta {\rm d} t )^2
    + q_{x x} {\rm d} x^2
    + q_{y y} {\rm d} y^2
  \end{equation}
now has the spatial components
  \begin{equation}
    q_{\theta \theta} = \frac{\Bar{a}^2}{\varepsilon}
    \quad , \quad
    q_{x x} = e^{2 \Bar{W}} \varepsilon
    \quad , \quad
    q_{y y} = e^{- 2 \Bar{W}} \varepsilon
    \ .
    \label{eq:ADM line element - Gowdy - new variables}
\end{equation}
The new variables therefore closely resemble the conventional choice used in
(\ref{eq:ADM line element - Gowdy - classical - conventional variables}).

\subsection{Symmetries and observables}

Given a potentially large class of modifications, it is useful to impose
guiding principles such as the preservation of important symmetries of the
classical system. For the models considered here, there are discrete as well
as continuous symmetries.

\subsubsection{Discrete symmetry}
\label{sec:Discrete symmetry: Even parity}

The constraints (\ref{eq:Constraints - Gowdy - Classical}) are symmetric under
the exchange $E^x \leftrightarrow E^y , K_x \leftrightarrow K_y$, while the
full line element (\ref{eq:ADM line element - Gowdy - classical}) has the same
symmetry provided the coordinates are exchanged too, $x \leftrightarrow y$.
The complete discrete transformation is then given by
\begin{equation}
    E^x \leftrightarrow E^y
    \ ,\
    K_x \leftrightarrow K_y
    \ ,\
    x \leftrightarrow y
    \ .
    \label{eq:Discrete symmetry - original variables}
\end{equation}
This important symmetry implies the existence of an $x-y$ plane of wave
fronts, in which the two independent directions are interchangeable (while we
do not have isotropy in this plane unless $E^x=E^y$).  The modified theory
should therefore retain this symmetry as an important characterization of the
polarized Gowdy system.  In the new variables, the discrete transformation
takes the form
\begin{equation}
    P_{\Bar{W}} \to - P_{\Bar{W}}
    \ ,\
    \bar{W} \to - \bar{W}
    \ ,\
    x \leftrightarrow y
    \ ,
    \label{eq:Discrete symmetry - new variables}
\end{equation}
which is a symmetry of the system (\ref{eq:Diffeomorphism constraint - Gowdy - Classical - new variables})--(\ref{eq:ADM line element - Gowdy - new variables}).

\subsubsection{Continuous symmetries and related observables}

\paragraph{Field  observable:}
Another advantage of the new variables is that the constraint
(\ref{eq:Hamiltonian constraint - Gowdy - Classical - new variables}) is
manifestly invariant under the transformation $\bar{W} \to \bar{W} + \omega$
where $\omega$ is a constant.
Therefore, the phase-space functional
\begin{eqnarray}
    G [\omega] = \int {\rm d} \theta\ \omega P_{\bar{W}}
    \ ,
    \label{eq:Gowdy observable - classical}
\end{eqnarray}
is a symmetry generator:
\begin{equation}
    \{G[\omega] , H[N]\} = \{G[\omega] , H_{\theta}[N^{\theta}]\} = 0
\end{equation}
where we neglect boundary terms.

This property in turn implies that $G[\omega]$ is a conserved global charge
because $\dot{G}[\omega] = \{G[\omega] , H[N] + H_{\theta}[N^{\theta}]\} = 0$.
Furthermore, as discussed in \cite{EmergentScalar}, the boundary terms that
survive under the transformation of the local charge take the form
$\dot{G} = - \partial_a J^a$, which takes the form
$\partial_\mu J^\mu = \nabla_\mu J^\mu = 0$ of a covariant conservation law
for a space-time densitized 4-current with components
\begin{eqnarray}
    J^t = G = P_{\bar{W}}
    \ , \hspace{1cm}
    J^a = - \left( N \frac{\partial H}{\partial \bar{W}'} \right)'
    = - 2 \varepsilon^{3/2} \frac{\Bar{W}'}{\Bar{a}}
    \ .
    \label{eq:Gowdy conserved current - Classical}
\end{eqnarray}

\paragraph{Mass observable:}
In the limit of $P_{\bar{W}}=\bar{W}= 0$, the expression
\begin{eqnarray}
    \mathcal{M}
    &=&
    \frac{\sqrt{\varepsilon}}{2} \left(
    K^2
    - \left(\frac{\varepsilon'}{2 \bar{a}}\right)^2
    + \frac{\Lambda \varepsilon}{3} \right)
    \label{eq:Gravitational Dirac observable}
\end{eqnarray}
is a Dirac observable.

\subsection{Analogy with spherical symmetry}
\label{s:SphSymm}

In the new variables, the constraint (\ref{eq:Hamiltonian constraint - Gowdy -
  Classical - new variables}) is close to the spherically symmetric constraint
coupled to a scalar field.
In this subsection we will point out in detail how the two models are related.

In a spherically symmetric model, the space-time line element can always be
written as
\begin{equation}\label{eq:ADM line element - spherical}
    {\rm d} s^2 = - N^2 {\rm d} t^2 + q_{x x}^{\rm sph} ( {\rm d} x + N^x {\rm d} t )^2
    + q_{\vartheta \vartheta}^{\rm sph} {\rm d} \Omega^2
  \end{equation}
  where ${\rm d}\Omega^2={\rm d}\vartheta^2+\sin^2\vartheta {\rm d}\varphi^2$
  in spherical coordinates.
As initially developed for models of loop quantum gravity
\cite{SymmRed,SphSymm,SphSymmHam}, it is convenient to parameterize the
metric components $q_{xx}^{\rm sph}$ and $q_{\vartheta\vartheta}^{\rm sph}$ as
\begin{equation}
    q_{x x}^{\rm sph} = \frac{(E^\varphi)^2}{E^x}
    \quad ,\quad
    q_{\vartheta \vartheta}^{\rm sph} = E^x
    \ , \label{eq:Metric components - clssical spherical}
\end{equation}
where $E^x$ and $E^\varphi$ are the radial and angular densitized-triad
components, respectively. We assume $E^x>0$, fixing the orientation of space.

The canonical pairs for spherically symmetric classical gravity are given by
$(K_\varphi , E^\varphi)$ and $(K_x , E^x)$ where $2K_x$ and $K_{\varphi}$ are
components of extrinsic curvature.  We have a further canonical pair
$(\phi,P_\phi)$ if scalar matter is coupled to the gravitational system. The
basic Poisson brackets are given by
\begin{subequations}
\begin{equation}
    \{ K_x (x) , E^x (y)\} = \{ K_\varphi(x) , E^\varphi (y) \} = \{ \phi(x) ,
  P_\phi (y) \} = \delta (x-y) . 
\end{equation}
\end{subequations}
(Compared with other conventions, our scalar phase-space variables are divided
by $\sqrt{4\pi}$, absorbing the remnant of a spherical integration. We use
units in which Newton's constant, $G$, equals one. This convention is
  formally different from what we are using in Gowdy models, where $2G/\pi$
  equals one. The discrepancy is necessary in order to take into account the
  difference in coordinate areas for the symmetry orbits, given by $4\pi^2$ in
  the toroidal Gowdy model and $4\pi$ in spherical symmetry, as well as the
  varying multiplicity of independent degrees of freedom in the homogeneous
  directions.)

The Hamiltonian constraint is
given by
\begin{eqnarray}
    H^{\rm sph}
    &=&
    - \frac{\sqrt{E^x}}{2} \Bigg[
    E^\varphi  \left(
    \frac{1}{E^x}
    + \frac{K_\varphi^2}{E^x}
    + 4 K_\varphi \frac{K_x}{E^\varphi} \right)
    - \frac{1}{4 E^x} \frac{((E^x)')^2}{E^\varphi}
    + \frac{(E^x)' (E^\varphi)'}{(E^\varphi)^2}
    - \frac{(E^x)''}{E^\varphi}
    \Bigg]
    \nonumber\\
    &&\qquad
    + \frac{1}{2} \left(
    \frac{\sqrt{q^{xx}_{\rm sph}}}{E^x} P_\phi{}^2
    + E^x \sqrt{q^{xx}_{\rm sph}} (\phi')^2 + \sqrt{q_{xx}^{\rm sph}} E^x V (\phi) \right)
    \ ,
    \label{eq:Hamiltonian constraint - spherical symmetry - Scalar field - Classical}
\end{eqnarray}
with a scalar potential $V(\phi)$ (or $\frac{1}{2}V(\phi)$, depending on conventions), and
\begin{eqnarray}
    H_x^{\rm sph}
    &=& E^\varphi K_\varphi' - K_x (E^x)'
    + P_\phi \phi'
    \ ,
    \label{eq:Diffeomorphism constraint - spherical symmetry- Scalar field}
\end{eqnarray}
is the diffeomorphism constraint.
The primes denote derivatives with respect to the radial coordinate $x$, which
is unrelated to the coordinates of the Gowdy model.
These constraints are first class and have Poisson brackets of hypersurface-deformation form,
\begin{subequations}
\label{eq:Hypersurface deformation algebra - spherical - Scalar field1}
\begin{eqnarray}
    \{ H_x^{\rm sph} [N^x] , H_x^{\rm sph}[M^x] \} &=& H_x^{\rm sph} [N^x {M^x}' - {N^x}' M^x]
    \ , \\
    \{ H^{\rm sph} [N] , H_x^{\rm sph} [M^x] \}&=& - H^{\rm sph} [M^r N'] 
    \ , \\
    \{ H^{\rm sph} [N] , H^{\rm sph}[M] \}&=& H_x^{\rm sph} \left[ q^{x
                                              x}_{\rm sph} \left( N M' - N' M \right)\right]
\end{eqnarray}
\end{subequations}
with the structure function $q^{x x}_{\rm sph} = E^x/(E^\varphi)^2$ equal to the
inverse radial component of the space-time metric.

The off-shell gauge transformations for lapse and shift
\begin{equation}\label{eq:Off-shell gauge transformations for lapse and shift - spherical}
    \delta_\epsilon N = \dot{\epsilon}^0 + \epsilon^r N' - N^r (\epsilon^0)' \quad,\quad
    \delta_\epsilon N^r = \dot{\epsilon}^r + \epsilon^r (N^r)' - N^r (\epsilon^r)' + q^{x x}_{\rm sph} \left(\epsilon^0 N' - N (\epsilon^0)' \right)
\end{equation}
together with the realization of covariance conditions for space-time,
\begin{equation}
    \frac{\partial H^{\rm sph}}{\partial K_x'} \bigg|_{\text{O.S.}}
    = \frac{\partial H^{\rm sph}}{\partial K_x''} \bigg|_{\text{O.S.}}
    = \dotsi
    = 0
\label{eq:Covariance condition angular component - spherical}
\end{equation}
and
\begin{eqnarray}
    \frac{\partial \left(\{ q^{x x}_{\rm sph} , H^{\rm sph}[\epsilon^0] \}\right)}{\partial (\epsilon^0)'} \bigg|_{\text{O.S.}}
    = \frac{\partial \left(\{ q^{x x}_{\rm sph} , H^{\rm sph}[\epsilon^0] \}\right)}{\partial (\epsilon^0)''} \bigg|_{\text{O.S.}}
    = \cdots
    = 0
    \ ,
\label{eq:Covariance condition - spherical}
\end{eqnarray}
which have been derived in \cite{HigherCov} and are clearly satisfied, ensures
that the line element \eqref{eq:ADM line element - spherical} is
invariant. Its coefficients then form a covariant metric tensor in the sense
that its canonical gauge transformations reproduce space-time diffeomorphisms
on-shell:
\begin{equation}
    \delta_\epsilon g_{\mu \nu} \big|_{\text{O.S.}} =
    \mathcal{L}_\xi g_{\mu \nu}
    \,.
\end{equation}
The gauge functions $(\epsilon^0,\epsilon^r)$ on the left-hand side are related to
the 2-component vector generator $\xi^\mu = (\xi^t,\xi^r)$ of the
diffeomorphism on the right-hand side by
\begin{equation}
    \xi^\mu = \epsilon^0 n^\mu + \epsilon^x s^\mu = \xi^t t^\mu + \xi^x s^\mu
\end{equation}
with
\begin{equation}
    \xi^t = \frac{\epsilon^0}{N}
    \quad , \quad
    \xi^x = \epsilon^x - \frac{\epsilon^0}{N} N^x
    \ .
    \label{eq:Diffeomorphism generator projection - spherical}
\end{equation}

In addition, the realization of the covariance conditions for matter \cite{EmergentScalar},
\begin{equation}
    \frac{\partial H^{\rm sph}}{\partial P_\phi'}
    = \frac{\partial H^{\rm sph}}{\partial P_\phi''}
    = \cdots
    = 0
    \ ,
    \label{eq:Matter covariance condition - spherical}
\end{equation}
ensures that the matter field transforms as a space-time scalar in the
sense that its canonical gauge transformations reproduce space-time
diffeomorphisms on-shell:
\begin{equation}
    \delta_\epsilon \phi \big|_{\text{O.S.}} =
    \mathcal{L}_{\xi} \phi
    \ .
    \label{eq:Covariance condition of phi - modified1}
\end{equation}

Finally, we note that the spherically symmetric system in the absence of a
scalar potential permits the global symmetry generator
\begin{eqnarray}
    G^{\rm sph} [\alpha] = \int {\rm d} x\ \alpha P_\phi
    \ ,
    \label{eq:Symmetry generator of real scalar field - spherical}
\end{eqnarray}
with constant $\alpha$.
The gravitational mass observable is
\begin{equation}
    \mathcal{M}^{\rm sph} =
    \frac{\sqrt{E^x}}{2} \left(
    1 + K_{\varphi}^2
    - \left(\frac{(E^x)'}{2 E^\varphi}\right)^2 - \frac{\Lambda}{3} E^x \right)
    \,,
    \label{eq:Gravitational observable - spherical}
\end{equation}
which is a Dirac observable in the vacuum limit, $\phi = P_\phi = 0$.

We are now ready to identify the analog relationship between the Gowdy and the
spherically symmetric models. 
By inspection, we find that relabeling the canonical pairs according to
\begin{equation}
    (\mathcal{A} , \varepsilon) \to (K_x , E^x)
    \quad,\quad
    (K, \Bar{a}) \to (K_\varphi , E^\varphi)
    \quad,\quad
    (\Bar{W}, P_{\Bar{W}}) \to (\phi , P_\phi)
    \label{eq:Gowdy-Spherical analog phas-space}
\end{equation}
turns the Gowdy constraints (\ref{eq:Diffeomorphism constraint - Gowdy - Classical - new variables}) and (\ref{eq:Hamiltonian constraint - Gowdy - Classical - new variables}) into
\begin{eqnarray}
    H
    &=& - \sqrt{E^x} \Bigg[ 
    E^\varphi \left( \frac{K_\varphi^2}{4 E^x}
    + K_\varphi \frac{K_x}{E^\varphi} \right)
    - \frac{1}{4 E^x} \frac{((E^x)')^2}{E^\varphi}
    + \frac{(E^\varphi)' (E^x)'}{(E^\varphi)^2}
    - \frac{(E^x)''}{E^\varphi}
    \Bigg]
    \nonumber\\
    &&\qquad
    + \frac{\sqrt{q^{\theta\theta}}}{4 E^x} P_\phi{}^2
    + E^x \sqrt{q^{\theta\theta}} (\phi')^2
    \ ,
    \label{eq:Hamiltonian constraint - Gowdy-Spherical analog}
\end{eqnarray}
and
\begin{equation}
    H_\theta = E^\varphi K_\varphi' - K_x (E^x)' + P_\phi \phi'
    \ ,
\end{equation}
respectively,
and the Gowdy metric components (\ref{eq:ADM line element - Gowdy - new variables}) become
\begin{equation}
    q_{\theta \theta} = \frac{(E^\varphi)^2}{E^x}
    \quad , \quad
    q_{x x} = e^{2 \phi} E^x
    \quad , \quad
    q_{y y} = e^{- 2 \phi} E^x
    \ .
    \label{eq:Metric components - Gowdy-Spherical analog}
\end{equation}

Up to a few numerical factors, all the terms in the Gowdy constraint
(\ref{eq:Hamiltonian constraint - Gowdy-Spherical analog}) match those of the
spherically symmetric constraint (\ref{eq:Hamiltonian constraint - spherical
  symmetry - Scalar field - Classical}) except for the first and last terms of
the latter: The inverse triad $1/E^x$ and the scalar potential $V$ do
not appear in the former.
In the general modified constraints of the spherically symmetric system
\cite{EmergentScalar} these two terms are just the classical limits of
modification functions that are in principle allowed to be different from what
the classical dynamics requires. (The scalar potential may always be set equal
to zero in order to define a specific model, while the $1/E^x$-term is a
special case of the dilaton potential that would be a free function of $E^x$
if the spherically symmetric model were generalized to 2-dimensional dilaton
gravity.)  We thus conclude that the modified Gowdy constraint is equivalent
to the spherically symmetric one up to the choice of modification functions.
In arriving at this conclusion, we have implicitly assumed that all of the
conditions imposed in \cite{EmergentScalar} to obtain the general constraints
apply to the Gowdy system as well.
We now show that this is indeed the case.

The conditions for the modified theory considered in \cite{EmergentScalar} are
the following ones.
1) Anomaly-freedom, 2) covariance conditions, 3) existence of a conserved
matter-current, and 4) existence of a vacuum mass observable.
Anomaly-freedom of the Gowdy model takes exactly the same form as in spherical
symmetry because the structure function of the former, (\ref{eq:Metric
  components - Gowdy-Spherical analog}), is equivalent to that of the latter,
(\ref{eq:Metric components - clssical spherical}).
The covariance conditions of the Gowdy system, (\ref{eq:Covariance condition on
  homogeneous components - Gowdy})--(\ref{eq:Covariance condition on
  inhomogeneous component - Gowdy}), are also equivalent to the spherically
symmetric ones, (\ref{eq:Covariance condition - spherical}) and
(\ref{eq:Matter covariance condition - spherical}), upon using the analog identification (\ref{eq:Gowdy-Spherical analog phas-space}).
Finally, the Gowdy symmetry generator (\ref{eq:Gowdy observable - classical})
is identical to the spherically symmetric one (\ref{eq:Symmetry generator of
  real scalar field - spherical}) under the same identification, while the
Dirac observables (\ref{eq:Gravitational Dirac observable}) and
(\ref{eq:Gravitational observable - spherical}) are identical up to one term
that in the modified theory is given by the classical limit of a modification
function.
Therefore, all the classes of general modified constraints obtained in
\cite{EmergentScalar} are also the results of applying these conditions to the
Gowdy system, if we only invert the correspondence.

In fact, there is one additional condition of the Gowdy system that the spherically symmetric one does not have: The discrete symmetry discussed in Section~\ref{sec:Discrete symmetry: Even parity}.
In this sense the Gowdy system is more restricted than the spherically
symmetric one.
Therefore, we can simply take the final results of \cite{EmergentScalar} and impose the discrete symmetry on them.

\section{Linear combination of the constraints}
\label{sec:Linear combination}

Before discussing general modifications, an interesting restricted case is
given by linear combinations of the classical constraints with suitable
phase-space dependent coefficients. By construction, this class of 
theories preserves the classical constraint surface but modifies gauge
transformations and the dynamics, implying a non-classical emergent space-time
metric if the covariance conditions are fulfilled.

\subsection{Anomaly-free linear combination}

We define a new candidate for the Hamiltonian constraint as
\begin{eqnarray}
    H^{\text{(new)}} = B H^{\text{(old)}} + A H_\theta
    \label{eq:Linear combinations of constraints - generic - new variables}
\end{eqnarray}
with suitable phase-space functions $A$ and $B$,
using the original constraints $H^{\text{(old)}}$ and $H_\theta$ of the
classical theory and keeping the latter unchanged.  We consider the
phase-space dependence $B= B(K , \varepsilon , \Bar{W})$. (For more details
about the individual steps, see \cite{EmergentScalar}.)

The Leibniz rule allows us to reduce the new bracket
$\{H^{({\rm new})}[\epsilon_1],H^{({\rm new})}[\epsilon_2]\}$ to Poisson
brackets of the old constraints with the functions $A$ and $B$. Using the
derivative terms of the classical constraints, Poisson brackets relevant for
the anomaly-freedom and covariance conditions can be expanded by finitely many
terms with different orders of $\theta$-derivatives of the gauge
functions. For instance, we can write
\begin{eqnarray}
    \{ B , H^{\text{(old)}} [\bar{\epsilon}^0] \} 
    \big|_{\text{O.S.}}
    &=&
    {\cal B} \bar{\epsilon}^0
    + {\cal B}^\theta  \partial_\theta \bar{\epsilon}^0
    \big|_{\text{O.S.}}
    \label{eq:Transformation of B - Geometric condition - Linear combination - new variables}
\end{eqnarray}
with
\begin{eqnarray}
    {\cal B}^\theta 
    &=&
    \sqrt{\varepsilon} \frac{\varepsilon'}{\Bar{a}^2} \frac{\partial B}{\partial K}\,.
\end{eqnarray}
Anomaly freedom of the new constraints, using hypersurface-deformation
brackets for the old constraints, then requires
\begin{eqnarray}
    A  &=& - {\cal B}^\theta 
    = - \sqrt{\varepsilon} \frac{\varepsilon'}{\Bar{a}^2} \frac{\partial B}{\partial K}
    \label{eq:A coefficient - Linear combination - new variables}
\end{eqnarray}
because any term in $\{H^{({\rm new})}[\epsilon_1],H^{({\rm
    new})}[\epsilon_2]\}$ that is not proportional to the diffeomorphism
constraint must cancel out.

Similarly, we can write
\begin{eqnarray}
    \{ A , H^{\text{(old)}} [\bar{\epsilon}^0] \}
    &=&
    {\cal A}^0 \bar{\epsilon}^0
    + {\cal A}^{\theta} \partial_\theta \bar{\epsilon}^0
    \label{eq:Transformation of A - Geometric condition - Linear combination - new variables}
\end{eqnarray}
in which anomaly-freedom together with (\ref{eq:A coefficient - Linear
  combination - new variables}) implies 
\begin{eqnarray}
    {\cal A}^{\theta}
    &=&
    - \frac{\varepsilon}{\Bar{a}^2} \left( K \frac{\partial B}{\partial K}
    + \frac{(\varepsilon')^2}{\Bar{a}^2} \frac{\partial^2 B}{\partial K^2} \right)\,.
\end{eqnarray}
Using this new function, the bracket
\begin{eqnarray}
    \{ \mathcal{A}^{\theta} , H^{\text{(old)}} [\bar{\epsilon}^0] \} &=&
    \Lambda^0 \bar{\epsilon}^0
    + \Lambda^{\theta} \partial_\theta \bar{\epsilon}^0
\end{eqnarray}
requires
\begin{eqnarray}
    \Lambda^{\theta}
    &=&
    - \frac{\varepsilon^{3/2} \varepsilon'}{\Bar{a}^4} \left( \frac{\partial B}{\partial K} + 3 K \frac{\partial^2 B}{\partial K^2}
    + \frac{(\varepsilon')^2}{\Bar{a}^2} \frac{\partial^3 B}{\partial K^3} \right)
    \,.
\end{eqnarray}

The new structure function,
\begin{eqnarray}
    q^{\theta \theta}_{\rm (new)} &=&
    B^2 q^{\theta \theta} + B {\cal A}^\theta
    \,,
    \label{eq:New structure function - Linear combination - new variables}
\end{eqnarray}
follows from collecting all terms in the Poisson bracket of two Hamiltonian
constraints that can contribute to the diffeomorphism constraint.

\subsection{Covariant modified theory}

Using the new structure function as an inverse spatial metric, the covariance
condition is given by
\begin{eqnarray}
    \mathcal{C}
    &\equiv&
    \Lambda^{\theta}
    - B^{-1} {\cal B}^\theta {\cal A}^{\theta}
    \big|_{\text{O.S.}}
    \nonumber\\
    &=&
    - \frac{\varepsilon^{3/2} \varepsilon'}{\Bar{a}^4} \left( \frac{\partial B}{\partial K}
    + 3 K \frac{\partial^2 B}{\partial K^2}
    + \frac{(\varepsilon')^2}{\Bar{a}^2} \frac{\partial^3 B}{\partial K^3} \right)
    \nonumber\\
    &&
    + \frac{\varepsilon^{3/2} \varepsilon'}{\Bar{a}^4} B^{-1} \frac{\partial B}{\partial K} \left( K \frac{\partial B}{\partial K}
    + \frac{(\varepsilon')^2}{\Bar{a}^2} \frac{\partial^2 B}{\partial K^2} \right)
    = 0
    \ .
    \label{eq:Covariance condition - Gowdy - Linear combination - new variables}
\end{eqnarray}
We separate this condition into derivative terms,
\begin{eqnarray}
    \mathcal{C}
    &=&
    \mathcal{C}_{\varepsilon} \varepsilon'
    + \mathcal{C}_{\varepsilon \varepsilon \varepsilon} (\varepsilon')^3\,,
\end{eqnarray}
which must vanish individually.
The equation $\mathcal{C}_{\varepsilon} = 0$ implies
\begin{equation}
    K \left( \frac{\partial B}{\partial K} \right)^2 + B \left(K
      \frac{\partial^2 B}{\partial K^2} - \frac{\partial B}{\partial K}
    \right) = 0
  \end{equation}
  and is solved by
  \begin{equation} \label{B1}
 B = c_1 \sqrt{c_2 \pm K^2} 
    \,.
\end{equation}
The equation  $\mathcal{C}_{\varepsilon\varepsilon\varepsilon} = 0$ implies
\begin{equation}
    B \frac{\partial^3 B}{\partial K^3} + 3 \frac{\partial B}{\partial K} \frac{\partial^2 B}{\partial K^2}
    = 0
  \end{equation}
  and is solved by
  \begin{equation}
    B= \tilde{c}_1 \sqrt{ \tilde{c}_2 \pm K^2 + \tilde{c}_3 K}
    \,.
\end{equation}

In these solutions,  $c_i$ and $\tilde{c}_i$ are free functions of $\Bar{W}$ and $\varepsilon$.
Their mutual consistency requires
\begin{eqnarray} \label{eq:Covariant linear combination - classical - vacuum}
    B_s (K , \Bar{W} , \varepsilon) &=& \lambda_0 \sqrt{ 1 - s \lambda^2 K^2}
    \ ,
\end{eqnarray}
which then implies
\begin{eqnarray}\label{eq:Covariant linear combination - classical - vacuum2}
    A_s &=&
    \lambda_0 \sqrt{\varepsilon} \frac{\varepsilon'}{\Bar{a}^2} \frac{s \lambda^2 K}{\sqrt{ 1 - s \lambda^2 K^2}}
\end{eqnarray}
such that
\begin{eqnarray} \label{qnew}
    q^{\theta \theta}_{\text{(new)}} &=& 
    \lambda_0^2 \left( 1 + \frac{s \lambda^2}{1- s \lambda^2 K^2} \frac{(\varepsilon')^2}{\Bar{a}^2}\right) \frac{\varepsilon}{\Bar{a}^2}
    \,.
\end{eqnarray}
There are two remaining free functions,
$\lambda_0 = \lambda_0 (\Bar{W} , \varepsilon)$ and
$\lambda = \lambda(\Bar{W} , \varepsilon)$, and we have separated the sign
$s=\pm1$ from the original solution, (\ref{B1}).  Reality requires that
$1 - s \lambda^2 K^2 \geq 0$, which may place an upper bound on $K$
depending on $s$ and $\lambda$.  Finally, the discrete symmetry requires that
both modification functions are even in $\Bar{W}$:
$\lambda_0 (\Bar{W} , \varepsilon) = \lambda_0 (-\Bar{W} , \varepsilon)$ and
$\lambda (\Bar{W} , \varepsilon) = \lambda (-\Bar{W} , \varepsilon)$.

Since we now have complete solutions for $A$ and $B$, we can derive the
modified Hamiltonian constraint from (\ref{eq:Linear combinations of
  constraints - generic - new variables}): 
\begin{eqnarray}
    H^{\rm (new)}
    &=&
    - \lambda_0 \sqrt{\varepsilon} \sqrt{ 1 - s \lambda^2 K^2} \Bigg[ 
    \Bar{a} \left( \frac{K^2}{4 \varepsilon}
    - \frac{1}{4 \varepsilon} \frac{P_{\Bar{W}}^2}{\Bar{a}^2}
    + \frac{\mathcal{A}}{\Bar{a}} K \right)
    - \varepsilon \frac{(\Bar{W}')^2}{\Bar{a}}
    - \frac{1}{4 \varepsilon} \frac{(\varepsilon')^2}{\Bar{a}}
    + \frac{\Bar{a}' \varepsilon'}{\Bar{a}^2}
    - \frac{\varepsilon''}{\Bar{a}}
    \nonumber\\
    &&
    - \frac{\varepsilon'}{\Bar{a}^2} \frac{s \lambda^2 K}{1 - s \lambda^2 K^2} \left( \Bar{a} K' + P_{\Bar{W}} \Bar{W}'
    - \mathcal{A} \varepsilon' \right)
    \Bigg]
    \ .
    \label{eq:Covariant Hamiltonian - Linear combination - new variables}
\end{eqnarray}
The case $s=1$, together with a reality condition for the constraint, implies a curvature bound $K<1 / \lambda$.
The case $s=-1$, implies a possibility of signature change where
$q^{xx}_{({\rm new})}$ changes sign. (The inverse spatial metric is then
determined by the absolute value of (\ref{eq:Linear combinations of
  constraints - generic - new variables}).)

\subsubsection{Canonical transformations}

For the case $s = 1$, a natural canonical transformation is
\begin{eqnarray} \label{canonical}
    K \to
    \frac{\sin (\lambda K)}{\lambda}
    \quad &,&
    \quad
    \Bar{a} \to \frac{\Bar{a}}{\cos (\lambda K)}
    \nonumber\\
    \Bar{W} \to
    \Bar{W}
    \quad &,&
    \quad
    P_{\Bar{W}} \to P_{\Bar{W}}
    - \frac{\Bar{a}}{\cos (\lambda K)} \frac{\partial}{\partial \Bar{W}} \left(\frac{\sin (\lambda K)}{\lambda}\right)
    \nonumber\\
    \varepsilon \to \varepsilon
    \quad &,&
    \quad
    \mathcal{A} \to
    \mathcal{A}
    + \frac{\Bar{a}}{\cos (\lambda K)} \frac{\partial}{\partial \varepsilon} \left(\frac{\sin (\lambda K)}{\lambda}\right)
\end{eqnarray}
under which the modified Hamiltonian constraint becomes
\begin{eqnarray}
    H^{\rm (c)}
    &=&
    - \lambda_0 \sqrt{\varepsilon} \Bigg[ 
    \Bar{a} \bigg( \frac{1}{4 \varepsilon} \frac{\sin^2 (\lambda K)}{\lambda^2}
    - \frac{1}{4 \varepsilon} \cos^2 (\lambda K) \left( \frac{P_{\Bar{W}}}{\Bar{a}}
    - \frac{\partial \ln \lambda}{\partial \Bar{W}} K
    + \frac{\tan (\lambda K)}{\lambda} \frac{\partial \ln \lambda}{\partial \Bar{W}} \right)^2
    \nonumber\\
    &&
    + \frac{\sin (2 \lambda K)}{2 \lambda} \left( \frac{\mathcal{A}}{\Bar{a}}
    + \frac{\partial \ln \lambda}{\partial \varepsilon} K
    - \frac{\tan (\lambda K)}{\lambda} \frac{\partial \ln \lambda}{\partial \varepsilon} \right) \bigg)
    \nonumber\\
    &&
    - \varepsilon \frac{(\Bar{W}')^2}{\Bar{a}} \cos^2 (\lambda K)
    + \frac{\Bar{a}' \varepsilon'}{\Bar{a}^2} \cos^2 (\lambda K)
    - \lambda^2 \frac{\sin (2 \lambda K)}{2 \lambda} \left( \frac{P_{\Bar{W}}}{\Bar{a}} - \frac{\partial \ln \lambda}{\partial \Bar{W}} K \right) \frac{\Bar{W}' \varepsilon'}{\Bar{a}}
    \nonumber\\
    &&
    - \left( \frac{\cos^2 (\lambda K)}{4 \varepsilon}
    - \lambda^2 \frac{\sin (2 \lambda K)}{2 \lambda} \left( \frac{\mathcal{A}}{\Bar{a}}
    + \frac{\partial \ln \lambda}{\partial \Bar{\varepsilon}} K \right) \right) \frac{(\varepsilon')^2}{\Bar{a}}
    - \frac{\varepsilon''}{\Bar{a}} \cos^2 (\lambda K)
    \Bigg]
    \ .
    \label{eq:Covariant Hamiltonian - Linear combination - new holonomy variables}
\end{eqnarray}
A second canonical transformation,
\begin{eqnarray}
    K \to
    \frac{\Bar{\lambda}}{\lambda} K
    \quad &,&
    \quad
    \Bar{a} \to \frac{\lambda}{\Bar{\lambda}} \Bar{a}
    \nonumber\\
    \Bar{W} \to
    \Bar{W}
    \quad &,&
    \quad
    P_{\Bar{W}} \to P_{\Bar{W}}
    + \frac{\Bar{\lambda}}{\lambda} \Bar{a} \frac{\partial \ln \lambda}{\partial \Bar{W}} K
    \nonumber\\
    \varepsilon \to \varepsilon
    \quad &,&
    \quad
    \mathcal{A} \to
    \mathcal{A}
    - \frac{\Bar{\lambda}}{\lambda} \Bar{a} \frac{\partial \ln \lambda}{\partial \varepsilon} K
\end{eqnarray}
with constant $\Bar{\lambda}$, renders the modified Hamiltonian constraint periodic in $K$:
\begin{eqnarray}
    H^{\rm (cc)}
    &=&
    - \lambda_0\frac{\Bar{\lambda}}{\lambda} \sqrt{\varepsilon} \Bigg[ 
      \frac{\Bar{a}}{4 \varepsilon} \frac{\sin^2 (\Bar{\lambda} K)}{\Bar{\lambda}^2}
    - \frac{\Bar{a}}{4 \varepsilon} \cos^2 (\lambda K) \left( \frac{P_{\Bar{W}}}{\Bar{a}}
    + \frac{\tan (\Bar{\lambda} K)}{\Bar{\lambda}} \frac{\partial \ln \lambda}{\partial \Bar{W}} \right)^2
    \nonumber\\
    &&
    + \frac{\sin (2 \Bar{\lambda} K)}{2 \Bar{\lambda}}  \left( \mathcal{A}
    - \Bar{a}\frac{\tan (\Bar{\lambda} K)}{\Bar{\lambda}} \frac{\partial \ln
       \lambda}{\partial \varepsilon} \right) \nonumber\\
  &&
    - \varepsilon \cos^2 (\Bar{\lambda} K)
     \frac{(\Bar{W}')^2}{\Bar{a}} 
    +\left( \frac{\cos^2 (\Bar{\lambda} K)}{\Bar{\lambda}} \frac{\partial \lambda}{\partial \Bar{W}} - \Bar{\lambda}^2 \frac{\sin (2 \Bar{\lambda} K)}{2 \Bar{\lambda}} \frac{P_{\Bar{W}}}{\Bar{a}} \right) \frac{\Bar{W}' \varepsilon'}{\Bar{a}}
    \nonumber\\
    &&
    -  \left( \frac{\cos^2 (\Bar{\lambda} K)}{4 \varepsilon} \left(1 - 4 \varepsilon \frac{\partial \ln \lambda}{\partial \varepsilon} \right)
    - \Bar{\lambda}^2 \frac{\sin (2 \Bar{\lambda} K)}{2 \Bar{\lambda}} \frac{\mathcal{A}}{\Bar{a}} \right) \frac{(\varepsilon')^2}{\Bar{a}}\nonumber\\
  &&  + \cos^2 (\Bar{\lambda} K) \left( \frac{\Bar{a}' \varepsilon'}{\Bar{a}^2}
    - \frac{\varepsilon''}{\Bar{a}} \right)
    \Bigg]
    \ .
    \label{eq:Covariant Hamiltonian - Linear combination - new holonomy variables - Periodic}
\end{eqnarray}
(The term $(\partial\ln\lambda/\partial\epsilon)K$ in (\ref{eq:Covariant
  Hamiltonian - Linear combination - new holonomy variables}) then
disappears.)  Unlike the phase-space coordinates in (\ref{eq:Covariant
  Hamiltonian - Linear combination - new variables}), the holonomy-like
coordinates of (\ref{eq:Covariant Hamiltonian - Linear combination - new
  holonomy variables}) imply a finite constraint at the curvature bound,
implying a dynamics that can cross such a hypersurface of maximum curvature.

The holonomy-like object
\begin{equation} \label{sinK}
  \sin (\bar{\lambda} K) = \sin \left(\bar{\lambda} (E^x K_x + E^y K_y) / \sqrt{E^x E^y}
  \right)
\end{equation}
in the periodic version of the constraint always requires a non-trivial
dependence on the densitized triads, in contrast to what appears in
holonomy-like terms of spherically symmetric models, or of a restricted Gowdy
system in which $E^x=E^y$ and $K_x=K_y$. In the full polarized Gowdy model,
some densitized-triad dependence always remains even if the initial function
$\lambda$, which may depend on $\varepsilon$ as well as $\bar{W}$, has been replaced by a constant
$\bar{\lambda}$ using a canonical transformation. The specific phase-space
function in (\ref{sinK}) can be related to the $(x,y)$-contribution to the
trace of the momentum tensor $K_a^i$ canonically conjugate to $E^a_i$, given
by $K_a^iE^a_i/\sqrt{|\det E|}=e^a_iK_a^i$. In general, however, the
expression in emergent modified Gowdy models is not equal to the trace of
extrinsic curvature in the resulting emergent space-time for two
reasons. First, the phase-space expressions $E^a_i$ and $K_a^i$ have
modified geometrical meanings compared with the classical densitized triad and
extrinsic curvature of spatial slices because the geometry is determined by
the emergent metric. Secondly, the momentum tensor $K_a^i$ with components
that appear in (\ref{sinK}) has been altered by several canonical
transformations applied in our derivations. Relating the $K$-terms in
(\ref{eq:Covariant Hamiltonian - Linear combination - new holonomy variables -
  Periodic}) or (\ref{eq:Covariant Hamiltonian - Linear
  combination - new holonomy variables}) to traditional holonomy modifications
in models of loop quantum gravity therefore requires some care.

\subsubsection{Interpretation as holonomy modifications}
\label{s:Hol}

Any appearance of triad components in holonomy-like terms in models of loop
quantum gravity is usually motivated as a volume or area dependence of the
coordinate length of a holonomy used to construct the Hamiltonian
constraint. In particular, dynamical solutions lead to large symmetry orbits,
such as all of space in homogeneous models of an expanding universe or
spherical orbits in non-rotating black-hole models. As a consequence,
extrinsic-curvature components, given by linear combinations of time
derivatives of the metric or triad components, can be large even in classical
regimes. Their appearance in holonomies is then in danger of violating the
classical limit on large length scales. This problem can be solved in an
ad-hoc manner by using a length parameter for holonomies that decreases with
the size of increasing symmetry orbits, such that holonomy modifications are
negligible even when some extrinsic-curvature or connection components become
large. Heuristically, such a dependence can be motivated by lattice refinement
\cite{InhomLattice}, relating the holonomy length to a lattice structure in
space that is being subdivided as the symmetry orbit expands, maintaining
sufficiently short geometrical lengths of its edges.

Comparing with this motivation, the specific form of holonomy-like terms of
the form (\ref{sinK}) found here, required for covariance, is crucially
different: The coefficient functions of $K_x$ and $K_y$ can both be expressed
in terms of $E^x/E^y=\sqrt{q_{yy}/q_{xx}}=e^{-2\bar{W}}$, which describes the geometrical
anisotropy in the $x-y$ plane but is independent of its area
$\sqrt{q_{xx}q_{yy}}=\varepsilon$. Analyzing the general form of potential
physical implications of this difference requires us to perform a detailed
analysis of canonical transformations used here to arrive at the expression
(\ref{sinK}).

In this context, it is useful to consider possible forms and interpretations
of holonomy modifications for models of loop quantum gravity in the strictly
isotropic context \cite{IsoCosmo,LivRev}, in which spatial homogeneity
eliminates the non-trivial covariance conditions. (See also
\cite{EmergentMubar} for a related discussion in spherical symmetry.)
Extrinsic curvature (or a connection with its associated holonomies) reduced
to isotropy has a single independent component, $k$, canonically conjugate to
the independent densitized-triad component $p$. (We assume $p>0$, fixing the
orientation of space.) Classically, using the scale factor $a$, we have
$k\propto\dot{a}$ and $p\propto a^2$. Holonomies for ${\rm U}(1)$, or suitable
components of holonomies for ${\rm SU}(2)$, are then of the form
$\exp(i\ell k)$ with the coordinate length $\ell$ of a spatial curve, derived
from the general ${\cal P}\exp(\int A_a^i\sigma_i{\rm d}x^a)$ for an isotropic
$A_a^i\propto \delta_a^i$, with generators $\sigma_i$ of the gauge group. The
geometrical length of this curve in an expanding universe increases like
$\ell a$ and may therefore reach macroscopic values after a suitable amount of
time. Similarly, $k\propto \dot{a}=aH$ with the Hubble parameter $H$ is an
approximately linear function of $a$ in a universe dominated by dark energy or
during inflation. The exponent $\ell k$ is then large in a macroscopic
universe, such that modifications would be noticeable on low curvature scales
and contradict cosmological observations.

This problem can be solved by using a coordinate length or holonomy parameter
$\ell\propto a^{-1}\propto p^{-1/2}$, such that the geometrical length is
constant in an expanding universe. The relevant phase-space function
$\exp(i\bar{\ell} k/\sqrt{p})$, with a constant $\bar{\ell}$, then depends on
extrinsic-curvature and densitized-triad components.  It is easier to quantize
this expression if one first applies a canonical transformation that turns
$k/\sqrt{p}$ into a basic canonical variable. Classically, this ratio is
proportional to the Hubble parameter $H$, and the map from $(k,p)$ to $H$ can
be completed to a canonical transformation by using the volume $V\propto a^3$
of some region in space, whose precise form does not matter thanks to
homogeneity and isotropy. It is then possible to quantize $\exp(i\bar{\ell}H)$
to a simple translation operator in $V$.

Different versions of holonomy modifications are obtained by introducing
periodic functions depending on different variables, such as $k$ or $H$. The
classical contribution to the isotropic Hamiltonian constraint can be written
as $\sqrt{p}k^2=p^{3/2}(k/\sqrt{p})^2\propto VH^2$. Holonomy modifications may
then be introduced for $k$ or $H$ (or any function of the form $p^qk$ with
some exponent $q$), leading to dynamically inequivalent modifications of the
form ${\cal H}_1=\sqrt{p} \sin^2(\bar{\ell}k)/\bar{\ell}^2$ and
${\cal H}_2=V\sin^2(\bar{\ell}H)/\bar{\ell}^2$, respectively. The latter can be
transformed back to $k$-variables, implying a term proportional to
$p^{3/2}\sin^2(\bar{\ell}k/\sqrt{p})/\bar{\ell}^2$ in which the decreasing length
scale $\ell=\bar{\ell}/\sqrt{p}$ appears. Independently of canonical
transformations, the different types of holonomy modifications can also be
identified by analyzing equations of motion for small $\bar{\ell}$. From
${\cal H}_1$, we obtain $\dot{p}\propto \sqrt{p}k$ or $k\propto
\dot{p}/\sqrt{p}\propto \dot{a}$, while ${\cal H}_2$ implies $\dot{V}\propto
VH$ or $H\propto\dot{V}/V\propto\dot{a}/a$. Therefore, we do not have to know
which canonical transformations may have been applied in order to determine how
a given classical or modified constraint implies small or large values of
holonomy modifications in classical regimes.

In isotropic models, the appearance of a scale-dependent holonomy length can
be seen in two alternative ways: A dependence on the scale factor may directly
appear in periodic functions, as in $\sin(\bar{\ell}k/\sqrt{p})$, or it may be
implied by equations of motion that tell us whether an expression such as $H$
in $\exp(i\bar{\ell}H)$ equals the classical basic phase-space variable $k$ in
the limit of small $\bar{\ell}$, or a different function such as $H$ in which
the potential growth of $k$ as a function of $a$ in some dynamical solutions
is reduced.

More generally, the different status of a modification with
scale-factor dependent $\ell$ can be seen in coefficients of the Hamiltonian
constraint. In isotropic models, a holonomy modification can be implemented by
directly replacing the classical $k$ in the Hamiltonian constraint with
$\ell^{-1}\sin(\ell k)$. For constant $\ell=\bar{\ell}$, the $k$-independent
coefficient of this term retains its classical dependence on $p$. If $\ell$
depends on $p$, or if $H$ is used instead of $k$, the $p$-dependence of the
coefficient is modified along with the $k$-dependence. From the point of view
of canonical structures, there is no difference between $(k,p)$ and $(H,V)$ if
the relationship between $k$ or $H$ and classical extrinsic curvature is
ignored (or unknown if one considers a generic modified theory). A modification of the form
\begin{equation}
  \frac{\sin(\ell
    k_1)}{\ell}=\frac{\sin(\bar{\ell}k_2)}{\ell}=\frac{\bar{\ell}}{\ell}
  \frac{\sin(\bar{\ell}k_2)}{\bar{\ell}} 
\end{equation}
with triad-dependent
$\ell/\bar{\ell}$, such that the map from $k_1$ to $k_2$ is part of a
canonical transformation with $\ell k_1=\bar{\ell}k_2$, can therefore be
interpreted in two different ways, depending on whether $k_1$ or $k_2$ is
closely related to classically reduced extrinsic curvature. If $k_1$ is
extrinsic curvature, we have a triad-dependent holonomy length $\ell$, and the
small-$k_1$ limit reproduces the classical dependence of the coefficients
because $\ell^{-1}\sin(\ell k_1)= k_1(1+O(\ell^2 k_1^2))$. If $k_2$ is
extrinsic curvature, we have constant holonomy length $\bar{\ell}$, and the
classical triad-dependent coefficients of $k_2$ are modified because
\begin{equation} \label{k2}
\frac{\sin(\bar{\ell}k_2)}{\ell}=\frac{\bar{\ell}}{\ell}
k_2(1+O(\bar{\ell}^2k_2^2))=k_1(1+O(\bar{\ell}^2k_2^2))\,.
\end{equation}
Instead of reducing the growing holonomy length in an expanding universe, the
model is made compatible with the classical limit, producing the same $k_1$ to
leading order, by modifying the triad-dependent coefficients of $k_2$-holonomy
terms in the Hamiltonian constraint by factors of $\bar{\ell}/\ell$.

However,
this classical limit, assuming small $\bar{\ell}k_2$, is in general only formal because
it may not be guaranteed that this product is indeed small in expected
classical regimes, such as a large isotropic universe. The limit is suitable
as a classical one if $k_2=H$, but not if $k_2=k$. In isotropic models, the
$H$-variable is therefore preferred. Therefore, $k_1$ rather than $k_2$ can be
identified with extrinsic curvature in the classical limit, necessitating the application of a
non-constant holonomy function $\lambda$. Whether a canonical variable behaves like $k$ or
like $H$ (or possibly a different version) follows from equations of motion
generated by the modified Hamiltonian constraint.

The possibility of applying canonical transformations in isotropic models is
comparable to some of the steps in our derivation of covariant modifications
of polarized Gowdy models. We have constant holonomy parameters
$\bar{\lambda}$ or $\bar{\ell}$ in one form, and triad-dependent functions
$\lambda$ or $\ell$ in another one. In each case, both versions, if they are
related by (\ref{canonical}) or the canonical transformation that includes the
mapping between $H$ and $k$, are dynamically equivalent. But the two versions are not
equivalent if the holonomy function, constant or non-constant, multiplies the
same phase-space function without applying a canonical transformation. Since
the general form of a modified theory is defined only by its Hamiltonian
constraint and does not contain an independent specification of what canonical
transformations may have been used compared with the standard classical phase
space, we should consider equations of motion in order to determine whether
holonomy modifications depending on $K$ in a polarized Gowdy model can include an
area-dependent holonomy length. 

For non-constant $\lambda/\bar{\lambda}$, the $\varepsilon$-dependence of the
coefficients in (\ref{eq:Covariant Hamiltonian - Linear combination - new
  holonomy variables - Periodic}) differs from the classical one in the limit
of $\bar{\lambda}\to0$. As in (\ref{k2}), these terms signal deviations of $K$
from the original component of extrinsic curvature: The equation of motion for
$\varepsilon$ (which is canonically conjugate to ${\cal A}$) is given by
\begin{equation}
  \dot{\varepsilon}=\{\varepsilon,H^{(cc)}\}=-\lambda_0\frac{\bar{\lambda}}{\lambda}
  \sqrt{\epsilon}\: \frac{\sin(2\bar{\lambda}K)}{2\bar{\lambda}} (1+\bar{\lambda}^2(\varepsilon')^2/\bar{a}^2)
\end{equation}
and implies
\begin{equation} \label{Kepsdot}
  K\sim-\frac{\lambda}{\bar{\lambda}\lambda_0}
  \frac{\dot{\varepsilon}}{\sqrt{\varepsilon}}
\end{equation}
for small $\bar{\lambda}$ if we assume
$\lambda=\bar{\lambda}h(\varepsilon)$ with a $\bar{\lambda}$-independent
holonomy function $h(\varepsilon)$. (In some classes of modified theories,
$h$ may also depend on the anisotropy parameter $\bar{W}$.)
The formal classical limit
requires small $\bar{\lambda}K$, but a specific regime in which classical
behavior is expected may well imply large $\bar{\lambda}K$ if the area
$\varepsilon$ of symmetry orbits is large. If one chooses a $\lambda$ that
decreases with $\varepsilon$ sufficiently quickly, (\ref{Kepsdot}) implies
that the corresponding $K$ of the modified theory increases less strongly than
in a model with constant $\lambda=\bar{\lambda}$. 
It is easier to study
the classical limit if the inverse of the canonical transformation
(\ref{canonical}) is applied, such that all holonomy terms now depend on
$\lambda K$ with an explicit decreasing coefficient $\lambda$ as a function of $\varepsilon$.
As shown in the transition from a
Hamiltonian constraint of the form (\ref{eq:Covariant Hamiltonian - Linear
  combination - new holonomy variables - Periodic}) to an expression
(\ref{eq:Covariant Hamiltonian - Linear combination - new holonomy
  variables}), $\varepsilon$-dependent modifications $\lambda/\bar{\lambda}$
of coefficients in the Hamiltonian constraint are then replaced with
holonomy-like terms with an $\varepsilon$-dependent function $\lambda$.

In isotropic models, these two versions, given by triad-dependent coefficients
and triad-dependent holonomy length, respectively, are equivalent. In polarized Gowdy
models, in which modifications are strongly restricted by covariance
conditions, only the first viewpoint is available if a strict definition of
holonomy modifications as periodic functions is used: Only (\ref{eq:Covariant
  Hamiltonian - Linear combination - new holonomy variables - Periodic}), in
which the coefficient functions are modified in their
$\varepsilon$-dependence, is periodic in $K$, while (\ref{eq:Covariant
  Hamiltonian - Linear combination - new holonomy variables}), in which the
function $\lambda(\varepsilon)$ appears in some of the holonomy terms, also
contains non-periodic contributions linear in $K$ such as
$K \partial\ln\lambda/\partial\varepsilon$. Effects of an $\varepsilon$-dependent holonomy
length can therefore be inferred only indirectly when equations of motion are
used, turning it into an on-shell property. Assigning an
$\varepsilon$-dependent holonomy length directly to off-shell properties of
the Hamiltonian constraint, as in isotropic models, is not possible unless one
weakens the strict periodicity condition on holonomy modifications.

Another difference between isotropic and polarized Gowdy models appears in the
specific form (\ref{sinK}) interpreted in terms of components $K_x$ and $K_y$
of the momentum that appear in the phase-space function $K$.  The only triad
dependence allowed in this combination refers to anisotropy in the
$(x,y)$-plane rather than its area. This specific dependence, just as
properties of how an $\varepsilon$-dependent $\lambda$ may appear in holonomy
modifications, is implied by general covariance. The anisotropy dependence of
holonomy modifications is therefore unavoidable, and unlike
$\lambda(\varepsilon)$ it cannot be moved to coefficient functions. Moreover,
holonomy modifications can only be implemented for the specific combination of
$K_x$ and $K_y$ given by (\ref{sinK}), but not separately for the two
components $K_x$ and $K_y$ because all modified constraints allowed by
covariance depend polynomially on the second phase-space variable,
$P_{\bar{W}}$, that together with $K$ represents $K_x$ and $K_y$ after our
first canonical transformation.

Therefore, unlike in spherically symmetric models, the Hamiltonian constraint
is not built out of basic holonomy operators that depend only on momentum
components canonically conjugate to the densitized triad. There is always a
necessary triad dependence given by the specific form of $K$ that may appear
in periodic terms as a linear combination of $K_x$ and $K_y$ with
triad-dependent coefficients, derived from the covariance conditions. In loop
quantum gravity, curvature (or connection) components and the triad are
instead separated into basic holonomy and flux operators, which were used as
building blocks of the first proposed operators for the Hamiltonian constraint
\cite{AnoFree}. More recent versions
\cite{AnoFreeWeak,AnoFreeWeakDiff,ConstraintsG} use triad-dependent shift
vectors in order to construct detailed properties of hypersurface deformations
from operators, which is somewhat reminiscent of but conceptually unrelated to
the triad dependence of holonomy-type expressions found here.

\section{General modified theory}
\label{s:General}

Linear combinations of the classical constraints with phase-space dependent
coefficients have revealed interesting properties of possible modifications of
polarized Gowdy models. More generally, one may expect that individual terms
in the Hamiltonian constraint can receive independent modifications. We now
analyze this possibility within a setting of effective field theory in which
we expand a generic Hamiltonian constraint in derivatives up to
second order. The resulting expressions then determine gravitational theories
of polarized Gowdy models compatible with the symmetry of general covariance,
taking into account the possibility that the space-time metric is not
fundamental but rather emergent. New modifications are then possible even at
the classical order of derivatives.

\subsection{Constraint ansatz and the emergent space-time metric}

We consider modifications to the Gowdy system with phase-space variables
$(\bar{W} , P_{\Bar{W}})$, $(K , \Bar{a})$, and $(\mathcal{A} , \varepsilon)$.
If we modify the Hamiltonian constraint, then the constraint brackets
\eqref{eq:H_t,H_t bracket}--\eqref{eq:H,H bracket} determine the
inhomogeneous component of the spatial metric via
$\tilde{q}_{\theta \theta} = 1 / \tilde{q}^{\theta \theta}$, while the homogeneous components
of the metric cannot be obtained in this way because they do not appear in
the structure functions.
The emergent space-time line element is then given by
\begin{equation}
    {\rm d} s^2 = - N^2 {\rm d} t^2 + \tilde{q}_{\theta \theta} ( {\rm d} \theta + N^\theta {\rm d} t )^2
    + \alpha_\varepsilon (\varepsilon) e^{2 f_W (\varepsilon, \Bar{W})} {\rm d} x^2
    + \alpha_\varepsilon (\varepsilon) e^{- 2 f_W (\varepsilon, \Bar{W})} {\rm d} y^2
    \ ,
\label{eq:ADM line element - Gowdy - modified}
\end{equation}
with $\tilde{q}_{\theta \theta}$ to be determined by anomaly-freedom of the
hypersurface deformation brackets, while we have partially chosen the form of
the homogeneous components $\tilde{q}_{x x}$ and $\tilde{q}_{yy}$ based on
their classical forms. We will discuss the free functions
$\alpha_{\varepsilon}$ and $f_W$ in due course. (If the structure function is
negative in some regions, the inhomogeneous metric component is determined by
the inverse of its absolute value, while $-N^2{\rm d}t^2$ is replaced by
$-\sigma N^2{\rm d}t^2$ where $\sigma$ is the sign of the structure function
relative to the classical function,
making the 4-dimensional line element Euclidean in regions where
$\sigma=-1$. For more details, see \cite{HigherCov}.)

We consider the following ansatz for the Hamiltonian constraint:
\begin{eqnarray}
    \tilde{H} &=& a_0
    + e_{\Bar{W} \Bar{W}} (\Bar{W}')^2
    + e_{\Bar{a} \Bar{a}} (\Bar{a}')^2
    + e_{\varepsilon \varepsilon} (\varepsilon')^2
    + e_{\Bar{W} \Bar{a}} \Bar{W}' \Bar{a}'
    + e_{\Bar{W} \varepsilon} \Bar{W}' \varepsilon'
    + e_{\Bar{a} \varepsilon} \Bar{a}' \varepsilon'
    \notag\\
    &&
    + p_{\Bar{a} \varepsilon} K' \varepsilon'
    + r_{\Bar{a} \Bar{a}} \Bar{a}' K'
    + e_{2 \Bar{a}} \Bar{a}''
    + p_{2 \Bar{a}} K''
    + e_{2 \varepsilon} \varepsilon''
    \label{eq:Hamiltonian constraint ansatz - Gowdy}
\end{eqnarray}
where $a_0$, $e_{i j}$, $e_{2 i}$, $p_{i j}$, and $p_{2 i}$ are all functions
of the phase-space variables, but not of their derivatives. For the sake of
tractability, we have omitted some terms that would be possible at second
order in derivatives, such as terms containing $\bar{W}''$.  Spatial
derivatives of $\mathcal{A}$ and $P_{\bar W}$ have been omitted in
anticipation of covariance condition (\ref{eq:Covariance condition on
  homogeneous components - Gowdy - modified}), to be discussed shortly, which
requires the constraint to be independent of them.  Because of the discrete
symmetry $\tilde{H} (\Bar{W},P_{\Bar{W}}) = \tilde{H} (-\Bar{W},-P_{\Bar{W}})$,
we see that all the functions obey this even symmetry, except for
$e_{\Bar{W} \varepsilon}$ which should be odd.

Starting from this constraint ansatz we will obtain the conditions for it to
satisfy the hypersurface-deformation brackets, \eqref{eq:H_t,H_t
  bracket}--\eqref{eq:H,H bracket}, with a possibly modified structure
function, $\tilde{q}^{\theta \theta}$.
We will then apply the covariance conditions in order to make sure that the
new structure function can play the role of an inverse metric component in space-time.

\subsubsection{Canonical transformations I}

In order to obtain distinct classes of Hamiltonian constraints for possible
modified theories, it is crucial that we factor out canonical transformations
that preserve the diffeomorphism constraint.  If we do not take this extra
care we risk obtaining equivalent versions of the same theory that differ only
in a choice of the phase-space coordinates.  Two constraints differing only by
a canonical transformation will look different and even the space-time metric
will do so too kinematically, but they in fact describe the same physical
system.

We will therefore consider the following set of canonical transformations that preserves the diffeomorphism constraint:
\begin{subequations}
\label{eq:Diffeomorphism-constraint-preserving canonical transformations - Spherical - general}
\begin{eqnarray}
    &&\bar{W} = f_c^{\bar{W}} (\varepsilon , \tilde{\bar{W}})
    \quad,\quad
    P_{\bar{W}} = \tilde{P}_{\bar{W}} \left( \frac{\partial f_c^{\bar{W}}}{\partial \tilde{\bar{W}}} \right)^{-1}
    - \tilde{\bar{a}} \frac{\partial f_c^{K}}{\partial \tilde{\bar{W}}} \left( \frac{\partial f_c^{K}}{\partial \tilde{K}} \right)^{-1}
    \ ,
    \label{eq:Diffeomorphism-constraint-preserving canonical transformations - W - Gowdy - general}
    \\
    &&K = f_c^{K} (\varepsilon , \tilde{\bar{W}}, \tilde{K})
    \quad,\quad
    E^\varphi = \tilde{E}^\varphi \left( \frac{\partial f_c^{K}}{\partial \tilde{K}} \right)^{-1}
    \ ,
    \label{eq:Diffeomorphism-constraint-preserving canonical transformations - K - Gowdy - general}
    \\
   && {\cal A} = \frac{\partial (\alpha_c^2 \varepsilon)}{\partial \varepsilon} \tilde{\cal A}
    + \tilde{E}^\varphi \frac{\partial f_c^{K}}{\partial \varepsilon} \left( \frac{\partial f_c^{K}}{\partial \tilde{K}} \right)^{-1}\!\!\!\!\!
    + \tilde{P}_{\bar{W}} \frac{\partial f_c^{\bar{W}}}{\partial \varepsilon} \left( \frac{\partial f_c^{\bar{W}}}{\partial \tilde{\bar{W}}} \right)^{-1}\!\!
    ,\;
    \tilde{\varepsilon} = \alpha_c^2 (\varepsilon) \varepsilon
    ,
    \label{eq:Diffeomorphism-constraint-preserving canonical transformations - A - Gowdy - general}
\end{eqnarray}
\end{subequations}
where the new phase-space variables are written with a tilde.  A
transformation with $f_c^K=\tilde{K}$,
$f_c^{\bar{W}} (\varepsilon,\tilde{\bar{W}})$, and $\alpha_c (\varepsilon)$
can always be used to transform the homogeneous components of the metric in
(\ref{eq:ADM line element - Gowdy - modified}) from potentially modified
expressions to their classical ones $\tilde{q}_{xx}=\varepsilon e^{2\bar{W}}$
and $\tilde{q}_{yy}=\varepsilon e^{-2\bar{W}}$.  If we fix the classical form
for these components, the residual canonical transformations are given by
\begin{subequations}\label{eq:Diffeomorphism-constraint-preserving canonical transformations - Gowdy - residual}
\begin{eqnarray}
    &&\bar{W} = \tilde{\bar{W}}
    \quad,\quad
    P_{\bar{W}} = \tilde{P}_{\bar{W}}
    - \tilde{\bar{a}} \frac{\partial f_c^{K}}{\partial \tilde{\bar{W}}} \left( \frac{\partial f_c^{K}}{\partial \tilde{K}} \right)^{-1}
    \ ,
    \label{eq:Diffeomorphism-constraint-preserving canonical transformations - W - Gowdy - residual}
    \\
    &&K = f_c^{K} (\varepsilon , \tilde{\bar{W}}, \tilde{K})
    \quad,\quad
    E^\varphi = \tilde{E}^\varphi \left( \frac{\partial f_c^{K}}{\partial \tilde{K}} \right)^{-1}
    \ ,
    \label{eq:Diffeomorphism-constraint-preserving canonical transformations - K - Gowdy - residual}
    \\
   && {\cal A} = \tilde{\cal A}
    + \tilde{E}^\varphi \frac{\partial f_c^{K}}{\partial \varepsilon} \left( \frac{\partial f_c^{K}}{\partial \tilde{K}} \right)^{-1}
    \quad,\quad
    \tilde{\varepsilon} = \varepsilon
    \label{eq:Diffeomorphism-constraint-preserving canonical transformations - A - Gowdy - residual}
\end{eqnarray}
\end{subequations}
where the new phase-space variables are again written with a tilde.

Under the previous canonical transformation, the emergent space-time metric simplifies to
\begin{equation}
    {\rm d} s^2 = - N^2 {\rm d} t^2 + \tilde{q}_{\theta \theta} ( {\rm d} \theta + N^\theta {\rm d} t )^2
    + \varepsilon e^{2 \Bar{W}} {\rm d} x^2
    + \varepsilon e^{- 2 \Bar{W}} {\rm d} y^2
    \ ,
\end{equation}
while the constraint ansatz (\ref{eq:Hamiltonian constraint ansatz - Gowdy}) does not acquire any new derivative terms.

\subsubsection{Anomaly-freedom and covariance conditions}

Starting with a Hamiltonian constraint of the form (\ref{eq:Hamiltonian
  constraint ansatz - Gowdy}) we impose anomaly-freedom by requiring that,
together with the unmodified diffeomorphism constraint, it reproduces the
hypersurface-deformation brackets (\ref{eq:H_t,H_t bracket})-(\ref{eq:H,H
  bracket}) up to a potentially modified structure function:
\begin{eqnarray}
    \{ H_\theta [N^\theta] , H_\theta [M^\theta] \}&=& - H_\theta [ M^\theta (N^\theta)' - N^\theta (M^\theta)']
    \ , 
    \label{eq:H_t,H_t bracket-modified} \\
    \{ \tilde{H} [N] , H_\theta [M^\theta] \}&=& - \tilde{H}[M^\theta N'] 
    \ ,
    \label{eq:H,H_theta bracket-modified}
    \\
    \{ \tilde{H} [N] , \tilde{H} [M] \}&=& - H_\theta \left[ \tilde{q}^{\theta \theta} \left( M N' - N M' \right)\right]
    \label{eq:H,H bracket-modified}
    \ .
\end{eqnarray}
Doing so restricts the functions in (\ref{eq:Hamiltonian constraint ansatz -
  Gowdy}) by a set of partial differential equations for the modification
functions. The same procedure reveals the dependence of the structure function
$\tilde{q}^{\theta \theta}$ on the phase-space variables.  Furthermore, the
modified brackets (\ref{eq:H_t,H_t bracket-modified})-(\ref{eq:H,H
  bracket-modified}) imply gauge transformation of the shift
vector (\ref{eq:Gauge transformation for shift - Gowdy}) according to
\begin{equation}
    \delta_\epsilon N^\theta = \dot{\epsilon}^\theta + \epsilon^\theta (N^\theta)' - N^\theta (\epsilon^\theta)' + \tilde{q}^{\theta \theta} \left(\epsilon^0 N' - N (\epsilon^0)' \right) ,
    \label{eq:Gauge transformation for shift - Gowdy - modified}
\end{equation}
which now involves the modified structure function as it should if it were to
play the role of a component of the emergent space-time metric. 

Once the structure function has been obtained from the imposition of
anomaly-freedom, we have the full expression of a candidate space-time metric
as in
(\ref{eq:ADM line element - Gowdy - modified}).  Because the homogeneous
metric components retain their classical forms, the covariance conditions
(\ref{eq:Covariance condition on homogeneous components - Gowdy}) remain
unchanged. In the new phase-space variables they are given by
\begin{equation}
    \frac{\partial \tilde{H}}{\partial \mathcal{A}'}
    = \frac{\partial \tilde{H}}{\partial \mathcal{A}''}
    = \frac{\partial \tilde{H}}{\partial P_{\bar{W}}'}
    = \frac{\partial \tilde{H}}{\partial P_{\bar{W}}''}
    = 0
    \ .
    \label{eq:Covariance condition on homogeneous components - Gowdy - modified}
\end{equation}
The inhomogeneous component (\ref{eq:Covariance condition on inhomogeneous
  component - Gowdy}) turns into highly non-trivial conditions,
\begin{equation}
    \frac{\partial \left(\{ \tilde{q}^{\theta \theta} , \tilde{H}[\epsilon^0] \}\right)}{\partial (\epsilon^0)'} \bigg|_{\text{O.S.}}
    = \frac{\partial \left(\{ \tilde{q}^{\theta \theta} , \tilde{H}[\epsilon^0] \}\right)}{\partial (\epsilon^0)''} \bigg|_{\text{O.S.}}
    = \dotsi
    = 0
    \,.
\label{eq:Covariance condition on inhomogeneous component - Gowdy2}
\end{equation}

In addition to general covariance, we will require that the modified system
retains the types of conserved quantities of the classical theory.   We therefore impose
the preservation of the symmetry generator $G[\omega]$ in (\ref{eq:Gowdy observable -
  classical}), such that it commutes with the modified constraint,
\begin{equation}
    \{G[\omega] , \tilde{H}[N]\} = 0
    \ .
\end{equation}
We will also demand that for $P_{\bar{W}},\bar{W}\to 0$ a
gravitational Dirac observable exists with (\ref{eq:Gravitational Dirac
  observable}) as its classical limit.

\subsubsection{Canonical transformations II and additional guiding conditions}

The imposition of anomaly-freedom, covariance, and the gravitational
symmetries all restrict the generic Hamiltonian constraint
(\ref{eq:Hamiltonian constraint ansatz - Gowdy}) by providing a large set of
partial differential equations.  These equations can be considerably
simplified by a choice of phase-space coordinates, fixing the residual
canonical transformations (\ref{eq:Diffeomorphism-constraint-preserving
  canonical transformations - Gowdy - residual}).

So far, all of the conditions we impose in the new variables are identical to
those we chose when coupling scalar matter in spherical symmetry
\cite{EmergentScalar}.  While these conditions are quite restrictive, they
still do not allow a complete exact solution of the partial differential
equations for modification functions. We therefore refer to additional
conditions, most of which have also been used in the scalar case.
\begin{description}
\item[Classical $\bar{W}$ limit.]  One such condition applied in
  \cite{EmergentScalar} was the compatibility of the constraint with a limit
  in which the corresponding class of modifications contains models with the
  classical equations of motion for the scalar matter, corresponding to the
  Klein-Gordon equation on a curved emergent space-time.  We can apply the
  same condition to the Gowdy model, thanks to its correspondence with the
  spherically symmetric system, by imposing compatibility of the
  constraint with a limit in which the class of modifications contains models
  with the classical equations of motion for $\bar{W}$ on an emergent
  background $\tilde{q}_{\theta\theta}$.
\item[Classical constraint surface as a limit.]  Another additional condition
  considered in \cite{EmergentScalar} was the compatibility of the constraint
  with a non-trivial limit in which the constraint surface took its classical
  form.  This is the case if there is a limit in which the constraint contains
  the modifications from linear combinations of the classical constraints, as
  derived in the previous section.
\item[Classes of constraints.]  With these conditions, we are in the position
  to obtain explicit expressions for the modified Hamiltonian constraint.
  The conditions of anomaly-freedom, covariance, implementation of symmetries,
  and the factoring out of canonical transformations imply a set of
  differential equations that can be solved exactly if the additional
  conditions just described are considered.

  However, if we implement the essential conditions we are left with some
  ambiguities.  If one is not interested in the classical $\bar{W}$ limit,
  several of these ambiguities are removed, but one modification function
  remains unresolved.  The vanishing of this function then leads to the
  compatibility of the classical constraint surface as a limit, thus
  describing our first class of constraints.

  On the other hand, using a non-trivial choice for this modification function
  leads us to another class of constraints that is no longer compatible with
  the classical constraint surface as a limit, nor with the classical
  $\bar{W}$ limit. Lacking a positive characterization of these models, we
  simply call this set of modified theories the class of the
  second kind.

  Finally, a third class of constraints can be obtained by imposing
  compatibility with the classical $\bar{W}$ limit.
  
  These are precisely the three classes of constraints obtained in
  \cite{EmergentScalar} for the spherically symmetric system coupled to scalar
  matter.  We can then simply import the results and reinterpret the
  Hamiltonian constraint by its correspondence with the Gowdy system.  The
  modified theory allows for modification functions that can be redefined and
  adapted to the Gowdy system.
\item[Discrete symmetry.]  The Gowdy system has one further symmetry that is
  not obvious in the
  spherically symmetric system coupled to scalar matter.  This is the
  discrete symmetry $P_{\bar{W}} \to - P_{\bar{W}} , \bar{W} \to - \bar{W}$.
  We will implement this symmetry in the classes of constraints imported from
  \cite{EmergentScalar} and, therefore, obtain slightly simpler expressions.
\end{description}

\subsection{Emergent modified gravity as a basis for quantization}

Our generic Hamiltonian constraint contains only up to second-order spatial
derivatives and uses the classical phase space, which implies that the
equations of motion do not have higher time derivatives.  Theories of this
form, even though they are modified compared with general relativity, may
therefore be considered classical gravitational systems that can be used as a
basis for canonical quantization. Several additional conditions are then
useful for different procedures of finding suitable constraint operators.

\subsubsection{Partial Abelianization}
\label{sec:Partial Abelianization}

Gravitational theories with a description as space-time geometry require
constraints that generate hypersurface deformations. The presence of structure
functions is then a well-known obstacle toward quantization because an
operator-valued structure function implies severe ordering problems in
commutators of the constraints. This problem can be simplified if the original
constraints can be replaced by linear combinations that replace the structure
function by a constant, and perhaps setting it equal to zero in a partial
Abelianization. In spherical symmetry, such a procedure has been proposed in
\cite{LoopSchwarz}, and then generalized in \cite{HigherCov} by making it
fully local. 

For a systematic derivation of partial Abelianizations we make use of the
procedure described in Section~\ref{sec:Linear combination}, introducing a new
phase-space function as a linear combination of the (now already modified)
Hamiltonian constraint and the classical diffeomorphism constraint that
replaces the Hamiltonian constraint of hypersurface-deformation brackets. The
new constraints therefore have brackets that differ from the classical
gravitational ones, and their gauge transformations do not
correspond to hypersurface deformations. However, they have the same
constraint surface as the original system, which can therefore be turned into
a quantum description by this procedure.

We will make use of definition (\ref{eq:Linear
  combinations of constraints - generic - new variables}) for the constraint function
\begin{eqnarray}
    \tilde{H}^{\rm (A)} = B \tilde{H} + A H_\theta
    \,.
    \label{eq:Abelian constraint - new variables}
\end{eqnarray}
Poisson brackets of $B$ and $A$ with the old Hamiltonian constraint are given
by (\ref{eq:Transformation of B - Geometric condition - Linear combination -
  new variables}) and (\ref{eq:Transformation of A - Geometric condition -
  Linear combination - new variables}), and the latter
is related to the former by
\begin{eqnarray}
    A = - {\cal B}^\theta(B)
\end{eqnarray}
as in (\ref{eq:A coefficient - Linear combination - new variables}).
The only difference with the procedure in Section~\ref{sec:Linear combination}
is that we are not seeking a new covariant modified theory, but rather a partial
Abelianization of the brackets of $H^{\rm (A)}$ and $H_\theta$.
Therefore, we impose the condition that the new structure function
(\ref{eq:New structure function - Linear combination - new variables})
vanishes:
\begin{eqnarray}
    \tilde{q}^{\theta\theta}_{\rm (A)} &=&
    B^2 \tilde{q}^{\theta \theta} + B {\cal A}^\theta
    = 0
    \ .
    \label{eq:Abelianization condition structure function - new variables}
\end{eqnarray}
We will apply this condition to the three classes of constraints derived
below.

\subsubsection{Point holonomies}

In \cite{EmergentScalar}, it was possible to include point holonomies of the
scalar field $\phi$, given by periodic modification functions depending on
this variable. Like partial Abelianizations, this property may be useful for
quantizations because some of the basic fields can be represented by bounded
operators, akin to a loop quantization.

Invoking the correspondence between spherical symmetry and the polarized Gowdy
model, this result can be translated to point holonomies of $\bar{W}$ in the
latter case.  Recall that the relation between this variable and the
original phase-space degrees of freedom is given by $\bar{W} = \ln \sqrt{E^y / E^x}$.
This function depends on densitized-triad components rather than their
momenta, classically related to extrinsic curvature or a connection, and the
dependence is logarithmic. 

Given the logarithmic dependence on triad components instead of linear
combinations of extrinsic curvature, periodic modification functions of
$\bar{W}$, or polymerizations of this variable, are rather different
from what is usually assumed in models of loop quantum gravity, even compared
with a polymerization of $K$ in (\ref{sinK}) which already showed several
deviating features.  But while a polymerization of $\bar{W}$ may not be directly
motivated by traditional loop quantum gravity, we include this possibility
here for completeness of the correspondence with spherical symmetry. New
canonical quantizations could still be constructed in this way by exploiting
the boundedness of operators quantizing a periodic function of $\Bar{W}$.

\section{Classes of constraints}
\label{s:Classes}

As derived in detail in \cite{EmergentScalar}, we consider different classes
of constraints and the modified structure functions they imply, depending on
which conditions are chosen in order to make the consistency equations
explicitly solvable.

\subsection{Constraints compatible with the classical constraint surface}

Modified constraints that are compatible with the classical constraint surface
in a suitable limit are direct generalizations of the models constructed
in Section~\ref{sec:Linear combination} from linear combinations of the
classical constraints. 

\subsubsection{General constraint}

As always, the expression for Hamiltonian constraints compatible with certain
symmetry conditions may depend on modification functions that distinguish
different cases of consistent constraints, but also on free functions that
represent the freedom to apply canonical transformations. Here, we fix the
latter choice by working with partially periodic modification functions in the
phase-space variable $K$.  In this class, the general expression of the
Hamiltonian constraint is given by
\begin{eqnarray}
    \tilde{H} &=&
    - \frac{\bar{\lambda}}{\lambda} \lambda_0 \cos^2 (\bar{\nu} \bar{W}) \sqrt{\varepsilon} \Bigg[
    \bar{a} \bigg( - \frac{\lambda^2}{\bar{\lambda}^2} \Lambda_0
    + \left( \frac{\alpha_2}{4 \varepsilon} c_f
    + \frac{1}{2} \frac{\partial c_f}{\partial \varepsilon} \right) \frac{\sin^2 (\bar{\lambda} K)}{\bar{\lambda^2}}
    \nonumber\\
    &&
    + \left( \frac{\mathcal{A}}{\bar{a}} - \frac{P_{\bar{W}}}{\bar{a}} \frac{\tan (\bar{\nu}\bar{W})}{\bar{\nu}} \frac{\partial \ln \nu}{\partial \varepsilon}
    - \frac{\tan (\bar{\lambda} K)}{\bar{\lambda}} \frac{\partial \ln \lambda}{\partial \varepsilon} \right) c_f \frac{\sin (2\bar{\lambda} K)}{2\bar{\lambda}}
    \nonumber\\
    &&
    - \left( \frac{P_{\bar{W}}}{\bar{a} \cos (\bar{\nu} \bar{W})} + \frac{\tan (\bar{\lambda} K)}{\bar{\lambda}} \left(\frac{\bar{\nu}}{\nu} c_{h3} + \frac{\partial \ln \lambda}{\partial \bar{W}}\right) \right)^2 \frac{\alpha_3}{4 \varepsilon} \frac{\nu^2}{\bar{\nu}^2} c_f \cos^2 (\bar{\lambda} K)
    \bigg)
    \nonumber\\
    &&
    + \frac{(\varepsilon')^2}{\bar{a}} \Bigg( 
    \bar{\lambda}^2 \frac{\mathcal{A}}{\bar{a}} \frac{\sin (2 \bar{\lambda} K)}{2 \bar{\lambda}}
    \nonumber\\
    &&
    + \cos^2 (\bar{\lambda} K) \left( \frac{\partial \ln \lambda}{\partial \varepsilon} - \frac{\alpha_2}{4 \varepsilon} - \frac{\sin (\bar{\nu}\bar{W})}{\bar{\nu}} \frac{\partial \ln \nu}{\partial \varepsilon} \left( \frac{\bar{\nu}}{\nu} c_{h3} + \frac{\partial \ln \lambda}{\partial \bar{W}}
    + \frac{\sin (\bar{\nu}\bar{W})}{\bar{\nu}} \frac{\partial \ln \nu}{\partial \varepsilon} \frac{\varepsilon}{\alpha_3} \right) \right)
    \Bigg)
    \nonumber\\
    &&
    + \left( \frac{\varepsilon' \bar{a}'}{\bar{a}^2} - \frac{\varepsilon''}{\bar{a}} \right) \cos^2 (\bar{\lambda} K)
    \nonumber\\
    &&
    + \cos^2 (\bar{\lambda} K) \Bigg( 
    - \frac{1}{\bar{a}} \left(\left(\frac{\sin (\bar{\nu}\bar{W})}{\bar{\nu}}\right)'\right)^2 \frac{\varepsilon}{\alpha_3}
    + \frac{\varepsilon'}{\bar{a}} \left(\frac{\sin (\bar{\nu}\bar{W})}{\bar{\nu}}\right)' \Bigg( \frac{2 \varepsilon}{\alpha_3} \frac{\sin (\bar{\nu}\bar{W})}{\bar{\nu}} \frac{\partial \ln \nu}{\partial \varepsilon}
    \nonumber\\
    &&
    + \frac{\bar{\nu}}{\nu} c_{h3}
    + \frac{\partial \ln \lambda}{\partial \bar{W}}
    - \frac{P_{\bar{W}}}{\bar{a} \cos (\bar{\nu} \bar{W})} \bar{\lambda}^2 \frac{\tan (\bar{\lambda} K)}{\bar{\lambda}}
    \Bigg)\Bigg)
    \Bigg]
    \label{eq:Hamiltonian constraint - periodic - CCSL - scalar polymerized}
\end{eqnarray}
with the structure function
\begin{eqnarray}
    \tilde{q}^{\theta \theta}
    =
    \left( c_f
    + \left(\frac{\bar{\lambda} \varepsilon'}{\bar{a}} \right)^2 \right) \cos^2 \left(\bar{\lambda} K\right)
    \frac{\bar{\lambda}^2}{\lambda^2} \lambda_0^2 \cos^4 (\bar{\nu} \bar{W}) \frac{\varepsilon}{\bar{a}^2}
    \,.
    \label{eq:Structure function - periodic - CCSL - scalar polymerized}
\end{eqnarray}

All the non-classical parameters are undetermined functions of $\varepsilon$
only, except for $\bar{\lambda}$ and $\bar{\nu}$, which are constants, and
$\lambda_0$ and $\lambda$, which can depend on both $\varepsilon$ and
$\bar{W}$. (This is the only class of constraints that allows $\lambda$ to
depend on $\bar{W}$.)
The constraint (\ref{eq:Hamiltonian constraint - periodic - CCSL - scalar
  polymerized}) and its structure function (\ref{eq:Structure function -
  periodic - CCSL - scalar polymerized}) are symmetric under the discrete
transformation $\bar{W} \to - \bar{W}$, $P_{\bar{W}} \to - P_{\bar{W}}$ only
if the $\lambda_0$ and $\lambda$ dependence on $\bar{W}$ is restricted by the discrete
symmetry (\ref{eq:Discrete symmetry - new variables}) to be of the form
$\lambda_0 (\varepsilon , \bar{W})=\lambda_0 (\varepsilon , -\bar{W})$, and
only if $c_{h3} = 0$ because it is independent of $\bar{W}$. (Alternatively,
the discrete transformation could be redefined as $\bar{W} \to - \bar{W}$,
$P_{\bar{W}} \to - P_{\bar{W}}$, and $c_{h3} \to - c_{h3}$, in which case the
constraint and structure function are symmetric even for non-zero $c_{h3}$.)
The classical limit can be taken in different ways, the simplest one given by
$\lambda\to \bar{\lambda}$ and $\nu\to\bar{\nu}$ followed by
$\lambda_0 , c_f , \alpha_2,\alpha_3 \to 1$,
$\bar{\lambda} , \bar{\nu} \to 0$ and $\Lambda_0 \to \Lambda$.

The inhomogeneous-field observable in this class is given by
\begin{eqnarray}
    G [\omega] &=& \int {\rm d} \theta\ \omega \frac{\nu}{\bar{\nu}} \left( \frac{P_{\bar{W}}}{\cos(\bar{\nu}\bar{W})}
    + \bar{a} \frac{\tan (\bar{\lambda} K)}{\bar{\lambda}} \frac{\partial \ln \lambda}{\partial \bar{W}} \right)
    \ ,
    \label{eq:Scalar field symmetry generator - periodic - CCSL - scalar polymerized}
\end{eqnarray}
where $\omega$ is a constant.
The associated conserved current $J^\mu$ has the components
\begin{eqnarray}
    J^t &=& \frac{\nu}{\bar{\nu}} \left( \frac{P_{\bar{W}}}{\cos(\bar{\nu}\bar{W})}
    + \bar{a} \frac{\tan (\bar{\lambda} K_\varphi)}{\bar{\lambda}} \frac{\partial \ln \lambda}{\partial \bar{W}} \right)
    \ , \\
    J^{\theta} &=& - \frac{\nu}{\bar{\nu}}
    \frac{\bar{\lambda}}{\lambda} \lambda_0 \sqrt{\varepsilon}
    \cos^2 (\bar{\nu} \bar{W}) \cos^2 (\bar{\lambda} K) \Bigg( 
    - \frac{2}{\bar{a}} \left(\frac{\sin (\bar{\nu}\bar{W})}{\bar{\nu}}\right)' \frac{\varepsilon}{\alpha_3}
    \label{eq:Conserved matter current - CCSL}\\
    &&
    + \frac{\varepsilon'}{\bar{a}} \Bigg( \frac{2 \varepsilon}{\alpha_3} \frac{\sin (\bar{\nu}\bar{W})}{\bar{\nu}} \frac{\partial \ln \nu}{\partial \varepsilon}
    + \frac{\bar{\nu}}{\nu} c_{h3}
    + \frac{\partial \ln \lambda}{\partial \bar{W}}
    - \frac{P_{\bar{W}}}{\bar{a} \cos (\bar{\nu} \bar{W})} \bar{\lambda}^2 \frac{\tan (\bar{\lambda} K)}{\bar{\lambda}}
    \Bigg)\Bigg)
    \ .\nonumber
\end{eqnarray}

When $P_{\bar{W}},\bar{W}\to0$, the homogeneous mass observable
associated to (\ref{eq:Hamiltonian constraint - periodic - CCSL - scalar polymerized}) is given by
\begin{eqnarray}
    \mathcal{M}
    &=&
    d_0
    + \frac{d_2}{8} \left(\exp \int {\rm d} \varepsilon \ \left(\frac{\alpha_2}{2 \varepsilon} - \frac{\partial \ln \lambda^2}{\partial \varepsilon}\right)\right)
    \left(
    c_f \frac{\sin^2\left(\bar{\lambda} K\right)}{\bar{\lambda}^2}
    - \cos^2 (\bar{\lambda} K) \left(\frac{\varepsilon'}{\bar{a}}\right)^2
    \right)
    \notag\\
    &&
    + \frac{d_2}{4} \int {\rm d} \varepsilon \ \left(\frac{\lambda^2}{\bar{\lambda}^2} \Lambda_0 \exp \int {\rm d} \varepsilon \ \left(\frac{\alpha_2}{2 \varepsilon} - \frac{\partial \ln \lambda^2}{\partial \varepsilon}\right)\right)
    \ ,
    \label{eq:Gravitational weak observable - CCSL - scalar polymerization - periodic}
\end{eqnarray}
where $d_0$ and $d_2$ are constants with classical limits given by $d_0\to0$ and $d_2 \to 1$.

\subsubsection{Partial Abelianization}

Following Section~\ref{sec:Partial Abelianization} and using the constraint (\ref{eq:Hamiltonian constraint - periodic - CCSL - scalar polymerized}) with the structure function (\ref{eq:Structure function - periodic - CCSL - scalar polymerized}) we obtain 
\begin{eqnarray}
    \tilde{q}^{\rm (A)} = Q_0 + Q_\varepsilon (\varepsilon')^2
    = 0
\end{eqnarray}
for the Abelianization condition (\ref{eq:Abelianization condition structure
  function - new variables}), where $Q_0$ and $Q_\varepsilon$ are functions of
$B$, as it appears in $ \tilde{H}^{\rm (A)} = B \tilde{H} + A H_\theta$, and
of the phase-space variables, but
not of their derivatives.
Therefore, these two coefficients must vanish independently.
The condition $Q_0=0$ implies the equation
\begin{eqnarray}
    B - \frac{\sin(2 \bar{\lambda} K)}{2 \bar{\lambda}} \frac{\partial B}{\partial K}
    = 0
    \ ,
\end{eqnarray}
with the general solution
\begin{eqnarray}
    B = B_0 \frac{\tan (\bar{\lambda} K)}{\bar{\lambda}}
    \label{eq:Partial Abelianization solution to first equation - CCSL}
\end{eqnarray}
for an undetermined function $B_0 (\varepsilon , \bar{W})$.
The condition $Q_\varepsilon = 0$ implies the equation
\begin{eqnarray}
    2 \bar{\lambda}^2 B + \bar{\lambda} \sin (2\bar{\lambda} K) \frac{\partial B}{\partial K} - 2 \cos^2 (\bar{\lambda} K) \frac{\partial^2 B}{\partial K^2}
    = 0
    \ .
\end{eqnarray}
By direct substitution we find that (\ref{eq:Partial Abelianization solution to first equation - CCSL}) solves this equation too.
The Abelianized constraint is then given by
\begin{eqnarray}
    \frac{\tilde{H}^{\rm (A)}}{B_0} &=&
    - \frac{\tan (\bar{\lambda} K)}{\bar{\lambda}} \frac{\bar{\lambda}}{\lambda} \lambda_0 \cos^2 (\bar{\nu} \bar{W}) \sqrt{\varepsilon} \Bigg[
    \bar{a} \bigg( - \frac{\lambda^2}{\bar{\lambda}^2} \Lambda_0
    + \left( \frac{\alpha_2}{4 \varepsilon} c_f
    + \frac{1}{2} \frac{\partial c_f}{\partial \varepsilon} \right) \frac{\sin^2 (\bar{\lambda} K)}{\bar{\lambda^2}}
    \nonumber\\
    &&
    + \left( \frac{\mathcal{A}}{\bar{a}} - \frac{P_{\bar{W}}}{\bar{a}} \frac{\tan (\bar{\nu}\bar{W})}{\bar{\nu}} \frac{\partial \ln \nu}{\partial \varepsilon}
    - \frac{\tan (\bar{\lambda} K)}{\bar{\lambda}} \frac{\partial \ln \lambda}{\partial \varepsilon} \right) c_f \frac{\sin (2\bar{\lambda} K)}{2\bar{\lambda}}
    \nonumber\\
    &&
    - \left( \frac{P_{\bar{W}}}{\bar{a} \cos (\bar{\nu} \bar{W})} + \frac{\tan (\bar{\lambda} K)}{\bar{\lambda}} \left(\frac{\bar{\nu}}{\nu} c_{h3} + \frac{\partial \ln \lambda}{\partial \bar{W}}\right) \right)^2 \frac{\alpha_3}{4 \varepsilon} \frac{\nu^2}{\bar{\nu}^2} c_f \cos^2 (\bar{\lambda} K)
    \bigg)
    \nonumber\\
    &&
    + \frac{(\varepsilon')^2}{\bar{a}} \Bigg( 
    \bar{\lambda}^2 \frac{\mathcal{A}}{\bar{a}} \frac{\sin (2 \bar{\lambda} K)}{2 \bar{\lambda}}
    \nonumber\\
    &&
    + \cos^2 (\bar{\lambda} K) \left( \frac{\partial \ln \lambda}{\partial \varepsilon} - \frac{\alpha_2}{4 \varepsilon} - \frac{\sin (\bar{\nu}\bar{W})}{\bar{\nu}} \frac{\partial \ln \nu}{\partial \varepsilon} \left( \frac{\bar{\nu}}{\nu} c_{h3} + \frac{\partial \ln \lambda}{\partial \bar{W}}
    + \frac{\sin (\bar{\nu}\bar{W})}{\bar{\nu}} \frac{\partial \ln \nu}{\partial \varepsilon} \frac{\varepsilon}{\alpha_3} \right) \right)
    \Bigg)
    \nonumber\\
    &&
    + \left( \frac{\varepsilon' \bar{a}'}{\bar{a}^2} - \frac{\varepsilon''}{\bar{a}} \right) \cos^2 (\bar{\lambda} K)
    \nonumber\\
    &&
    + \cos^2 (\bar{\lambda} K) \Bigg( 
    - \frac{1}{\bar{a}} \left(\left(\frac{\sin (\bar{\nu}\bar{W})}{\bar{\nu}}\right)'\right)^2 \frac{\varepsilon}{\alpha_3}
    + \frac{\varepsilon'}{\bar{a}} \left(\frac{\sin (\bar{\nu}\bar{W})}{\bar{\nu}}\right)' \Bigg( \frac{2 \varepsilon}{\alpha_3} \frac{\sin (\bar{\nu}\bar{W})}{\bar{\nu}} \frac{\partial \ln \nu}{\partial \varepsilon}
    \nonumber\\
    &&
    + \frac{\bar{\nu}}{\nu} c_{h3}
    + \frac{\partial \ln \lambda}{\partial \bar{W}}
    - \frac{P_{\bar{W}}}{\bar{a} \cos (\bar{\nu} \bar{W})} \bar{\lambda}^2 \frac{\tan (\bar{\lambda} K)}{\bar{\lambda}}
    \Bigg)\Bigg)
    \Bigg]
    \nonumber\\
    &&
    - \frac{\bar{\lambda}}{\lambda} \lambda_0 \cos^2 (\bar{\nu} \bar{W}) \sqrt{\varepsilon} \varepsilon' \left( \bar{a} K' + P_{\bar{W}} \bar{W}' - {\cal A} \varepsilon' \right)
    \ .
    \label{eq:Abelianized constraint - periodic - CCSL - scalar polymerized}
\end{eqnarray}

The first line in (\ref{eq:Abelianized constraint - periodic - CCSL - scalar
  polymerized}) has a kinematical divergence at $K = \pi / (2 \bar{\lambda})$
due to the overall tangent factor.
This divergence can be removed if 
\begin{eqnarray}
    \frac{\alpha_2}{4 \varepsilon} c_f
    + \frac{1}{2} \frac{\partial c_f}{\partial \varepsilon}
    - \lambda^2 \Lambda_0 = 0
\end{eqnarray}
is satisfied because the relevant terms then combine to produce a $\cos^2$-factor and hence cancel
the divergence of the tangent.
If we interpret this condition as an equation for $c_f$, we must restrict
$\lambda$ to be a function of $\varepsilon$ only.
However, the solution to this equation is not compatible with the classical
limit $c_f \to 1$. (For instance, for $\Lambda_0=0$ we have $c_f\propto \varepsilon^{-\alpha_2/2}$.)
Therefore, we have to weaken the condition by neglecting the first term, and hence leave it as a divergent term of the Abelian constraint.
The resulting equation for the partial resolution of the divergence,
\begin{eqnarray}
    \frac{1}{2} \frac{\partial c_f}{\partial \varepsilon}
    - \lambda^2 \Lambda_0 = 0
    \ ,
\end{eqnarray}
can now be directly integrated, yielding the modification function
\begin{equation}
    c_f = 2 \int \lambda^2 \Lambda_0 {\rm d} \varepsilon
    \ .
\end{equation}

If we choose the classical value of the cosmological constant
$\Lambda_0 = \Lambda$, and $\lambda^2 = \Delta / \varepsilon$ with a constant
$\Delta$ (sometimes used in models of loop quantum gravity), we obtain
\begin{equation}
    c_f = 1 + 2 \Lambda \Delta \ln \left(\frac{\varepsilon}{c_0}\right)
    \ ,
    \label{eq:MONDian-LQG function}
\end{equation}
where $c_0$ is an integration constant. The correct classical limit is obtained for $\Delta\to0$.
The logarithmic dependence on $\epsilon$ is relevant on intermediate scales
far from black-hole or cosmological horizons. It may then be related to
MOND-like effects as shown in \cite{HigherMOND}.

\subsection{Constraints of the second kind}

A second class of explicit modified constraints is obtained from a specific
choice for one of the modification functions, so far without a detailed
physical motivation. Nevertheless, this case is interesting because it can be
used to show the variety of possible covariant theories.

\subsubsection{General constraint}

Again referring to more detailed derivations for a scalar field in spherical
symmetry \cite{EmergentScalar}, we now have the modified Hamiltonian constraint
\begin{eqnarray}
    \tilde{H} &=&
    - \frac{\bar{\lambda}}{\lambda} \lambda_0 \cos^2 (\bar{\nu} \bar{W}) \sqrt{\varepsilon} \Bigg[
    \bar{a} \left(
    - \frac{\lambda^2}{\bar{\lambda}^2}\Lambda_0
    + \frac{\sin^2 (\bar{\lambda} K)}{\bar{\lambda}^2} \left( \left(\frac{\alpha_2}{4 \varepsilon} - \frac{\partial \ln \lambda}{\partial \varepsilon}\right) c_f + \frac{1}{2} \frac{\partial c_f}{\partial \varepsilon} \right)\right)
    \nonumber\\
    &&\qquad
    + \bar{a} \frac{\sin (2 \bar{\lambda} K)}{2 \bar{\lambda}} \left( \left(\frac{\alpha_2}{2 \varepsilon} - \frac{\partial \ln \lambda}{\partial \varepsilon}\right) \frac{\lambda}{\bar{\lambda}} q
    + \frac{\lambda}{\bar{\lambda}} \frac{\partial q}{\partial \varepsilon} \right)
    \nonumber\\
    &&\qquad
    + \left( \mathcal{A} + \frac{P_{\bar{W}}}{\cos (\bar{\nu}\bar{W})} \left( \frac{\nu}{\bar{\nu}} c_{h3}
    - \frac{\sin (\bar{\nu}\bar{W})}{\bar{\nu}} \frac{\partial \ln \nu}{\partial \varepsilon} \right) \right) \left(c_f \frac{\sin (2 \bar{\lambda} K)}{2 \bar{\lambda}}
    + \frac{\lambda}{\bar{\lambda}} q \cos(2 \bar{\lambda} K)\right)
    \nonumber\\
    &&\qquad
    - \frac{\nu^2}{\bar{\nu}^2} \frac{P_{\bar{W}}{}^2}{\bar{a} \cos^2 (\bar{\nu}\bar{W})} \frac{\alpha_3}{4 \varepsilon} \left( c_f \cos^2 (\bar{\lambda} K)
    - 2 \frac{\lambda}{\bar{\lambda}} q \bar{\lambda}^2 \frac{\sin (2
       \bar{\lambda} K)}{2 \bar{\lambda}} \right) \nonumber\\
&&\qquad    + \left( \frac{\varepsilon' \bar{a}'}{\bar{a}^2} - \frac{\varepsilon''}{\bar{a}} \right) \cos^2 (\bar{\lambda} K)
    \nonumber\\
    &&\qquad
    - \frac{(\varepsilon')^2}{\bar{a}} \Bigg( \left(\frac{\alpha_2}{4 \varepsilon} - \frac{\partial \ln \lambda}{\partial \varepsilon}\right) \cos^2 (\bar{\lambda} K)\nonumber\\
&&\qquad\qquad    - \left( \frac{\mathcal{A}}{\bar{a}}
    + \frac{P_{\bar{W}}}{\bar{a} \cos (\bar{\nu}\bar{W})} \left( \frac{\nu}{\bar{\nu}} c_{h3}
    - \frac{\sin (\bar{\nu}\bar{W})}{\bar{\nu}} \frac{\partial \ln \nu}{\partial \varepsilon} \right) \right) \bar{\lambda}^2 \frac{\sin (2 \bar{\lambda} K)}{2 \bar{\lambda}}
    \nonumber\\
    &&\qquad\qquad
    + \frac{\nu^2}{\bar{\nu}^2} \frac{P_{\bar{W}}{}^2}{\bar{a}^2 \cos^2 (\bar{\nu}\bar{W})} \bar{\lambda}^2 \frac{\alpha_3}{4 \varepsilon} \cos^2 (\bar{\lambda} K) \Bigg)
    \nonumber\\
    &&\quad
    - \frac{1}{\bar{a}} \frac{\bar{\nu}^2}{\nu^2} \left(  \left(\frac{\sin (\bar{\nu}\bar{W})}{\bar{\nu}}\right)'
    + \varepsilon' \left( \frac{\nu}{\bar{\nu}} c_{h3}
    - \frac{\sin (\bar{\nu}\bar{W})}{\bar{\nu}} \frac{\partial \ln \nu}{\partial \varepsilon} \right) \right)^2 \frac{\varepsilon}{\alpha_3}
    \Bigg]
    \ ,
    \label{eq:Hamiltonian constraint - DF - scalar polymerization}
\end{eqnarray}
with structure function 
\begin{eqnarray}
    \tilde{q}^{\theta \theta}
    =
    \frac{\bar{\lambda}^2}{\lambda^2} \lambda_0^2 \left( \left( c_f
    + \left(\frac{\bar{\lambda} \varepsilon'}{\bar{a}}\right)^2 \right) \cos^2\left(\bar{\lambda} K\right)
    - 2 \bar{\lambda}^2 \frac{\lambda}{\bar{\lambda}} q \frac{\sin \left(2 \bar{\lambda} K\right)}{2 \bar{\lambda}} \right) \cos^4 (\bar{\nu} \bar{W}) \frac{\varepsilon}{\bar{a}^2}
    \label{eq:Structure function - DF - scalar polymerization}
\end{eqnarray}
All the non-classical parameters are undetermined functions of $\varepsilon$
only, except for the parameters $\bar{\lambda}$ and $\bar{\nu}$ which are
constants, and $\lambda_0$ which can depend on both $\varepsilon$ and
$\bar{W}$.
The constraint (\ref{eq:Hamiltonian constraint - DF - scalar polymerization})
and its structure function (\ref{eq:Structure function - DF - scalar
  polymerization}) are symmetric under the discrete transformation
$\bar{W} \to - \bar{W}$, $P_{\bar{W}} \to - P_{\bar{W}}$ only if the
$\lambda_0$ dependence on $\bar{W}$ is restricted by the discrete symmetry
(\ref{eq:Discrete symmetry - new variables}) to the form
$\lambda_0 (\varepsilon , \bar{W})=\lambda_0 (\varepsilon , -\bar{W})$, and
only if $c_{h3} = 0$ because it is independent of $\bar{W}$. (Alternatively,
the discrete transformation can be redefined as $\bar{W} \to - \bar{W}$,
$P_{\bar{W}} \to - P_{\bar{W}}$, and $c_{h3} \to - c_{h3}$, in which case the
constraint and structure function are symmetric even for non-zero $c_{h3}$.)
The classical limit can be taken in different ways, the simplest one given by
$\lambda\to \bar{\lambda}$ and $\nu\to\bar{\nu}$, followed by
$\lambda_0 , c_f , \alpha_2,\alpha_3 \to 1$,
$q, \bar{\lambda} , \bar{\nu} \to 0$, and $\Lambda_0 \to \Lambda$.

The inhomogeneous-field observable is
\begin{eqnarray}
    G [\omega] &=& \int {\rm d}\theta\ \omega \frac{\nu}{\bar{\nu}} \frac{P_{\bar{W}}}{\cos(\bar{\nu}\bar{W})}
    \ ,
    \label{eq:Scalar field symmetry generator - DF}
\end{eqnarray}
where $\omega$ is a constant.
The associated conserved current $J^\mu$ has the components
\begin{eqnarray}
    J^t &=& \frac{\nu}{\bar{\nu}} \frac{P_{\bar{W}}}{\cos(\bar{\nu}\bar{W})}
    \ , \\
    J^{\theta}
    &=& \frac{\bar{\nu}}{\nu} \frac{\bar{\lambda}}{\lambda} \lambda_0 \cos^2 (\bar{\nu} \bar{W}) \frac{2 \varepsilon^{3/2}}{\alpha_3 \bar{a}}  \left( \left(\frac{\sin (\bar{\nu}\bar{W})}{\bar{\nu}}\right)'
    + \varepsilon' \left( \frac{\nu}{\bar{\nu}} c_{h3}
    - \frac{\sin (\bar{\nu}\bar{W})}{\bar{\nu}} \frac{\partial \ln \nu}{\partial \varepsilon} \right) \right)
    \ .
    \label{eq:Conserved matter current - DF}
\end{eqnarray}
When $P_{\bar{W}},\bar{W}\to0$, the homogeneous mass observable
associated to (\ref{eq:Hamiltonian constraint - periodic - CWL}) is given by
\begin{eqnarray}
    \mathcal{M}
    &=&
    d_0
    + \frac{d_2}{8} \left(\exp \int {\rm d} \varepsilon \
        \left(\frac{\alpha_2}{2 \varepsilon} - \frac{\partial \ln
        \lambda^2}{\partial \varepsilon}\right)\right)\nonumber\\
  &&\qquad\quad\times
    \left(
    c_f \frac{\sin^2\left(\bar{\lambda} K\right)}{\bar{\lambda}^2}
    + 2 \frac{\lambda}{\bar{\lambda}} q \frac{\sin \left(2 \bar{\lambda}  K\right)}{\bar{\lambda}}
    - \cos^2 (\bar{\lambda} K) \left(\frac{\varepsilon'}{\bar{a}}\right)^2
    \right)
    \notag\\
    &&
    + \frac{d_2}{4} \int {\rm d} \varepsilon \ \left(\frac{\lambda^2}{\bar{\lambda}^2} \Lambda_0 \exp \int {\rm d} \varepsilon \ \left(\frac{\alpha_2}{2 \varepsilon} - \frac{\partial \ln \lambda^2}{\partial \varepsilon}\right)\right)
    \ ,
    \label{eq:Gravitational weak observable - SF - scalar polymerization - periodic}
\end{eqnarray}
where $d_0$, and $d_2$ are constants with classical limits given by $d_0\to0$
and $d_2 \to 1$. Most of these properties are similar to those in the first class, but
explicit solutions in solvable cases, such as spatially homogeneous ones, can
reveal crucial differences, as we will see in Section~\ref{sec:Homogeneous
  spacetime}.

A key difference between the first and second class can be seen easily in the
structure functions (\ref{eq:Structure function - periodic - CCSL - scalar
  polymerized}) and (\ref{eq:Structure function - DF - scalar
  polymerization}), respectively. The first one is even in the curvature
component $K$, while the second one contains an odd term, multiplied by the
modification function $q$. The same behavior was possible in spherical
symmetry \cite{HigherCov} where it may have far-reaching implications for
various particle effects \cite{Axion}. In the Hamiltonian constraint, the
$q$-terms show that there may be modifications linear in $K$ (if a Taylor
expansion is used for the trigonometric functions). Such terms can be more
relevant than the classical quadratic terms as the curvature scale is
increased. The second class of modified constraints for polarized Gowdy models
shows that these interesting features are not restricted to spherical
symmetry.

\subsubsection{Partial Abelianization}

The partial Abelianization of this constraint follows the same procedure as
the last one. It requires exactly the same $B$-factor (\ref{eq:Partial
  Abelianization solution to first equation - CCSL}) and the associated
$A$. However, a partial Abelianization is subject to the additional condition
that the modification function $q$ vanishes.

\subsection{Constraints compatible with the classical-$\bar{W}$ limit}

The third class has modified constraints that have a limit in which the field
$\bar{W}$ behaves like a classical scalar field on the emergent (and
non-classical) space-time. This case is useful because it allows us to make
comparisons between the propagation speed of $\bar{W}$ as a polarized
gravitational wave and the speed of a massless scalar field that may be
coupled minimally.

\subsubsection{General constraint}

The modified Hamiltonian constraint is given by
\begin{eqnarray}
    \tilde{H} &=&
    - \frac{\bar{\lambda}}{\lambda} \lambda_0 \cos^2 (\bar{\nu} \bar{W}) \sqrt{\varepsilon} \Bigg[
    \bar{a} \left(
    \frac{\lambda^2}{\bar{\lambda}^2} \Lambda_0
    + \left( c_f \left(\frac{\alpha_2}{4 \varepsilon} - \frac{\partial \ln \lambda}{\partial \varepsilon}\right) + \frac{1}{2} \frac{\partial c_{f}}{\partial \varepsilon} \right) \frac{\sin^2 \left(\bar{\lambda} K\right)}{\bar{\lambda}^2} \right)
    \nonumber\\
    && \qquad
    + \bar{a} \left( \frac{q}{2} \left(\frac{\alpha_2}{\varepsilon} - 2 \frac{\partial \ln \lambda}{\partial \varepsilon}\right) + \frac{\lambda}{\bar{\lambda}} \frac{\partial q}{\partial \varepsilon} \right) \frac{\sin \left(2 \bar{\lambda} K\right)}{2 \bar{\lambda}}
    \nonumber\\
    &&\qquad
    + \left( \mathcal{A}  - P_{\bar{W}}\frac{\tan (\bar{\nu} \bar{W})}{\bar{\nu}} \frac{\partial \ln \nu}{\partial \varepsilon}\right) \left( c_f \frac{\sin (2 \bar{\lambda} K)}{2 \bar{\lambda}}
    + \frac{\lambda}{\bar{\lambda}} q \cos(2 \bar{\lambda} K)\right)
    \nonumber\\
    &&\qquad
    + \frac{(\varepsilon')^2}{\bar{a}} \left( \left(\frac{\partial \ln
       \lambda}{\partial \varepsilon}-\frac{\alpha_2}{4 \varepsilon}\right) \cos^2
       (\bar{\lambda} K)\right.\nonumber\\
  &&\left.\qquad
    + \bar{\lambda}^2 \left(\frac{\mathcal{A}}{\bar{a}} - \frac{P_{\bar{W}}}{\bar{a}} \frac{\tan (\bar{\nu} \bar{W})}{\bar{\nu}} \frac{\partial \ln \nu}{\partial \varepsilon}\right) \frac{\sin (2 \bar{\lambda} K)}{2 \bar{\lambda}} \right)
    \nonumber\\
    &&\qquad
    + \left( \frac{\varepsilon' \bar{a}'}{\bar{a}^2} - \frac{\varepsilon''}{\bar{a}} \right) \cos^2 (\bar{\lambda} K)
    \Bigg]
    \nonumber\\
    &&
    + \frac{\bar{\nu}^2}{\nu^2} \frac{\sqrt{q^{\theta\theta}}}{2} \left[
    \frac{P_{\bar{W}}{}^2}{\cos^2 (\bar{\nu} \bar{W})} \frac{\alpha_3}{2 \varepsilon}
    + \frac{2 \varepsilon}{\alpha_3} \left( \left(\frac{\sin (\bar{\nu} \bar{W})}{\bar{\nu}}\right)'
    - \frac{\sin (\bar{\nu} \bar{W})}{\bar{\nu}} \frac{\partial\ln \nu}{\partial \varepsilon} \varepsilon' \right)^2
       \right]
    \ ,
    \label{eq:Hamiltonian constraint - periodic - CWL}
\end{eqnarray}
with the structure function
\begin{eqnarray}
    \tilde{q}^{\theta\theta}
    =
    \left(
    \left( c_{f}
    + \left(\frac{\bar{\lambda} \varepsilon'}{\bar{a}} \right)^2 \right) \cos^2 \left(\bar{\lambda} K\right)
    - 2 q \frac{\lambda}{\bar{\lambda}} \bar{\lambda}^2 \frac{\sin \left(2 \bar{\lambda} K\right)}{2 \bar{\lambda}}\right)
    \frac{\bar{\lambda}^2}{\lambda^2}
    \lambda_0^2 \cos^4 (\bar{\nu} \bar{W}) \frac{\varepsilon}{\bar{a}^2}
    \label{eq:Structure function - periodic - CWL}
\end{eqnarray}
appearing explicitly in the last line.  All the non-classical parameters are
undetermined functions of $\varepsilon$ only, except for the parameters
$\bar{\lambda}$ and $\bar{\nu}$ which are constants, and $\lambda_0$ which can
depend on both $\varepsilon$ and $\bar{W}$.
The constraint (\ref{eq:Hamiltonian constraint - periodic - CWL}) and its
structure function (\ref{eq:Structure function - periodic - CWL}) are
symmetric under the discrete transformation $\bar{W} \to - \bar{W}$,
$P_{\bar{W}} \to - P_{\bar{W}}$ only if the $\lambda_0$ dependence on
$\bar{W}$ is restricted by the discrete symmetry (\ref{eq:Discrete symmetry -
  new variables}) to the form
$\lambda_0 (\varepsilon , \bar{W})=\lambda_0 (\varepsilon , -\bar{W})$.
The classical limit can be taken in different ways, the simplest one given by
$\lambda\to \bar{\lambda}$ and $\nu\to\bar{\nu}$, followed by
$\lambda_0 , c_f , \alpha_2,\alpha_3 \to 1$,
$q, \bar{\lambda} , \bar{\nu} \to 0$, and $\Lambda_0 \to \Lambda$.  The
classical-$\bar{W}$ limit is obtained for $\nu=\bar{\nu}\to0$ and
$\alpha_3\to1$. The last parenthesis in the Hamiltonian constraint then
approaches the form of a classical scalar field propagating on the emergent
space-time with inhomogeneous spatial component $\tilde{q}_{\theta\theta}$.

The $\bar{W}$-field observable is given by
\begin{eqnarray}
    G [\omega] &=& \int {\rm d} \theta\ \omega \frac{\nu}{\bar{\nu}} \frac{P_{\bar{W}}}{\cos(\bar{\nu}\bar{W})}
    \label{eq:Scalar field symmetry generator - CML - scalar polymerization - periodic}
\end{eqnarray}
where $\omega$ is a constant.
The associated conserved current $J^\mu$ has the components
\begin{eqnarray}
    J^t &=& \frac{\nu}{\bar{\nu}} \frac{P_{\bar{W}}}{\cos(\bar{\nu}\bar{W})}
    \ , \\
    J^{\theta} &=& 
    = \frac{\bar{\nu}}{\nu} \sqrt{q^{\theta \theta}}
    \frac{2 \varepsilon}{\alpha_3} \left( \left(\frac{\sin (\bar{\nu} \bar{W})}{\bar{\nu}}\right)'
    - \varepsilon' \frac{\sin (\bar{\nu} \bar{W})}{\bar{\nu}} \frac{\partial\ln \nu}{\partial \varepsilon} \right)
    \ .
    \label{eq:Conserved matter current - CML}
\end{eqnarray}
When $P_{\bar{W}},\bar{W}\to0$, the homogeneous mass observable
associated to (\ref{eq:Hamiltonian constraint - periodic - CWL}) is given by
\begin{eqnarray}
    \mathcal{M}
    &=&
    d_0
    + \frac{d_2}{8} \left(\exp \int {\rm d} \varepsilon \
        \left(\frac{\alpha_2}{2 \varepsilon} - \frac{\partial \ln
        \lambda^2}{\partial \varepsilon}\right)\right)\nonumber\\
  &&\qquad\quad\times
    \left(
    c_f \frac{\sin^2\left(\bar{\lambda} K\right)}{\bar{\lambda}^2}
    + 2 \frac{\lambda}{\bar{\lambda}} q \frac{\sin \left(2 \bar{\lambda}  K\right)}{2 \bar{\lambda}}
    - \cos^2 (\bar{\lambda} K) \left(\frac{\varepsilon'}{\bar{a}}\right)^2
    \right)
    \notag\\
    &&
    + \frac{d_2}{4} \int {\rm d} \varepsilon \ \left(\frac{\lambda^2}{\bar{\lambda}^2} \Lambda_0 \exp \int {\rm d} \varepsilon \ \left(\frac{\alpha_2}{2 \varepsilon} - \frac{\partial \ln \lambda^2}{\partial \varepsilon}\right)\right)
    \ ,
    \label{eq:Gravitational weak observable - CML - scalar polymerization - periodic}
\end{eqnarray}
where $d_0$, and $d_2$ are constants with classical limits given by $d_0\to0$ and $d_2 \to 1$.

\subsubsection{Partial Abelianization}

The partial Abelianization of this constraint follows the same procedure as
outlined in the first class of modified constraints. It implies the $B$-factor
(\ref{eq:Partial Abelianization solution to first equation - CCSL}) and the
associated $A$-factor, but in addition requires that the modification function
$q$ vanishes, as in the second class.

\section{Dynamical solutions with  homogeneous spatial slices}
\label{sec:Homogeneous spacetime}

First indications of possible physical effects of our modifications can be
obtained by looking at properties of spatially homogeneous solutions. In this
case, partial differential equations are replaced by ordinary ones that can
often be solved more easily.

\subsection{Classical constraint}

For the sake of comparison, we first present useful gauge conditions and
solutions in the classical case with vanishing cosmological constant. Strict
homogeneity then implies Kasner solutions, while an inhomogeneous solution for
$\bar{W}$ can also be allowed.

\subsubsection{Conformal gauge}

The coordinates of the conventional Gowdy metric (\ref{eq:ADM line element -
  Gowdy - classical - conventional variables}) are associated to the gauge choice
\begin{equation}
    N^\theta = 0 \quad ,\quad \varepsilon = T
\end{equation}
with the time coordinate $T$.
We impose this gauge and for now work with the classical constraint
(\ref{eq:Hamiltonian constraint - Gowdy - Classical - new variables}).
The remaining metric components can be expressed in terms of the lapse
function and the two fields
\begin{eqnarray}
    W=\ln\sqrt{E^y/E^x}=\Bar{W}
    \quad , \quad
    a = \ln \sqrt{E^x E^y / \varepsilon} = \ln \Bar{a} - \frac{1}{2}\ln \varepsilon
    \ .
    \label{eq:Idenitification conventional and new variables - Gowdy}
\end{eqnarray}

The on-shell conditions $H_\theta = 0$ and $H=0$ in this gauge become
\begin{eqnarray}
    H_\theta &=& P_{\Bar{W}} \Bar{W}' + \Bar{a} K' = 0
    \ , \\
    H &=& 
    \frac{P_{\Bar{W}}{}^2 - 4 T \Bar{a} K \mathcal{A} - \Bar{a}^2 K^2
    + 4 T^2 (\Bar{W}')^2}{4 \sqrt{T} \Bar{a}}
    = 0
    \ .
    \label{eq:Constraints - classical - Conformal gauge}
\end{eqnarray}
Using the latter expression, we obtain an equation of motion
\begin{eqnarray}
    \partial_T (\Bar{a} K) = \{ \Bar{a} K , H [N]\} = 
    - NH
\end{eqnarray}
which vanishes on-shell, such that $\Bar{a} K = \mu$ where $\mu$ is a constant.
The consistency equation
$\dot{\varepsilon}=\partial\varepsilon/\partial T = 1$ can be solved for the
lapse function
\begin{equation}
    N = \frac{1}{K \sqrt{T}}
    = \frac{\mu^{-1} \bar{a}}{\sqrt{\varepsilon}}
    = \mu^{-1} \sqrt{q_{\theta\theta}}
    \ .
    \label{eq:Lapse Gowdy solution - Classical}
\end{equation}
The equations of motion for $\Bar{a}$ and $\Bar{W}$, respectively, imply
\begin{eqnarray}
    \mathcal{A} &=& \mu \left( \frac{\dot{\Bar{a}}}{\Bar{a}} - \frac{1}{2 T}\right)
    \ , \nonumber\\
    P_{\Bar{W}} &=& 2 \mu T \dot{\Bar{W}}
    \ .
    \label{eq:Momenta dynamical solution - classical}
\end{eqnarray}
Using these results, the on-shell conditions $H_\theta=0$ and $H=0$ can be rewritten as
\begin{subequations}
    \label{eq:a EoM/constraints - classical - conformal gauge}
\begin{eqnarray}
    a' &=&
    2 T \Dot{W} W'
    \ , \\
    \Dot{a} &=&
    - \frac{1}{4 T} + T \left(\Dot{W}^2 + \frac{(W')^2}{\mu^2}\right)
    \ , \label{Dota}
\end{eqnarray}
\end{subequations}
where we have used the identification (\ref{eq:Idenitification conventional
  and new variables - Gowdy}). 

The equation of motion $\Ddot{W} = \{\{W , H[N]\} , H[N]\}$ requires some care
because the lapse function (\ref{eq:Lapse Gowdy solution - Classical}) is
phase-space dependent. In a first-order equation of motion, using a single
Poisson bracket with $H[N]$, any term resulting from a non-zero Poisson
bracket with $N$ would be multiplied by $H$ and therefore vanish
on-shell. However, this argument does not apply to iterated Poisson brackets
of some phase-space function with $H[N]$, where non-zero on-shell terms may
contribute. 
Duely taking into account the phase-space dependence of $N$, such that the
second $\{\cdot,H[N]\}$ acts on this function contained in the first $H[N]$,
we obtain the second-order equation of motion
\begin{eqnarray}
    0 &=&
    \Ddot{W} + \frac{\Dot{W}}{T} - \frac{W''}{\mu^2}
    \ .
    \label{eq:W EoM - classical - conformal gauge}
\end{eqnarray}
It can be checked that this equation is equivalent to what would be obtained
in standard general relativity. (If $N$ were treated as phase-space
independent, we would instead obtain the bracket
$\{\{W , H[N]\} , H[N]\}= - \Dot{W}/T + W''/\mu^2 + \dot{W} \left(1/(4 T) - T
  \dot{W} - T W''/\mu^2 \right) $, using the lapse function (\ref{eq:Lapse
  Gowdy solution - Classical}) only after computing the brackets.  This
expression has
extra terms compared with the correct equation (\ref{eq:W EoM - classical - conformal gauge}).)

These are the equations of motion for the polarized Gowdy system in conventional variables, which have the general solution \cite{GowdyQuadratic}
\begin{eqnarray}
    W &=& 
    \alpha + \beta \ln T
    + \sum_{n=1}^\infty \left[ 
    a_n J_0 (n T) \sin (n \mu^{-2} \theta + \gamma_n)
    + b_n N_0 (n T) \sin (n \mu^{-2} \theta + \delta_n)
    \right]
    \ ,
    \label{eq:W general solution Gowdy T^3 - classical}
\end{eqnarray}
where $\alpha$, $\beta$, $a_n$, $b_n$, $\gamma_n$, and $\delta_n$ are real
constants, and $J_0$ and $N_0$ are Bessel and Neumann functions of the zeroth
order, respectively. 
Given the solution for $W$, direct integration of (\ref{eq:a EoM/constraints -
  classical - conformal gauge}) gives the expression for $a$. 
The space-time line element in this gauge becomes
\begin{eqnarray}
    {\rm d} s^2
    &=&
    \mu^{-2} e^{2 a} \left( - {\rm d} T^2
    + \mu^2 {\rm d} \theta^2 \right)
    + T \left( e^{-2 W} {\rm d} x^2
    + e^{2 W} {\rm d} y^2 \right)
    \ .
    \label{eq:General Gowdy solution - Classical}
\end{eqnarray}
The square of the lapse function, $N^2 = \Bar{a}^2 / (\mu^2 T)$, exactly
equals $q_{\theta\theta} = \Bar{a}^2 / T$ only if $\mu=\pm 1$.
Thus, setting $\mu=1$, imposing positivity of the lapse function in the region
$T>0$, we obtain a conformally flat metric for the $T-\theta$ components. With
this choice, the equation of motion (\ref{eq:W EoM - classical - conformal
  gauge}) implies that $W$-excitations travel at the speed of light.

Finally, note that using (\ref{eq:Momenta dynamical solution - classical}) and
(\ref{eq:W general solution Gowdy T^3 - classical}) and inserting this
solution in the symmetry generator (\ref{eq:Gowdy observable - classical})
with $\omega=1$ we find that it corresponds to 
\begin{eqnarray}
    G [1]
    &=&
    4 \pi \mu \beta
    \ ,
\end{eqnarray}
where we used the periodicity conditions in $\theta$.
This is clearly a conserved quantity.

\subsubsection{Homogeneous solution}

For a spatially homogenous background, we consider the special case $W'=0$.
From (\ref{eq:W general solution Gowdy T^3 - classical}), this implies that $a_n=b_n = 0$.
The equations of motion (\ref{eq:a EoM/constraints - classical - conformal
  gauge}) now have the solution 
\begin{eqnarray}
    a &=& \frac{4\beta^2-1}{4} \ln \left( \frac{T}{T_0}\right)
    \ , \\
    \bar{a} &=& T_0^{1/4-\beta^2} T^{\beta^2+1/4}
    \ ,
\end{eqnarray}
with a constant $T_0$, and hence
\begin{equation}
    K = \mu T_0^{\beta^2-1/4} T^{-\beta^2-1/4}
    \ .
\end{equation}
The space-time metric (\ref{eq:General Gowdy solution - Classical}) becomes
\begin{eqnarray}
    {\rm d} s^2
    &=&
    \mu^{-2} \left(\frac{T}{T_0}\right)^{2\beta^2-1/2} \left( - {\rm d} T^2
    + \mu^2 {\rm d} \theta^2 \right)
    + e^{2 \alpha} T^{1+2\beta} {\rm d} x^2
    + e^{-2 \alpha} T^{1-2\beta} {\rm d} y^2
    \ .
    \label{eq:Homogeneous Gowdy solution - Classical}
\end{eqnarray}
The constants $\mu$, $T_0$ and $\alpha$ can be absorbed in the definition of
coordinates. In proper time, defined by $\tau(T)\propto T^{\beta^2+3/4}$, we
then have the line element
\begin{equation}
  {\rm d}s^2 = -{\rm d}\tau^2+ \tau^{2p_1}{\rm d}\theta^2+ \tau^{2p_2}{\rm
    d}x^2+ \tau^{2p_3}{\rm d}y^2
\end{equation}
with exponents
\begin{equation} \label{Kasner}
  p_1=\frac{\beta^2-1/4}{\beta^2+3/4} \quad,\quad
  p_2=\frac{\beta+1/2}{\beta^2+3/4}\quad,\quad
  p_3=\frac{\beta-1/2}{\beta^2+3/4}
\end{equation}
that satisfy the Kasner relations $p_1+p_2+p_3=1=p_1^2+p_2^2+p_3^2$. The
background Kasner behavior is therefore determined by the observable
$G[1]$. If the periodic terms from a non-homogeneous $W$ in (\ref{eq:W general
  solution Gowdy T^3 - classical}) are included, they describe a polarized
gravitational wave travelling on a Kasner background.

\subsubsection{The flat Kasner solution}

As a special case, a flat solution within the Kasner class is defined by further taking
$\alpha=0$, $\beta = 1/2$, and $\mu=1$.
We obtain
\begin{eqnarray}
    a = 0
    \quad &,&\quad
    \bar{a} = \sqrt{T}
    \quad,\quad
    K = \mu / \sqrt{T}
    \ , \\
    \bar{W} = \frac{1}{2} \ln T
    \quad &,&\quad
    P_{\bar{W}} = 1
    \quad , \quad
    {\cal A} = 0
\end{eqnarray}
and the space-time metric (\ref{eq:General Gowdy solution - Classical}) becomes
\begin{eqnarray}
    {\rm d} s^2
    &=&
    - {\rm d} T^2
    + {\rm d} \theta^2
    + T^2 {\rm d} x^2
    + {\rm d} y^2
    \ .
    \label{eq:Homogeneous Kasner solution - Classical}
\end{eqnarray}

The Ricci scalar vanishes, which means that this expression may be considered
a vacuum solution. (The $T-x$ part is a 2-dimensional Milne model.)
Upon applying the coordinate transformation
\begin{equation}    \label{eq:Kasner-Minkowski coordinate transformation - classical}
    t_{\rm M} = T \cosh x
    \quad,  \quad x_{\rm M} = T \sinh x
\end{equation}
with inverse
\begin{equation}
  T^2 = t_{\rm M}^2 - x_{\rm M}^2
    \quad, \quad x = {\rm arctanh} \frac{x_{\rm M}}{t}
\end{equation}
such that
\begin{equation}
  - {\rm d} t_{\rm M}^2 + {\rm d} x_{\rm M}^2 = 
    - {\rm d} T^2
    + T^2 {\rm d} x^2
    \ , 
\end{equation}
the Kasner-like line element (\ref{eq:Homogeneous Kasner solution -
  Classical}) becomes Minkowskian,
\begin{eqnarray}
    {\rm d} s^2
    &=&
    - {\rm d} t_{\rm M}^2
    + {\rm d} \theta^2
    + {\rm d} x_{\rm M}^2
    + {\rm d} y^2
    \ .
\end{eqnarray}

While this result shows that the specific Kasner solution (\ref{eq:Homogeneous Kasner
  solution - Classical}) is a locally flat space-time, hypersurfaces of constant
time $T$ define a 3-dimensional space with non-vanishing extrinsic curvature
\begin{eqnarray}
    K_{a b} {\rm d} x^a {\rm d} x^b = \frac{1}{2} \{q_{a b} , H[1]\} {\rm d} x^a {\rm d} x^b 
    = T {\rm d} x^2
    \ .
\end{eqnarray}

\subsubsection{Homogeneous solution: Internal-time gauge}

In order to compare the results of the different classes of constraints, we
will evaluate them in the homogeneous case,
$P_{\bar{W}}'=\bar{W}'=\bar{a}'=\varepsilon'=K'=\mathcal{A}'=N'=0$, with vanishing
cosmological constant, $\Lambda=0$.
It will be convenient to work with coordinates adapted to the full range of
the curvature variable $K$, used as an internal time coordinate.
We define this internal-time gauge by
\begin{eqnarray}
    N^\theta = 0
    \ , \qquad
    K = T_K
    \ ,
    \label{eq:Internal time gauge}
\end{eqnarray}
with a new time coordinate $T_K$.

In the homogeneous case, as before, we obtain
\begin{eqnarray}
    \partial_{T_K}(\bar{a} K) = \{\bar{a} K , H[N] \} = - H N
    \ ,
\end{eqnarray}
which vanishes on-shell and implies
\begin{eqnarray}
    \bar{a} = \frac{\mu}{T_K} = \frac{\mu}{K}
    \ , \label{eq:a solution - internal time gauge - classical}
\end{eqnarray}
for some constant $\mu$. Because of homogeneity, the local version of the observable
(\ref{eq:Gowdy observable - classical}) is conserved,
$\dot{G}=2\pi\dot{P}_W=0$. Therefore, any $P_W$ in the constraints and equations
of motion is time-independent and can be set equal to
$P_W=2 \mu \beta$, defining the constant $\beta$ in the internal-time gauge.
Using this expression and the chain rule, we obtain the equations
\begin{eqnarray}
    \frac{{\rm d} \varepsilon}{{\rm d} K} &=& \frac{\dot{\varepsilon}}{\dot{K}}
    = - \frac{4 \varepsilon}{\left(1 + 4 \beta^2\right) K}
    \ , \\
    \frac{{\rm d} \bar{W}}{{\rm d} K} &=& \frac{\dot{\bar{W}}}{\dot{K}}
    = - \frac{4 \beta}{\left(1 + 4 \beta^2\right) K}
    \ ,
\end{eqnarray}
with solutions
\begin{eqnarray}
    \varepsilon &=& c_\varepsilon T_K^{- 4 / (1 + 4 \beta^2)}
    \ , \\
    \bar{W} &=& c_w - \frac{4 \beta}{1 + 4 \beta^2} \ln T_K
    = \ln \left( e^{c_w} T_K^{- 4 \beta/(1 + 4 \beta^2)} \right)
    \,.
\end{eqnarray}
The integration constants may be redefined as
\begin{equation}
    c_\varepsilon = \left(\mu T_0^{(4\beta^2-1)/4}\right)^{4/(4 \beta^2+1)}
    \quad , \quad
    e^{2 c_w} = e^{2\alpha} \left(\mu T_0^{(4\beta^2-1)/4}\right)^{8 \beta/(4 \beta^2+1)}
    \ .
    \label{eq:Integration constants of E and W}
\end{equation}

The on-shell condition $H_\theta=0$ is trivial in the homogeneous case, while
$H=0$ greatly simplifies and can be solved for
\begin{eqnarray}
    \mathcal{A} = \mu \frac{4 \beta^2 - 1}{4 \varepsilon}
    = \frac{4 \beta^2 - 1}{4 c_\varepsilon} \mu T_K^{4 \mu^2 / (\mu^2 + 4 \beta^2)}
    \ .
\end{eqnarray}

Finally, the lapse function is obtained by solving the consistency equation
$\dot{K}=\partial K/\partial T_K=1$,
\begin{eqnarray}
    N = - \frac{4}{1 + 4 \beta^2} \sqrt{c_\varepsilon} T_K^{-2 /(1 + 4 \beta^2) - 2}
    \ .
\end{eqnarray}
The negative value of the lapse function means that evolution runs from higher
to lower values of $T_K$, similar to what happens in Schwarzschild coordinates
in a black hole's interior.
The space-time metric (\ref{eq:ADM line element - Gowdy - new variables}) is then given by
\begin{eqnarray}
    {\rm d} s^2 &=&
    c_{\varepsilon}^{-1} T_K^{2 (1 - 4 \beta^2) / (4 \beta^2+1)} \left( - \left(\frac{4}{1 + 4 \beta^2}\right)^2 c_\varepsilon^2 T_K^{- 2 (5 + 4 \beta^2)/(4 \beta^2+1)} {\rm d} T_K^2
    + \mu^2 {\rm d} \theta^2 \right)
    \nonumber\\
    &&
    + c_\varepsilon \left( e^{2 c_w} T_K^{- 4 ( 2 \beta + 1 )/(4 \beta^2+1)} {\rm d} x^2
    + e^{-2 c_w} T_K^{4 (2 \beta - 1)/(4 \beta^2+1)} {\rm d} y^2 \right)
    \ .
    \label{eq:Homogeneous Gowdy solution - Classical - Internal time gauge}
\end{eqnarray}

The conventional time coordinate $T$ and our curvature time $T_K$ are related by
\begin{equation}
    T_K = \mu T_0^{\beta^2-1/4} T^{-\beta^2-1/4}
    \ .
\end{equation}
This coordinate transformation turns the metric (\ref{eq:Homogeneous Gowdy solution - Classical - Internal time gauge}) into (\ref{eq:Homogeneous Gowdy solution - Classical}).
The flat Kasner solution defined by $\alpha=0$, $\beta = 1/2$, and $\mu=1$, in
the present  gauge implies
\begin{eqnarray}
    \varepsilon &=& T_K^{- 2}
    \ , \\
    \bar{W} &=& - \ln T_K
    \ .
    \label{eq:Phase-space variables Kasner solution - Internal time - classical}
\end{eqnarray}
The coordinate transformation relating the two gauges is then simplified to 
\begin{equation}
    T_K = 1/\sqrt{T}
\end{equation}
and the space-time metric is given by
\begin{eqnarray}
    {\rm d} s^2 &=&
    - 4 T_K^{-6} {\rm d} T_K^2
    + {\rm d} \theta^2
    + T_K^{-4} {\rm d} x^2 + {\rm d} y^2
    \label{eq:Homogeneous Kasner solution - Classical - Internal time gauge}
\end{eqnarray}
with extrinsic curvature
\begin{equation}
    K_{a b} {\rm d} x^a {\rm d} x^b
    = T_K^{-2} {\rm d} x^2
\end{equation}
of constant-$T_K$-slices.  The coordinate singularity of (\ref{eq:Homogeneous
  Kasner solution - Classical}) at $T\to0_+$ is here given by
$T_K \to \infty$, while the coordinate singularity of (\ref{eq:Homogeneous
  Kasner solution - Classical - Internal time gauge}) at $T_K \to 0_+$
corresponds to $T\to \infty$.

\subsection{Singularity-free solutions}

We now consider the first two classes of modified theories, given by
constraints compatible with the limit of reaching the classical constraint
surface and constraints of the second kind. In both cases, the classical
singularity of homogeneous solutions is removed.

\subsubsection{New variables}

We first use the modified constraint (\ref{eq:Covariant Hamiltonian - Linear
  combination - new variables}) with constant $\lambda=\bar{\lambda}$ and $\lambda_0$,
choosing the gauge
\begin{equation}
    N^\theta = 0 \ , \hspace{1cm} \varepsilon = T
    \ .
\end{equation}
The on-shell conditions do not change under a linear combination of the
constraints, and thus we have the classical constraint surface given by
(\ref{eq:Constraints - classical - Conformal gauge}). 

The consistency equation $\dot{\varepsilon} = 1$ can be solved for the lapse function
\begin{equation}
    N = \frac{\lambda_0^{-1}}{\sqrt{\varepsilon}} \frac{1}{K \sqrt{1 - \bar{\lambda}^2 K^2}}
    \ .
\end{equation}
Using this, we obtain
\begin{eqnarray}
    \partial_T \left( \Bar{a} K \right) = \left\{ \Bar{a} K , \tilde{H} [N] \right\} \propto \tilde{H}
    \ ,
\end{eqnarray}
which vanishes on-shell. Thus, $\Bar{a} K = \mu$ with a constant $\mu$.

The $T-\theta$ part of the line element is no longer conformally flat because
we have the emergent metric component
\begin{eqnarray}
    \tilde{q}_{\theta\theta} &=& 
    \lambda_0^{-2} \frac{\bar{a}^2}{T} = \lambda_0^{-2} e^{2 a}
\end{eqnarray}
while
\begin{eqnarray}
    N^2
    = \frac{(\mu \lambda_0)^{-2}}{1 - \bar{\lambda}^2 K^2} \frac{\Bar{a}^2}{\varepsilon} = \frac{\mu^{-2}}{1 - \bar{\lambda}^2 K^2} \tilde{q}_{\theta \theta}
    = \mu^{-2} B^{-2} q_{\theta \theta}^{\rm (cl)}\,,
\end{eqnarray}
where $q_{\theta \theta}^{\rm (cl)} = \Bar{a}^2 / \varepsilon$ is the
classical expression and $B = \lambda_0 \sqrt{1 - \bar{\lambda}^2 K^2}$ is the factor in
the linear combination (\ref{eq:Covariant linear combination - classical -
  vacuum}).
Because $\varepsilon'=0$ in this gauge, the coefficient $A$ in the linear
combination vanishes.
Thus, in this gauge the full Hamiltonian generator is identical to the
classical one:
\begin{equation}
  \tilde{H}[N]+H_\theta[N^\theta]=\tilde{H}\left[B^{-1} \sqrt{\mu^{-1} q_{\theta
      \theta}^{\rm (cl)}}\right]= H\left[\sqrt{\mu^{-1} q_{\theta \theta}^{\rm (cl)}}\right]=
  H[N^{\rm (cl)}]\,.
\end{equation}
This means that all of the equations of motion are identical to the classical
ones, (\ref{eq:a EoM/constraints - classical - conformal gauge}) and (\ref{eq:W
  EoM - classical - conformal gauge}), obtaining the same classical solutions
(\ref{eq:W general solution Gowdy T^3 - classical}).

However, even though all the phase-space solutions retain their classical
forms, the resulting space-time geometry is non-classical because the structure
function differs by a constant factor of $\lambda_0^2$ from the classical one and
the lapse function differs from its classical expression more significantly.
The resulting emergent space-time line element is given by
\begin{eqnarray}
    {\rm d} s^2 &=&
    \frac{e^{2 a}}{\mu^{2} \lambda_0^2} \left( - \frac{{\rm d} T^2}{1-\bar{\lambda}^2 \mu^2 e^{- 2 a} / T} + \mu^{2} {\rm d} \theta^2 \right)
    + T \left( e^{2 W} {\rm d} x^2 + e^{- 2 W} {\rm d} y^2 \right)
    \ ,
\end{eqnarray}
where $a$ and $W$ are related to the phase-space variables by
(\ref{eq:Idenitification conventional and new variables - Gowdy}).
In the limit $\bar{\lambda}\to0$, $\lambda_0\to1$ we recover the classical
solution whose curvature invariant
$R_{\alpha \beta \mu \nu} R^{\alpha \beta \mu \nu}$ diverges as $T \to
0_+$. Seen from positive $T$, this singularity lies in the past where the
$x-y$ plane had collapsed to zero area $\varepsilon=T$.

In the modified case, the $T-\theta$ part of the metric is no longer
conformally flat. There is a new singularity at
$T =\bar{\lambda}^2 e^{-2 a} =: T_{\bar{\lambda}}$, a time later than the
classical singularity at $T=0_+$. This new singularity is therefore the
relevant one in the positive-$T$ branch, but it is not a physical singularity.
To see this in detail, we will use a new gauge in the next subsection.

For now, the special case of a (modified) flat Kasner solution can be analyzed
more easily. It is given by $\mu=1$, $\beta = 1/2$, $\alpha=a_n = b_n = 0$,
which implies $W = \frac{1}{2} \ln T$ and $a = 0$.
The line element is then equal to
\begin{eqnarray}
    {\rm d} s^2 &=&
    - \frac{{\rm d} T^2}{1-\bar{\lambda}^2 / T} + {\rm d} \theta^2
    + T^2 {\rm d} x^2 + {\rm d} y^2
    \ ,
    \label{eq:Kasner solution - CCSL}
\end{eqnarray}
where we have chosen $\lambda_0=1$ so as to recover the classical Kasner metric for large $T\gg \bar{\lambda}^2$.
In the classical case $\bar{\lambda} \to 0$ this is the flat Kasner solution
whose curvature invariants are finite. The solutions can be analytically
extended across $T=0$, but they are causally ill-behaved as they form closed
time-like curves.
However, for $\bar{\lambda} \neq 0$, the singularity $T=0_+$ is
hidden inside a region bounded by the new singularity at
$T= \bar{\lambda}^2 > 0$.
Curvature invariants can be used to support the expectation that this is only a coordinate singularity.

We first note that the Kasner solution (\ref{eq:Kasner solution -
  CCSL}) of the modified theory is not flat: It has the Ricci scalar
\begin{eqnarray}
    R = \frac{\bar{\lambda}^2}{T^3}
    \ ,
\end{eqnarray}
and the Kretschmann invariant
\begin{eqnarray}
    {\cal K} \equiv R_{\mu \nu \alpha \beta} R^{\mu \nu \alpha \beta}
    = - \frac{1}{2} \frac{\bar{\lambda}^4}{T^{10}} \left(1 + T^4 \left( 1 - \frac{\bar{\lambda}^2}{T} \right)^2\right)
    \ .
\end{eqnarray}
At $T= \bar{\lambda}^2$, both are finite. Moving across this value, a new
gauge must be chosen, in which, as we will see in what follows, the physical
singularity at $T=0$ no longer appears. (This construction is similar to the
non-singular black-hole models in \cite{SphSymmEff,SphSymmEff2}.)
Any hypersurface of constant time $T$ of the modified Kasner spacetime
(\ref{eq:Kasner solution - CCSL}) defines a 3-dimensional space with
non-vanishing extrinsic curvature
\begin{eqnarray}
    K_{a b} {\rm d} x^a {\rm d} x^b
    = T \sqrt{1 - \frac{\bar{\lambda}^2}{T}} {\rm d} x^2
\end{eqnarray}
unless $T=\bar{\lambda}$. This specific value implies a hypersurface of time-reflection
symmetry, which can be used to glue a time reverse of our solution at
$T=\bar{\lambda}$. The classical singularity at $T=0$ is then replaced by a
transition from collapse to expansion.

If the interpretation of the classical flat solution as a vacuum space-time is
extended to the modified theory, it could suggest a vacuum different from the
usual Minkowski one, being approximately flat only for $T\gg
\bar{\lambda}^2$. However, in Section~\ref{s:flat} we will show that drawing
such a conclusion only based on homogeneous models in a fixed gauge of
internal time would not be justified. For now, we continue our analysis of
homogeneous dynamics.

\subsubsection{Periodic variables}

We shall now use the modified constraint (\ref{eq:Covariant Hamiltonian -
  Linear combination - new holonomy variables}) with constant $\lambda_0$ and
$\lambda=\bar{\lambda}$, reproducing the above results because this version will
serve as a guide to obtaining the dynamical solutions of the other two
constraints.

We again choose the gauge
\begin{equation}
    N^\theta = 0 \ , \hspace{1cm} \varepsilon = T
    \ .
\end{equation}
The on-shell conditions are
\begin{eqnarray}
    0 &=&
    \frac{1}{4 \varepsilon} \frac{\sin^2 (\bar{\lambda} K)}{\bar{\lambda}^2}
    - \frac{\cos^2 (\bar{\lambda} K)}{4 \varepsilon} \frac{P_{\Bar{W}}{}^2}{\Bar{a}^2}
    + \frac{\sin (2 \bar{\lambda} K)}{2 \bar{\lambda}} \frac{\mathcal{A}}{\Bar{a}}
    - \varepsilon \frac{(\Bar{W}')^2}{\Bar{a}^2} \cos^2 (\bar{\lambda} K)
    \ , \\
    0 &=&
    \Bar{a} K' + P_{\Bar{W}} \Bar{W}'
\end{eqnarray}
and the consistency equation $\dot{\varepsilon} = 1$ can be solved for the
lapse function
\begin{equation}
    N = \frac{\lambda_0^{-1}}{\sqrt{\varepsilon}} \frac{2 \bar{\lambda}}{\sin (2 \bar{\lambda} K)}
    \ .
\end{equation}
Using this result, we obtain
\begin{eqnarray}
    \partial_T \left( \Bar{a} \frac{\tan (\bar{\lambda} K)}{\bar{\lambda}} \right) = \left\{ \Bar{a} \frac{\tan (\bar{\lambda} K)}{\bar{\lambda}} , \tilde{H} [N] \right\} \propto \tilde{H}
    \ ,
\end{eqnarray}
which vanishes on-shell. Thus,
$\Bar{a} \tan (\bar{\lambda} K) / \bar{\lambda} = \mu$ where $\mu$ is a
constant.  Because of the canonical transformation involved, the
identification (\ref{eq:Idenitification conventional and new variables -
  Gowdy}) changes to
\begin{eqnarray}
    \Bar{W} = W
    \quad , \quad
    a = \ln \sqrt{E^x E^y / \varepsilon} = \ln \left( \frac{\Bar{a}}{\cos (\bar{\lambda} K)} \right) - \ln \sqrt{\varepsilon}
    \ .
    \label{eq:Idenitification conventional and periodic variables - Gowdy}
\end{eqnarray}
Therefore,
\begin{equation}
  \sin(\bar{\lambda} K)=\frac{\bar{\lambda}\mu}{\bar{a}/\cos(\bar{\lambda} K)}=
  \frac{\bar{\lambda}\mu}{\sqrt{\varepsilon} e^{a}}\,.
\end{equation}
The equations of motion for $\Bar{a}$ and $\Bar{W}$ respectively give
\begin{eqnarray}
    \mathcal{A} &=& \mu \partial_T \left( \ln \left( \frac{\Bar{a}}{\cos(\bar{\lambda} K)} \right) - \ln \sqrt{T}\right)
    = \mu \Dot{a}
    \ , \\
    P_{\Bar{W}} &=& 2 \mu T \dot{\Bar{W}}
    \ .
\end{eqnarray}
Using these results, the constraints $H_\theta=0$ and $\tilde{H}=0$ can be rewritten as
\begin{equation}   \label{eq:a EoM/constraints - classical - conformal gauge - periodic variables}
    a' =
    2 T \Dot{W} W'
    \quad , \quad
    \Dot{a} =
    - \frac{1}{4 T} + T \left(\Dot{W}^2 + \frac{(W')^2}{\mu^2}\right)
    \ ,
\end{equation}
where we have used the identification (\ref{eq:Idenitification conventional
  and periodic variables - Gowdy}). The equation of motion $\Ddot{\Bar{W}} =
\{\{\Bar{W} , \tilde{H}[N]\} , \tilde{H}[N]\}$, such that the brackets
\emph{do act} on the lapse as discussed before, can be rewritten as 
\begin{eqnarray}
    0 &=&
    \Ddot{W} + \frac{\Dot{W}}{T} - \frac{W''}{\mu^2}
    \ .
    \label{eq:W EoM - periodic - conformal gauge}
\end{eqnarray}
These equations are identical to the classical ones.
The emergent line element is then given by
\begin{eqnarray}
    {\rm d} s^2 &=&
    \lambda_0^{-2} e^{2 a} \left( - \sec^2 (\bar{\lambda} K) {\rm d} T^2 + {\rm d} \theta^2 \right)
    + T \left( e^{2 W} {\rm d} x^2 + e^{- 2 W} {\rm d} y^2 \right)\nonumber\\
  &=&  \lambda_0^{-2} e^{2 a} \left( - \frac{ {\rm d} T^2}{1-\bar{\lambda}^2\mu^2/(T e^{2a})} + {\rm d} \theta^2 \right)
+ T \left( e^{2 W} {\rm d} x^2 + e^{- 2 W} {\rm d} y^2 \right)
    \,.
\end{eqnarray}

Modified Kasner models are obtained for $W=\beta\ln T$, which implies
$e^{2a}\propto T^{2\beta^2-1/2}$ as in the classical case. The line element
then equals
\begin{equation} \label{dsT}
  {\rm d}s^2 = \lambda_0^{-2}T^{2\beta^2-1/2} \left( - \frac{ {\rm d} T^2}{1-\bar{\lambda}^2T^{-2\beta^2-1/2}} + {\rm d} \theta^2 \right)
    + T^{1+2\beta} {\rm d} x^2 + T^{1-2\beta} {\rm d} y^2
\end{equation}
if we set $\mu^2$ equal to the proportionality factor in $e^{2a}$. For
constant but non-zero $\lambda$, proper time is now given by a hypergeometric
function of $T$, which complicates any further analysis of general Kasner
models. It is nevertheless possible to understand the general behavior.

To do so, we first introduce a new time coordinate $t=T^{\beta^2+3/4}$, such
that
\begin{equation}
 {\rm d}\tau\propto \frac{T^{\beta^2-1/4}}{\sqrt{1-\bar{\lambda}^2 T^{-2\beta^2-1/2}}}
   {\rm d}T= \frac{1}{\beta^2+3/4} \frac{{\rm d}t}{\sqrt{1-\bar{\lambda}^2 t^{-2(\beta^2+1/4)/(\beta^2+3/4)}}}
\end{equation}
according to the time component of (\ref{dsT}). For large $T$, $t$ and $\tau$
therefore proceed at almost the same rate, up to a constant rescaling. With
respect to proper time, we now have the line element
\begin{equation}
  {\rm d}s^2=- {\rm d}\tau^2+ \lambda_0^{-2} t(\tau)^{2p_1}{\rm
    d}\theta^2+ t(\tau)^{2p_2} {\rm d}x^2+t(\tau)^{2p_3}{\rm d}y^2
\end{equation}
with Kasner exponents $p_i$ as in (\ref{Kasner}), obeying the classical
relations, and the inverse of a hypergeometric 
function (times $t$) for $t(\tau)$. (For $\bar{\lambda}=1$, $t(\tau)$ is the inverse of $t
\,_2F_1(1/2,1/a;1+1/a;t^a)$ if $a=-2(\beta^2+1/4)/(\beta^2+3/4)\not=-1$. If
$a=-1$, $t(\tau)$ is the inverse of $\sqrt{(t-1)t}-\sinh^{-1}(t)$.) 

\begin{figure}
\begin{center}
  \includegraphics[width=16cm]{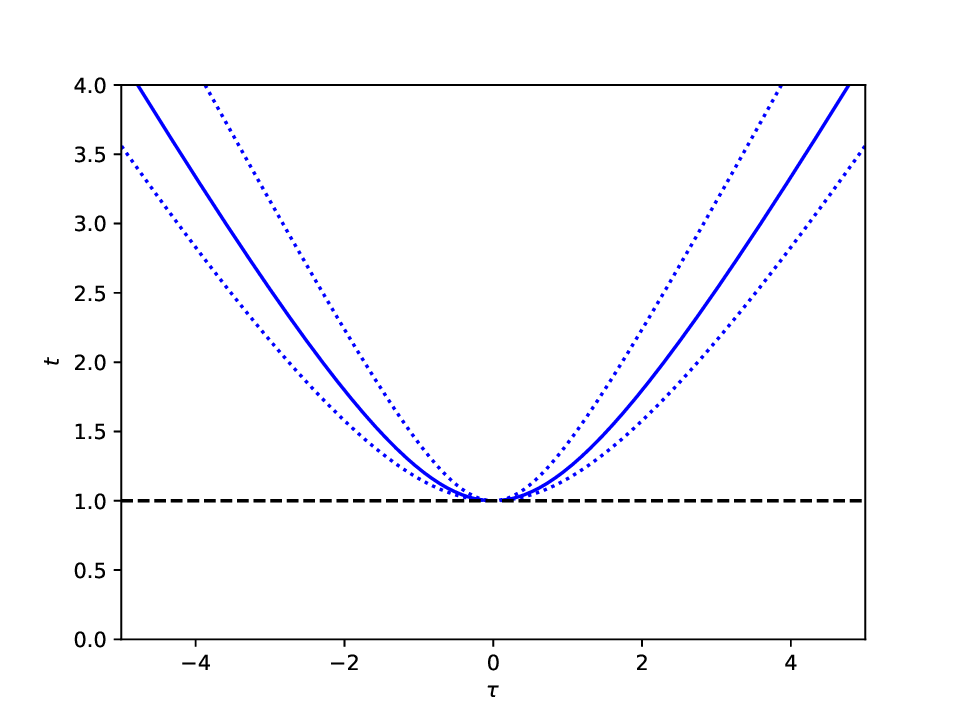}
  \caption{The function of $t(\tau)$ with $\bar{\lambda}=1$, obtained from the
    inverse of a hyperbolic 
    function, shows the transition from collapse to expansion in Kasner models
    of emergent modified gravity. Its dependence on different values of
    $\beta$ is shown here by the range of possible curves, with the upper
    bound given by large $\beta$ such that
    $-2(\beta^2+1/4)/(\beta^2+3/4)\approx-2$ and the lower bound given by
    $\beta=0$ such that $-2(\beta^2+1/4)/(\beta^2+3/4)=-2/3$ (dotted
    curves). The value $\beta=1/2$ (solid), such that
    $-2(\beta^2+1/4)/(\beta^2+3/4)=-1$, is close to the midpoint of this range.
    \label{f:ttau}}
\end{center}
\end{figure}

\begin{figure}
\begin{center}
  \includegraphics[width=16cm]{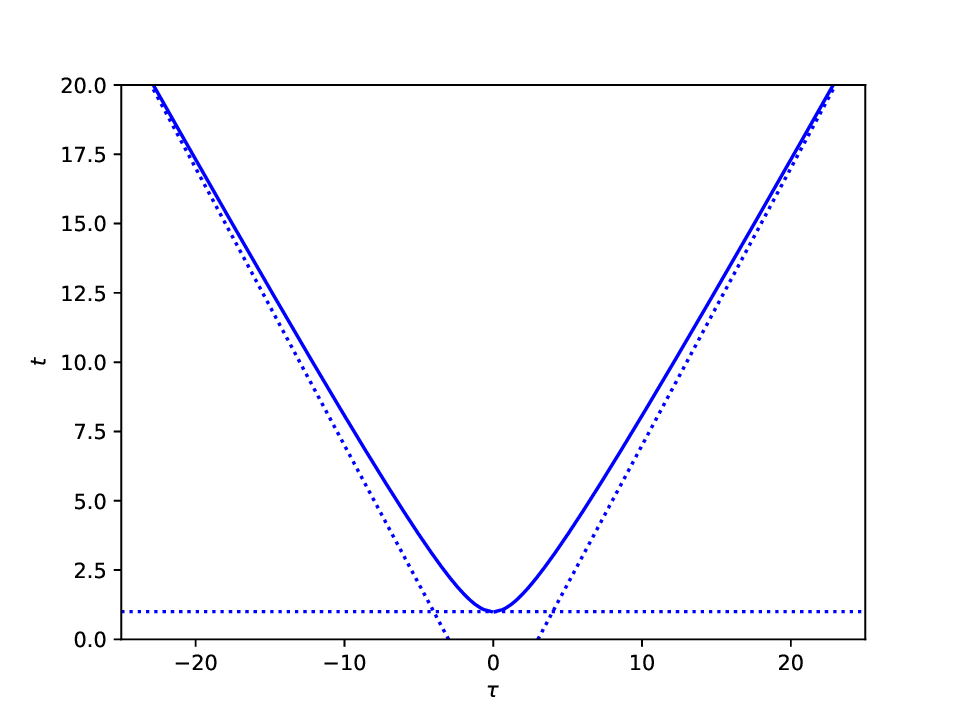}
  \caption{The asymptotic behavior of function of $t(\tau)$ for large $\tau$
    is close to $t(\tau)=\pm\tau+t_{\pm}$ if $\bar{\lambda}=1$. The value
    $\beta=1/4$ has been used for the solid curve.
    \label{f:ttaularge}}
\end{center}
\end{figure}

For large $t(\tau)$ such that
$\bar{\lambda} \ll T^{\beta^2+1/4}=t^{(\beta^2+1/4)/(\beta^2+3/4)}$, the behavior is
close to the classical Kasner dynamics with the same relationship
between the Kasner exponents and the conserved quantity $\beta$. For smaller
$t$, however, there is a new effect because the relationship between $t$ and
$\tau$ is not one-to-one, in contrast to the classical solutions. We have
\begin{equation}
  \frac{{\rm d}t}{{\rm d}\tau}= \lambda_0(\beta^2+3/4) \sqrt{1-\bar{\lambda}^2
    t^{-2(\beta^2+1/4)/(\beta^2+3/4)}}=0
\end{equation}
at
\begin{equation}
  t=t_{\bar{\lambda}}=\bar{\lambda}^{(\beta^2+3/4)/(\beta^2+1/4)}\,.
\end{equation}
At the same value of $t$,
\begin{eqnarray}
  \frac{{\rm d}^2t}{{\rm d}\tau^2}&=& \lambda_0 \bar{\lambda}^2(\beta^2+1/4)
  \frac{t^{-(3\beta^2+5/4)/(\beta^2+3/4)}}{\sqrt{1-\bar{\lambda}^2 
      t^{-2(\beta^2+1/4)/(\beta^2+3/4)}}} \frac{{\rm d}t}{{\rm d}\tau}\nonumber\\
&=& \lambda_0^2\bar{\lambda}^2(\beta^2+1/4)(\beta^2+3/4)
    t^{-(3\beta^2+5/4)/(\beta^2+3/4)}\nonumber\\
  &=&
\lambda_0^2(\beta^2+1/4)(\beta^2+3/4) t_{\bar{\lambda}}^{-1}>0
\end{eqnarray}
such that $t(\tau)$ has a local minimum at the value
$t(\tau)=t_{\bar{\lambda}}$. The full dynamics therefore describes non-singular
evolution of a collapsing Kasner model connected to an expanding Kasner model
with the same exponents. All three spatial directions transition from collapse
to expansion at the same time $\tau(t_{\bar{\lambda}})$. The behavior of
$t(\tau)$ is illustrated in Figs.~\ref{f:ttau} and \ref{f:ttaularge}.

The special case of the modified flat Kasner model is given by $\mu=1$,
$\beta = 1/2$, $\alpha=a_n = b_n = 0$, which implies $W = \frac{1}{2} \ln T$
and $a = 0$.
We have $\mathcal{A}=0$ and
\begin{eqnarray}
    \frac{\sin(\bar{\lambda} K)}{\bar{\lambda}}
    = \frac{1}{\sqrt{T}}
    \ ,
\end{eqnarray}
and the line element
\begin{eqnarray}
    {\rm d} s^2 &=&
    \lambda_0^{-2} \left( - \sec^2 (\bar{\lambda} K) {\rm d} T^2 + {\rm d} \theta^2 \right)
    + T^2 {\rm d} x^2
    + {\rm d} y^2
    \ .
\end{eqnarray}
Here, proper time $\tau(T)$ can be integrated more easily but its inversion to
$T(\tau)$ remains complicated.

\subsubsection{Homogeneous solution: Internal-time gauge}

Let us now use the inhomogeneous curvature component as an internal time, $T_K = K$.
The two time coordinates are related to each other by 
\begin{equation}
    \frac{\sin(\bar{\lambda} T_K)}{\bar{\lambda}}
    = \frac{1}{\sqrt{T}}
\end{equation}
such that
\begin{equation}
  - 2 \frac{\bar{\lambda}^3}{\sin^3 (\bar{\lambda} T_K)} \cos (\bar{\lambda} T_K) {\rm d} T_K = {\rm d} T
    \ .
\end{equation}
Substituting in the line element for the modified flat Kasner model, we obtain
\begin{eqnarray}
    {\rm d} s^2 &=&
    \lambda_0^{-2} \left( - 4 \frac{\bar{\lambda}^6}{\sin^6 (\bar{\lambda} T_K)} {\rm d} T_K^2 + {\rm d} \theta^2 \right)
    + \frac{\bar{\lambda}^4}{\sin^4(\bar{\lambda} T_K)} {\rm d} x^2
    + {\rm d} y^2
    \ ,
\end{eqnarray}
which is indeed regular at maximum curvature, $T_K = \pi / 2 \bar{\lambda}$,
defining a surface of reflection symmetry.

We now derive this result by directly solving the equations of motion in the
internal-time gauge (\ref{eq:Internal time gauge}), rather than performing a
coordinate transformation.  For the homogeneous model, we set
$P_{\bar{W}}'=\bar{W}'=\bar{a}'=\varepsilon'=K'=\mathcal{A}'=N'=0$ and assume
a vanishing cosmological constant, $\Lambda=0$.
We note that the modified constraints (\ref{eq:Hamiltonian constraint -
  periodic - CCSL - scalar polymerized}) and (\ref{eq:Hamiltonian constraint -
  DF - scalar polymerization}) are identical in the homogeneous case if the
classical values for the functions $c_f , \alpha_2,\alpha_3 \to 1$, $q \to 0$,
and $\Lambda_0 \to \Lambda \to 0$ are taken, with constant
$\lambda=\bar{\lambda}$, $\nu=\bar{\nu}$ and $\lambda_0$.
The results of the present and the following subsections then apply to both cases.

We first see that because of homogeneity the local version of the observable
(\ref{eq:Gowdy observable - classical}) is conserved,
$\dot{G}=2\pi \dot{P}_W=0$, and we will write the momentum as
$P_W=2 \mu \beta$ with constants $\mu$ and $\beta$.
The on-shell condition $H_\theta=0$ is trivially satisfied in this case, while $\tilde{H}=0$ is solved by
\begin{eqnarray}
    \mathcal{A} =
    - \frac{\bar{a}}{4 \varepsilon} \frac{\tan (\bar{\lambda} K)}{\bar{\lambda}}
    + \frac{\bar{\lambda} \cot (\bar{\lambda} K)}{4 \varepsilon} \frac{4 \mu^2 \beta^2}{\Bar{a}}
    \ .
\end{eqnarray}
We then obtain
\begin{eqnarray}
    \partial_T \left( \Bar{a} \frac{\tan (\bar{\lambda} K)}{\bar{\lambda}} \right) = \left\{ \Bar{a} \frac{\tan (\bar{\lambda} K)}{\bar{\lambda}} , \tilde{H} [N] \right\} \propto \tilde{H}
    \ ,
\end{eqnarray}
which vanishes on-shell, such that $\Bar{a} \tan (\bar{\lambda} K) / \bar{\lambda} = \mu$
where $\mu$ is a constant. Hence,
\begin{eqnarray}
    \mathcal{A} =
    \left( 4 \beta^2 - 1 \right) \frac{\mu}{4 \varepsilon}
    \ .
\end{eqnarray}

In combination with the chain rule we obtain the equations
\begin{eqnarray}
    \frac{{\rm d} \varepsilon}{{\rm d} K} &=& \frac{\dot{\varepsilon}}{\dot{K}}
    = - \frac{4 \bar{\lambda}}{(1+4 \beta^2) \tan (\bar{\lambda} K)} \varepsilon
    \ , \\
    \frac{{\rm d}}{{\rm d} K} \left(\frac{\sin(\bar{\nu}\bar{W})}{\bar{\nu}}\right) &=& - \frac{4 \beta \bar{\lambda}}{(1+4 \beta^2) \tan (\bar{\lambda} K)}
    \ ,
\end{eqnarray}
solved by
\begin{equation}
    \varepsilon = c_\varepsilon \left(\frac{\sin (\bar{\lambda} K)}{\bar{\lambda}}\right)^{-4/(1+4 \beta^2)}
\end{equation}
and
\begin{equation}
    \frac{\sin(\bar{\nu}\bar{W})}{\bar{\nu}} = \ln \left(e^{c_w} \frac{\sin (\bar{\lambda} K)}{\bar{\lambda}}\right)^{-4 \beta/(1+4 \beta^2)}\,.
    \label{eq:W solution - CCSL}
\end{equation}
For convenience, the integration constants may be redefined as
\begin{equation}
    c_\varepsilon = \left(\mu T_0^{(4\beta^2-1)/4}\right)^{4/(4 \beta^2+1)}
\end{equation}
and
\begin{equation}
  e^{2 c_w} = e^{2\alpha} \left(\mu T_0^{(4\beta^2-1)/4}\right)^{8 \beta/(4 \beta^2+1)}
    \ .
  \end{equation}
  
Finally, the lapse function is obtained by solving the consistency equation $\dot{K}=1$,
\begin{eqnarray}
    N = - \frac{4}{1 + 4 \beta^2} \frac{\sqrt{c_\varepsilon}}{\lambda_0} \sec^2 (\bar{\nu}\bar{W}) \left(\frac{\sin (\bar{\lambda} K)}{\bar{\lambda}}\right)^{-2 /(1 + 4 \beta^2) - 2}
    \ .
\end{eqnarray}
The space-time line element is then given by
\begin{eqnarray}
  {\rm d} s^2&=&
    \sec^4 (\bar{\nu} \bar{W}) \left(\frac{\sin (\bar{\lambda}
                 T_K)}{\bar{\lambda}}\right)^{4/(1+4 \beta^2)-2}\nonumber\\
  &&\qquad\times\left( - \left(\frac{4}{1 + 4 \beta^2}\right)^2 \frac{c_\varepsilon}{\lambda_0^2} \left(\frac{\sin (\bar{\lambda} T_K)}{\bar{\lambda}}\right)^{-8 /(1 + 4 \beta^2) - 2} {\rm d} T_K^2
    + \frac{\mu^2}{c_\varepsilon} {\rm d} \theta^2 \right)
    \nonumber\\
    &&
    + c_\varepsilon  e^{2 c_w} \left(\frac{\sin (\bar{\lambda}
       T_K)}{\bar{\lambda}}\right)^{- 4 ( 2 \beta + 1 )/(4 \beta^2+1)} {\rm d}
       x^2\nonumber\\ 
 &&   + c_{\varepsilon} e^{-2 c_w} \left(\frac{\sin (\bar{\lambda}
    T_K)}{\bar{\lambda}}\right)^{4 (2 \beta - 1)/(4 \beta^2+1)} {\rm d} y^2  
    \label{eq:Homogeneous Gowdy solution - CCSL - Internal time gauge}
\end{eqnarray}
where $\bar{W}$ is implicitly given by (\ref{eq:W solution - CCSL}).

\subsubsection{Modified flat Kasner solution}

The Kasner solution $\mu=1$, $\beta=1/2$ is given by the simpler metric
\begin{equation}
    {\rm d} s^2 =
    \sec^4 (\bar{\nu} \bar{W}) \left( - \frac{4}{\lambda_0^2} \left(\frac{\sin (\bar{\lambda} T_K)}{\bar{\lambda}}\right)^{-6} {\rm d} T_K^2
    + {\rm d} \theta^2 \right)
    + \left(\frac{\sin (\bar{\lambda} T_K)}{\bar{\lambda}}\right)^{- 4} {\rm d} x^2
    + {\rm d} y^2
    \ ,
    \label{eq:Homogeneous Kasner solution - CCSL - Internal time gauge}
\end{equation}
where $\bar{W}$ can be obtained from inverting its relation with $K$,
\begin{eqnarray}
    \frac{\sin (\bar{\lambda} T_K)}{\bar{\lambda}} = \exp \left(- \frac{\sin(\bar{\nu}\bar{W})}{\bar{\nu}} \right)
    \ .
\end{eqnarray}

Taking $\bar{\nu}\to 0$, the Kasner solution in this gauge (\ref{eq:Homogeneous Kasner solution - CCSL - Internal time gauge}) has the Ricci scalar
\begin{eqnarray} \label{Ricci}
    R = \bar{\lambda}^2 \left(\frac{\sin (\bar{\lambda} T_K)}{\bar{\lambda}}\right)^6
    \ ,
\end{eqnarray}
and the Kretschmann invariant
\begin{eqnarray}
    {\cal K} \equiv R_{\mu \nu \alpha \beta} R^{\mu \nu \alpha \beta}
    = - \frac{\bar{\lambda}^4}{8} \left(\frac{\sin (\bar{\lambda} T_K)}{\bar{\lambda}}\right)^{22}
    \ .
\end{eqnarray}
Both expressions are finite at $T_K= \pi / (2\bar{\lambda})$.
The model is approximately flat only for $T_K\ll \bar{\lambda}$, and both
curvature invariants vanish in the classical limit $\bar{\lambda}\to0$.

A hypersurface of constant time $T_K$ of the modified Kasner space-time
(\ref{eq:Homogeneous Kasner solution - CCSL - Internal time gauge}) defines a
3-dimensional space with extrinsic curvature
\begin{equation}
    K_{a b} {\rm d} x^a {\rm d} x^b
    = \lambda_0 \frac{\bar{\lambda}^2 \cos (\bar{\lambda} T_K)}{\sin^2 (\bar{\lambda} T_K)} {\rm d} \theta^2
\end{equation}
which vanishes only at the maximum-curvature hypersurface.

\subsubsection{Flat solution: Non-unique vacuum}
\label{s:flat}

From the above example one might conclude that the vacuum solution of this
theory is different from Minkowski space-time, as suggested for a similar case
for instance in \cite{KasnerFLat}. It is easy to see that such a statement is
incorrect because flat space-time, described by
\begin{eqnarray} \label{flat}
    && N = 1 \ ,\ N^\theta = 0
    \ ,\nonumber\\
    && \varepsilon = 1 \ , \ \bar{a} = 1 \ , \ \bar{W} = 0
    \ , \nonumber\\
    &&{\cal A} = 0 \ , \ K = 0 \ , \ P_{\bar{W}} = 0
    \ ,
\end{eqnarray}
is a solution to the same theory if, to be specific, we take the classical
values for all the modification functions except for $\lambda$ and $\nu$ which
we leave as arbitrary functions. This solution is excluded from the case of
Kasner-like line elements by the assumption that $\varepsilon$ can be used as
a time variable, such that $\varepsilon=1$ can be obtained only on one
spacelike hypersurface but not across an entire space-time region.
Nevertheless, flat Minkowski space-time is a solution of the same modified
theory in which we obtained our Kasner space-times. 

Minkowski space-time is relevant because its local behavior describes the
background space-time of vacuum states in quantum field theory. According to
the general meaning of ``vacuum'' in particle physics or general relativity,
all Kasner models are vacuum solutions because they do not include matter. The
case of $\beta=1/2$ is special only because, classically, it happens to be
locally equivalent to Minkowski space-time and just appears written in
non-Cartesian coordinates. The result that this correspondence is not realized
in a modified theory only means that there is no longer a Kasner model
related to flat Minkowski space-time. It does not mean that Minkowski
space-time itself is modified or no longer appears as a solution, as
demonstrated by the explicit counter-example of (\ref{flat}).

A distinguishing feature of (\ref{flat}) compared with any Kasner solution is
that it is not only devoid of matter, but also has a vanishing local
gravitational degrees of freedom described by $(\bar{W},P_{\bar{W}})$. We can
formalize this property by making use of the definition of an effective
stress-energy, obtained from the Einstein tensor of the emergent space-time
metric. We find that the flat solution (\ref{flat}) has a vanishing net
stress-energy tensor, while the modified flat Kasner solution has a
non-trivial one, as shown by the non-vanishing Ricci scalar (\ref{Ricci}).
Therefore, the solutions are distinguished from one another by their
effective gravitational energy content. From this perspective, the standard Minkowski
solution remains the preferred vacuum space-time also in a modified theory. For
$\bar{\lambda}\to0$, the effective stress-energy tensor vanishes, and the
modified flat Kasner solution approaches the strictly flat Minkowski
solution.

The correct identification of a candidate vacuum solution therefore requires
an extension of strict minisuperspace models to some inhomogeneity, which
tells us that the non-zero $\bar{W}$ and $P_{\bar{W}}$ in Kasner models are
homogeneous remnants of a propagating gravitational degree of freedom, and the
correct identification of a covariant space-time structure that defines
curvature and effective stress-energy. It is also important to have a
gauge-invariant treatment that is not built on a fixed gauge choice such as an
internal time, as such a choice might restrict the accessible solution
space. None of these ingredients had been available in previous models of
quantum cosmology.  With some choices of modification functions, it might happen that
strict Minkowski space-time is no longer a solution or that the zero-mass
limit of a black-hole solution differs from Minkowski space-time as seen
explicitly in an example in \cite{EmergentMubar}. But such a conclusion can not be
drawn in a reliable manner in theories based on restriced gauge choices or on
incomplete demonstrations of covariance properties.

\subsection{Constraints compatible with the classical-$\bar{W}$ limit}

We now use the constraint (\ref{eq:Hamiltonian constraint - periodic - CWL})
in the internal time gauge (\ref{eq:Internal time gauge}) for the homogeneous
case where $P_{\bar{W}}'=\bar{W}'=\bar{a}'=\varepsilon'=K'=\mathcal{A}'=N'=0$.
For simplicity we take the classical values for the following modification
functions and assume a vanishing cosmological constant, $c_f , \alpha_2,\alpha_3 \to 1$,
$q \to 0$, and $\Lambda_0 \to \Lambda \to 0$.
We also set $\lambda_0$, $\lambda=\bar{\lambda}$, and $\nu=\bar{\nu}$
constant. With these values, the inhomogeneous component of the emergent spatial metric
is given by
\begin{eqnarray} \label{struchom}
    \tilde{q}_{\theta\theta}
    = \lambda_0^{-2}\cos^{-4} (\bar{\nu} \bar{W})\frac{\bar{a}^2}{ \cos^2
  \left(\bar{\lambda} K\right)}
  \frac{1}{\varepsilon}\,.
\end{eqnarray}

As before, homogeneity implies that the local version of the observable
(\ref{eq:Gowdy observable - classical}) is conserved, $\dot{G}=0$, and we
shall write it as $G=4\pi \mu \beta$ such that $P_{\bar{W}}=2\mu\beta$,
anticipating an integration constant ($\mu$) that will be introduced in the
process of solving equations of motion. The value of $\beta$ then
parameterizes the momentum.

The relevant equations of motion for recovering the emergent space-time geometry are given by
\begin{eqnarray}
  &&  \frac{{\rm d} \ln \left(\bar{a}^2/\cos^2\left(\bar{\lambda} K\right)\right)}{{\rm d} \left( \sin \left(\bar{\lambda} K\right) / \bar{\lambda} \right)}\nonumber\\
    &=&
    - 2 \frac{\bar{\lambda}}{\sin \left(\bar{\lambda} K\right)} \left(
    \frac{\sin^2 (\bar{\lambda} K)}{\bar{\lambda}^2} \cos^2\left(\bar{\lambda} K\right) \frac{\bar{a}^2}{\cos^2\left(\bar{\lambda} K\right)}
    + 4 \mu^2 \beta^2 \frac{\cos\left(2\bar{\lambda} K\right)}{|\cos\left(\bar{\lambda} K\right)|} \right)
    \nonumber\\
    && \times
    \left( \frac{\sin^2(\bar{\lambda} K)}{\bar{\lambda}^2} \cos^2\left(\bar{\lambda} K\right) \frac{\bar{a}^2}{\cos^2\left(\bar{\lambda} K\right)} + 4 \mu^2 \beta^2 |\cos\left(\bar{\lambda} K\right)|\right)^{-1}
    \label{eq:a EoM - Homogeneous - CML}
\end{eqnarray}
as well as
\begin{eqnarray}
    \frac{{\rm d} \ln \varepsilon}{{\rm d} K}
    &=&
    - 4 \frac{\sin \left(2 \bar{\lambda} K\right)}{2 \bar{\lambda}}
    \left(
    \frac{\sin^2\left(\bar{\lambda} K\right)}{\bar{\lambda}^2} |\cos\left(\bar{\lambda} K\right)|
    + 4 \mu^2 \beta^2 \frac{\cos^2(\bar{\lambda} K)}{\bar{a}^2}
    \right)^{-1}
    \label{eq:E EoM - Homogeneous - CML}
\end{eqnarray}
and
\begin{eqnarray}
    &&\frac{{\rm d}}{{\rm d} K} \left(\frac{\sin
       (\bar{\nu}\bar{W})}{\bar{\nu}}\right)\nonumber\\ 
    &=&
    - 4 \mu \beta \frac{\cos^2\left(\bar{\lambda} K\right)}{\bar{a}}
    \Bigg(
    \frac{\sin^2\left(\bar{\lambda} K\right)}{\bar{\lambda}^2} |\cos\left(\bar{\lambda} K\right)|
    + 4 \mu^2 \beta^2 \frac{\cos^2\left(\bar{\lambda} K\right)}{\bar{a}^2} 
    \Bigg)^{-1}
    \label{eq:W EoM - Homogeneous - CML}
\end{eqnarray}
where we have chosen $K$ as an evolution parameter.
Equation (\ref{eq:a EoM - Homogeneous - CML}) can be solved exactly,
\begin{eqnarray}
    \frac{\bar{a}^2}{\cos^2 (\bar{\lambda} K)}
    &=&
    \frac{\mu^2}{4} \frac{\bar{\lambda}^2}{\sin^2(\bar{\lambda} K)}
    \left( 1-4 \beta^2 + \sqrt{ (1-4 \beta^2)^2 + \frac{16 \beta^2}{|\cos(\bar{\lambda} K)|}} \right)^2
    \,.
    \label{eq:a - Homogeneous - simple case - CML}
\end{eqnarray}
The integration constant $\mu^2$ and the sign of the square
root have been chosen for the solution to match the classical one, (\ref{eq:a
  solution - internal time gauge - classical}), in the limit
$\bar{\lambda}\to 0$.

The ratio (\ref{eq:a - Homogeneous - simple case - CML}) appears directly as a
factor in the emergent metric component $q_{\theta\theta}$, given by
(\ref{struchom}).  Near the maximum-curvature hypersurface, defined by
$K\to \pi / (2\bar{\lambda})$, this expression diverges as
$\sec (\bar{\lambda} K)$, and its internal-time derivative (\ref{eq:a EoM
  - Homogeneous - CML}) diverges as $\sec^2 (\bar{\lambda} K)$.
Using this, we find that the right-hand side of (\ref{eq:E EoM - Homogeneous - CML}) remains finite at the maximum-curvature hypersurface,
\begin{eqnarray}
    \frac{{\rm d} \ln \varepsilon}{{\rm d} K}
    &\approx& - 4 \frac{\bar{\lambda}}{\sin (\bar{\lambda} K)}
    \ ,
\end{eqnarray}
while that of (\ref{eq:W EoM - Homogeneous - CML}) vanishes.
We conclude that both $\varepsilon$ and $\sin (\bar{\nu}\bar{W})$ remain
finite, and hence the homogeneous components $q_{xx}$ and $q_{yy}$ are finite
too. 

We complete the gauge fixing by enforcing the consistency equation $\dot{K}=1$
and solve it for the lapse function,
\begin{eqnarray}
    N = - \frac{\sec^2 (\bar{\nu} \bar{W})}{\lambda_0} 4 \sqrt{\varepsilon} \left(\frac{\bar{a}^2}{\cos^2 (\bar{\lambda} K)} |\cos (\bar{\lambda} K)| \frac{\sin^2 (\bar{\lambda} K)}{\bar{\lambda}^2} + 4 \mu^2 \beta^2\right)^{-1} \frac{\bar{a}^2}{\cos^2 (\bar{\lambda} K)} |\cos (\bar{\lambda} K)|
    \ ,
\end{eqnarray}
which is finite at the maximum-curvature hypersurface provided $\bar{W} \neq -\pi/(2 \bar{\nu})$.

Because of the divergence of (\ref{eq:a - Homogeneous - simple case - CML})
and its derivative (\ref{eq:a EoM - Homogeneous - CML}), the emergent line
element has a singular $\theta$-component at the maximum-curvature
hypersurface,  and its time-derivatives are
singular there too.
Thus, neglecting the time-derivatives of the $q_{xx}$ and $q_{yy}$ components,
a homogeneous line element of the form
\begin{eqnarray}
    {\rm d} s^2 = - N^2 {\rm d} T_K^2 + \tilde{q}_{\theta\theta} {\rm d} \theta^2 +
  q_{xx} {\rm d} x^2 + q_{yy} {\rm d} y^2 
    \ ,
\end{eqnarray}
has the Ricci scalar
\begin{eqnarray}
    R \approx - \frac{\dot{\tilde{q}}_{\theta \theta}}{\tilde{q}_{\theta \theta}} \frac{\dot{N}}{N^3}
    - \frac{1}{2 N^2} \left( \left(\frac{\dot{\tilde{q}}_{\theta \theta}}{\tilde{q}_{\theta \theta}}\right)^2 - 2 \frac{\Ddot{q}_{\theta \theta}}{q_{\theta \theta}} \right)
    \ ,
\end{eqnarray}
which diverges as $R \sim \sec^2 (\bar{\lambda} T_K)$ near the maximum-curvature hypersurface, while the Kretschmann scalar takes the form
\begin{eqnarray}
    {\cal K}
    \approx - \frac{R^2}{2 \tilde{q}_{\theta \theta} N^2}
    \ ,
\end{eqnarray}
which diverges as ${\cal K} \sim \sec^3 (\bar{\lambda} T_K)$ near the maximum-curvature hypersurface
This constraint, unlike the other two versions considered in this paper, therefore
implies a singular geometry at the maximum-curvature hypersurface.

%%%%%%%%%%%%%%%%%%%%%%%%
%%%%%%%%%%%%%%%%%%%%%%%%
%%%%%%%%%%%%%%%%%%%%%%%%

\section{Discussion}
\label{sec:Discussion}

We have extended emergent modified gravity from spherically symmetric models
to polarized Gowdy systems, preserving most of the qualitative features
observed in previous publications. In particular, modification functions of
the same number and type remain in the classes of modified constraints derived
explicitly here, building on a relationship with models of a scalar field
coupled to spherically symmetric gravity. Emergent modified gravity therefore
is not restricted to spherical symmetry, and it is compatible with different
kinds of local degrees of freedom from matter or gravity.

One class of models, compatible with the classical limit of the local
gravitational degree of freedom, has a set of modification functions such that
polarized gravitational waves travel on an emergent space-time geometry just
like a minimally coupled scalar field. The existence of these models shows
that a non-trivial class of theories in emergent modified gravity has
gravitational waves and matter (a minimally coupled massless scalar field
propagating on the same geometry) travelling at the same speed. Emergent
modified gravity is therefore compatible with strong observational
restrictions on the difference of the two speeds
\cite{MultiMess1,MultiMess2,MultiMess3,MultiMess4}. Moreover, emergent
modified gravity does not require higher time derivatives for non-trivial
modifications, and is therefore free of related instabilities
\cite{OstrogradskiProblem}.

Compared with spherically symmetric models, polarized Gowdy systems have a
large class of homogeneous solutions that correspond to the full Kasner
dynamics of the Bianchi I model. We have derived consistent modifications of
this dynamics with the correct classical limit at large volume but different
behaviors at small volume. Some types of modifications lead to non-singular
evolution connecting collapsing and expanding Kasner dynamics, while models
compatible with the classical limit for the local gravitational degree of
freedom retain the classical big-bang singularity. In the non-singular case,
all three spatial directions transition from collapse to expansion at the same
time. We demonstrated that the modified Kasner family may no longer include
Minkowski space-time, but that a different gauge choice not based on an
internal time nevertheless shows that this geometry remains a solution of the
modified theory. Discussions of possible vacuum states in a modified theory
therefore require access to different gauge choices and cannot be made
reliably in a deparameterized setting, as often used in quantum cosmology.

The restrictions on inhomogeneous terms in the covariant constraints, imposing
the covariance requirement on an emergent space-time metric distinct from the
basic phase-space variables, demonstrates the non-trivial nature of
modifications or quantizations of the polarized Gowdy model. In particular, a
separate modification or quantization of a homogeneous Bianchi model coupled
to linearized classical-type inhomogeneity, as proposed for instance in hybrid
loop quantum cosmology \cite{Hybrid,Hybrid2,Hybrid3}, does not lead to
covariant space-time solutions because it is not contained in the general
class of consistent models derived here. Modifications of the background
dynamics, one of the key ingredients in cosmological models of loop quantum
gravity, instead have to be reflected in coefficients of the inhomogeneous
terms and in the corresponding emergent line element, as determined by strong
covariance conditions. Midisuperspace quantizations of
polarized Gowdy and related models, as in
\cite{PolCylVol,EinsteinRosenQuant,PlaneWavesLoop,PlaneWavesLoop2,GowdyCov,GowdyAbel,GowdyComplex,PlaneWavesLoop3},
would have to take into account the new holonomy behavior found in
equation~(\ref{sinK}) in order to be compatible with a covariant semiclassical
limit. The dependence of the holonomies on anisotropies (rather than
areas or volumes as previously assumed in models of loop quantum gravity) then
implies new phenomenological behaviors.
These applications indicate that emergent modified gravity has important
implications for classical as well as quantum models of gravity.

\section*{Acknowledgements}

The authors thank Manuel D\'{\i}az and Aidan Kelly for discussions.
This work was supported in part by NSF grant PHY-2206591.

%\bibliographystyle{../preprint}
%\bibliography{../Bib/QuantGra.bib}

\end{document}